\documentclass[11pt,a4paper]{article}

\usepackage{jheppub}

\usepackage{epsfig}
\usepackage{amssymb}
\usepackage{amsfonts}
\usepackage{amsbsy}
\usepackage[all]{xy}
\usepackage{amsmath}
\usepackage{enumerate}

\usepackage{amssymb,amscd}
\usepackage{mathrsfs}
\usepackage{amsmath,amsthm}
\usepackage{xspace}
\usepackage{framed}

\def\mod{{\rm mod}}

\def\IC{\mathbb{C}}

\def\IH{\mathbb{H}}

\def\IP{\mathbb{P}}

\def\IR{{\mathbb{R}}}

\def\IS{{\mathbb{S}}}

\def\IZ{{\mathbb{Z}}}

\def\fg{\mathfrak{g}}

\def\fL{\mathfrak{L}}
\def\fV{\mathfrak{V}}
\def\fP{\mathfrak{P}}

\def\CC {{\cal C}}

\def\CN{{\cal N}}
\def\CK {{\cal K}}
\def\CN {{\cal N}}
\def\CR {{\cal R}}
\def\CD {{\cal D}}

\def\CW {{\cal W}}

\def\CY {{\cal Y}}

\def\cY {{\cal Y}}
\def\CO {{\cal O}}

\def\CE {{\cal E}}

\def\CB {{\cal B}}
\def\CS {{\cal S}}
\def\cS {{\cal S}}
\def\CA{{\cal A}}
\def\CK{{\cal K}}
\def\cK{{\cal K}}

\def\CT{{\cal T}}

\def\half{\frac{1}{2}}

\renewcommand{\Re}{{\rm Re }}

\def\one{{\hbox{ 1\kern-.8mm l}}}

\def\p{\partial}

\def\be{\bar{e}}

\def\half{\frac{1}{2}}

\def\hk{hyperk\"ahler\xspace}

\newcommand{\abs}[1]{\lvert#1\rvert}

\newcommand{\ti}[1]{\textit{#1}}

\def\fb{\mathfrak{b}}
\def\fg{\mathfrak{g}}

\def\fv{\mathfrak{v}}

\def\p{\partial}

\def\mod{{\rm mod}}

\def\IC{\mathbb{C}}

\def\IZ{{\mathbb{Z}}}
\def\IR{{\mathbb{R}}}
\def\IP{\mathbb{P}}

\def\CB {{\cal B}}
\def\CC {{\cal C}}

\def\CN{{\cal N}}
\def\CK {{\cal K}}
\def\CN {{\cal N}}
\def\CR {{\cal R}}
\def\CD {{\cal D}}

\def\CW {{\cal W}}

\def\CY {{\cal Y}}
\def\CO {{\cal O}}

\def\CE {{\cal E}}

\def\CB {{\cal B}}
\def\CS {{\cal S}}
\def\CA{{\cal A}}
\def\CK{{\cal K}}

\def\CT{{\cal T}}

\def\half{\frac{1}{2}}

\def\be{ \begin{equation} }
\def\ee{ \end{equation}}
\def\wnet {{\cal W}}

\def\Yin{\emph{Yin}\xspace}
\def\Yang{\emph{Yang}\xspace}
\def\ab{{\rm ab}}

\newcommand\I{{\mathrm i}}
\newcommand{\inprod}[1]{\langle#1\rangle}
\newcommand\N{{\cal N}}

\newcommand\WKB{{\mathrm{WKB}}}

\newcommand\de{{\mathrm{d}}}

\DeclareMathOperator{\End}{{End}}

\DeclareMathOperator{\Hom}{{Hom}}

\newcommand{\insfig}[2]{\begin{figure}[htbp] \centering \includegraphics[scale=0.3]{figures/#1-crop.pdf} \caption{#2} \label{fig:#1} \end{figure}}
\newcommand{\insfigscaled}[3]{\begin{figure}[htbp] \centering \includegraphics[scale=#2]{figures/#1-crop.pdf} \caption{#3} \label{fig:#1} \end{figure}}

\hypersetup{pageanchor=false}

\title{Spectral Networks and Snakes}

\author[1,2]{Davide Gaiotto,}
\author[2]{Gregory W. Moore}
\author[3]{and Andrew Neitzke}

\affiliation[1]
{School of Natural Sciences, Institute for Advanced Study, \\
Princeton, NJ 08540, USA}
\affiliation[2]
{Perimeter Institute for Theoretical Physics,\\
Waterloo, Ontario, Canada N2L 2Y5}
\affiliation[3]
{NHETC and Department of Physics and Astronomy, Rutgers University,\\
Piscataway, NJ 08855--0849, USA}
\affiliation[4]
{Department of Mathematics, University of Texas at Austin,\\
Austin, TX 78712, USA}

\emailAdd{dgaiotto@ias.edu}
\emailAdd{gmoore@physics.rutgers.edu}
\emailAdd{neitzke@math.utexas.edu}

\abstract{We apply and illustrate the techniques of spectral networks
in a large collection of $A_{K-1}$ theories of class $S$, which we call
``lifted $A_1$ theories.''  Our construction
makes contact with Fock and Goncharov's work on higher Teichm\"uller theory.
In particular we show that the Darboux coordinates on moduli
spaces of flat connections which come from certain special spectral networks
coincide with the Fock-Goncharov coordinates.  We show, moreover, how these techniques can
be used to study the BPS spectra of lifted $A_1$ theories.
In particular, we determine the spectrum generators for all
the lifts of a simple superconformal field theory.}

\begin{document}

\maketitle

\newpage

\hypersetup{pageanchor=true}

\section{Introduction and Summary}

In the past few years there has been a
renaissance in the subject of four-dimensional
field theory with $\CN=2$ supersymmetry and its
mathematical applications. In part this development
has been driven by many new and important insights
into the spectrum of BPS states in these theories.
Nevertheless, explicit examples of the many new
mathematical structures that are emerging have
tended to focus on theories of class $S$ of $A_1$
type. In an effort to understand more deeply
theories of higher rank (for example, $SU(K)$
$\CN=2$ super-Yang-Mills theories with $K>2$), reference
\cite{Gaiotto2012} introduced a combinatorial object
called a \emph{spectral network}.  One of the virtues
of spectral networks is that they provide an algorithmic
approach to describing the BPS degeneracies of
all theories of class $S$ of $A$ type.  Another virtue
is that they have potentially interesting mathematical
applications to the theory of character varieties,
moduli spaces of flat connections on Riemann surfaces,
Hitchin systems, and higher Teichm\"uller theory
(as considered in \cite{MR2233852}.)

Spectral networks are unfamiliar objects and
\cite{Gaiotto2012} provided only a few examples.
The present paper is a continuation of \cite{Gaiotto2012}.
The purposes of this paper are:

\begin{enumerate}

\item To exhibit a rich set of examples of $A_{K-1}$ theories
of class $S$ with $K>2$, where the associated spectral
networks are nontrivial but tractable.  We call these theories
``lifted $A_1$ theories.''

\item To make a direct
connection to the work of Fock and Goncharov on
higher Teichm\"uller theory \cite{MR2233852}.
As explained in \cite{Gaiotto2012}, spectral networks
facilitate the construction of new coordinate systems on
moduli spaces of flat connections.
Here we will explain how some special cases of
these coordinates agree precisely with the coordinates introduced by
Fock and Goncharov.

\item To study the BPS degeneracies in special
regions of the Coulomb branch of the lifted
$A_1$ theories.

\end{enumerate}

We will now give a brief summary of the paper.
We will assume the reader is familiar with
the basic definitions of spectral networks and will
not review the subject here.  See
\cite{Gaiotto2012} and references therein for background material.

In \S \ref{sec:LevelKLifts} we define the \emph{lifted $A_1$ theories of
class $S$.}  A lifted theory consists of a special $A_{K-1}$ theory of class $S$
together with a distinguished ``lift locus'' on its Coulomb branch.
The central idea is that while the spectral networks of $A_{K-1}$ theories
are in general exceedingly complicated objects, near the lift locus
they can be tamed and brought under good control, because in
this region they are closely related to the
much simpler spectral networks of the parent $A_1$ theory.
The latter networks are simply dual to ideal triangulations of the ultraviolet curve $C$
used to define the theory.

In \S \ref{sec:LiftN1AD} we focus on the lifts of the
``Argyres-Douglas series'' of $A_1$ theories.  This is a series of
superconformal $d=4$, $\CN=2$ theories $AD_N$, $N = 1, 2, 3, \dots$,
studied in \cite{Argyres:1995xn,Argyres:1995jj,Shapere:1999xr}.
These theories have Seiberg-Witten
curves $\lambda^2 = z^N (\de z)^2$ at their superconformal points (often written in
local coordinates as $y^2 = z^N$.)
The theory $AD_1$ is somewhat trivial
from the four-dimensional viewpoint (it has no Coulomb branch, no
IR gauge fields and no BPS spectrum) and for this reason has not really been discussed in the literature before;
however it does admit interesting surface defects.
Such a defect can be viewed as a purely two-dimensional Landau-Ginzburg
theory with cubic superpotential \cite{Gaiotto:2011tf}.  All of the $AD_N$
theories are $A_1$ theories of class $S$ \cite{Shapere:1999xr,Gaiotto:2009hg}.
The reader is  urged not to confuse ``$A_1$ theory'' and ``$AD_1$ theory''
in what follows:  the former refers to the whole class of
theories $S[A_1, C, D]$ while the latter refers to a particular
theory in that class.

In \S \ref{sec:LiftN1AD} we spell out the spectral curves and the normalizable,
mass, and non-normalizable deformations of the level $K$ lifts of $AD_N$ theories.  We also
count the number of parameters in the Stokes matrices
of the associated Hitchin system on $C = \IC \IP^1$ with irregular
singular point at infinity.  A crucial result is \eqref{eq:WriteCoords}
which gives the rank of the lattice $\Gamma$ of charges.

In \S \ref{sec:LevelKAD1} we narrow our focus yet further, to the level $K$ lifts
of the $AD_1$ theory.  These theories are already nontrivial and provide
the foundation for our approach to the more general lifted $A_1$ theories.
We begin by defining a very special class of \emph{minimal spectral
networks.} These come in two types, which we dub \Yin\ and \Yang. These minimal
spectral networks are the very simplest spectral networks associated to the
level $K$ lifts of $AD_1$ theories and, while somewhat elaborate, are nevertheless
tractable, as the remainder of the paper amply demonstrates.
Indeed, in \S \ref{sec:FlagsFlatSections} we use these minimal spectral networks
to examine some special coordinate systems on moduli spaces of solutions of
Hitchin equations associated to the level $K$ lifts of $AD_1$ theories.
This section is really the heart of the paper:
it makes direct contact between spectral networks and
the work of Fock and Goncharov on higher Teichm\"uller theory. The
main result is equation \eqref{eq:MainResultOne}, identifying our coordinates
on the Hitchin moduli space with those of Fock-Goncharov.
In the course of proving this result we find that certain combinatorial
objects introduced by Fock and Goncharov, especially ``$(K-1)$-triangles of lines''
and ``snakes,'' arise extremely naturally from the viewpoint of spectral networks.

For the convenience of the reader we have reviewed those constructions of
Fock and Goncharov which are essential to our present story in
Appendix \ref{app:LinearAlgebra}. Although this appendix is a review,
we do add a slightly different approach to
the construction of the Fock-Goncharov coordinates on moduli spaces of three
flags in a complex vector space.  The relation between the new and old
approaches is a bit subtle and involves the comparison between
flags in a vector space $V$ and the corresponding dual flags in $V^*$.
This is described in \S \ref{subsubsec:Coord-Relation} and
\S \ref{subsec:rrcheck}. The new approach, based on compositions of
canonical homs around closed loops, is better suited to matching with
spectral networks.

In \S \ref{sec:Spin1-N1B} we illustrate some of the general reasoning
of \S \ref{sec:FlagsFlatSections} in the first nontrivial case, $K=3$.

In \S \ref{sec:Vary-Theta} we move on to show how spectral networks
allow us to determine, in principle, the BPS spectrum of the lifted $AD_1$
theories in the special region of the Coulomb branch where essentially
minimal spectral networks exist. The key result is an algorithm,
given in \eqref{eq:explicit-sg}, for determining the ``spectrum generator''
of a lifted $AD_1$ theory.
(The spectrum generator
is a particular symplectomorphism of a torus, whose factorization
in terms of Kontsevich-Soibelman symplectomorphisms
encodes the BPS spectrum of the theory.)

\S\ref{sec:Amalgamation}, \S\ref{sec:GeneralPicture}, and \S\ref{sec:SpinLift-WCF}
are of a more programmatic and telegraphic nature than the rest of the paper.
In \S \ref{sec:Amalgamation} we
sketch how one can use our detailed understanding of the level $K$ lifts of the
$AD_1$ theory to construct Darboux coordinates for the more general
level $K$ lifts of $A_1$ theories.  Curiously, the essential geometrical gluing
operation corresponds to a procedure in cluster algebra theory known
as \emph{amalgamation} \cite{amalg}.  Similarly, \S \ref{sec:GeneralPicture}
indicates how our understanding of the level $K$ lifts of the $AD_1$ theory
can be used to construct at least part of the BPS spectrum of the more general
level $K$ lifts of $A_1$ theories, again in a special region of the Coulomb branch.
Finally \S \ref{sec:SpinLift-WCF} just scratches the surface of what must be a
very rich combinatorial story concerning the wall-crossing identities of the
the level $K$ lifts of the $A_1$ theories.

Clearly, much more can be said about this subject. In \S \ref{sec:OpenProblems}
we indicate a few open problems which seem to us to be interesting and
feasible.

\section{The level \texorpdfstring{$K$}{K} lifts of \texorpdfstring{$A_1$}{A1} theories}\label{sec:LevelKLifts}

This paper is based on a very simple and general idea.
Suppose we have a solution $(A,\varphi)$ of Hitchin's equations for a compact Lie group $G$ and a homomorphism
$\rho: G \to G'$. Then we can define a corresponding solution of Hitchin's equations for $G'$ just by applying
$\rho$ to the transition functions of the bundle and to $(A, \varphi)$.\footnote{This construction is also natural
in the framework of Higgs bundles:  the holomorphic bundle $E\to C$ is
associated to a holomorphic $G_{\IC}$-bundle $P$; using the analytic continuation of $\rho$
we can make a holomorphic $G_{\IC}'$ bundle $P'$, and the image of the Higgs field under $\rho$
gives the new Higgs field.}

This general technique allows one to embed a Hitchin
moduli space as a special locus in another
Hitchin moduli space.  In this paper we narrow the focus considerably
and apply this procedure to the homomorphism
$\rho: SU(2) \to SU(K)$ given by
the $K$-dimensional irreducible representation of $SU(2)$, for any integer $K \ge 2$.

Let us write the corresponding spectral curves.
These will be the Seiberg-Witten (SW) curves of a ``lifted theory of class $S$.''
The SW curve of an $A_1$ theory of class $S$ (i.e. the
spectral curve of an $A_1$ Hitchin system) takes the form
\begin{equation} \label{eq:a1-curve}
\det(\lambda 1_2 -\varphi) = \lambda^2 + \phi_2 = 0,
\end{equation}
where $\varphi$ is a 1-form valued (locally) in $2 \times 2$ traceless
matrices, i.e. in $sl(2,\IC)$.  Applying $\rho$ we obtain a new curve:
\begin{equation}
\det(\lambda 1_K - \rho(\varphi) ) = 0.
\end{equation}
We will call this new curve the \emph{level $K$ lift} of the original curve \eqref{eq:a1-curve}.

We can write the lifted curve explicitly in terms of the
quadratic differential $\phi_2$.  Suppose we diagonalize $\varphi$:
\begin{equation}
\varphi \sim \begin{pmatrix} \sqrt{-\phi_2} & 0 \\ 0 & - \sqrt{-\phi_2} \end{pmatrix}.
\end{equation}
In the $K$-dimensional representation the eigenvalues of $\rho(\varphi)$ are
\begin{equation}
(K-1) \sqrt{-\phi_2},\ (K-3)\sqrt{-\phi_2},\ \dots,\ (3-K) \sqrt{-\phi_2},(1-K) \sqrt{-\phi_2},
\end{equation}
and hence if $K$ is even the lifted curve has the form
\begin{equation}\label{eq:rep-k-ev}
(\lambda^2 + \phi_2)(\lambda^2 + 9 \phi_2)(\lambda^2 + 25 \phi_2) \cdots (\lambda^2 + (K-1)^2 \phi_2) = 0,
\end{equation}
while if $K$ is odd the lifted curve has the form
\begin{equation}\label{eq:rep-k-odd}
\lambda (\lambda^2 + 4\phi_2) (\lambda^2 + 16 \phi_2) \cdots (\lambda^2 + (K-1)^2 \phi_2) = 0.
\end{equation}
The first few polynomials are explicitly:
\begin{equation}\label{eq:explicit}
\begin{split}
K=2: \qquad & \qquad \lambda^2 + \phi_2, \\
K=3: \qquad & \qquad \lambda^3 + 4 \phi_2 \lambda, \\
K=4: \qquad & \qquad \lambda^4 + 10 \phi_2 \lambda^2 + 9 \phi_2^2, \\
K=5: \qquad & \qquad \lambda^5 + 20 \phi_2 \lambda^3 + 64   \phi_2^2 \lambda, \\
K=6: \qquad & \qquad \lambda^6 + 35 \phi_2 \lambda^4 + 259\phi_2^2 \lambda^2 + 225 \phi_2^3, \\
K=7: \qquad & \qquad \lambda^7 + 56 \phi_2 \lambda^5 + 784\phi_2^2 \lambda^3 + 2304 \phi_2^3  \lambda, \\
K=8: \qquad & \qquad \lambda^8 + 84 \phi_2 \lambda^6 + 1974 \phi_2^2 \lambda^4 + 12916 \phi_2^3 \lambda^2 + 11025\phi_2^4.
\end{split}
\end{equation}

Of course, the curves \eqref{eq:rep-k-ev} and \eqref{eq:rep-k-odd} are highly reducible.
A generic small perturbation makes the curve irreducible.  A second key idea
in this paper is that WKB spectral networks\footnote{We use the term ``WKB spectral network'' to refer to
the particular spectral networks $\wnet_\vartheta$ which were the object of study in most of
\cite{Gaiotto2012}; a WKB spectral network is
uniquely determined by a choice of a point of the Coulomb branch and a phase $\vartheta$.}
associated to such irreducible
curves behave ``at long distance'' (on $C$) like WKB spectral networks of the $A_1$
theory.  That is, the $\CS$-walls of the lifted spectral network
will --- for parametrically small perturbations ---
coalesce to $\CS$-walls of the original $A_1$ spectral network. Thus $A_1$ spectral
networks, which are well understood, serve as a springboard to understand
the considerably more intricate $A_{K-1}$ spectral networks.

For example, consider the level $3$ lift of the SW curve $\lambda^2 + \phi_2=0$
of an $A_1$ theory.  After perturbation, the lifted curve has the form
\begin{equation}
\lambda^3 + (4 \phi_2 + \delta \phi_2) \lambda + \delta \phi_3 = 0.
\end{equation}
For $\delta \phi_2$ and $\delta \phi_3$ which are ``small'' in an
appropriate sense, each of the branch points of the $A_1$ curve splits into
three closely separated branch points for the lifted curve.
Far away from these branch points, the $\CS$-walls of the lifted spectral network
are ``close'' to the $\CS$-walls of the $A_1$ spectral network.

Now what is the physics of this construction?  We began with an $A_1$ theory and
produced a lifted $A_{K-1}$ theory, at a rather special locus of the Coulomb branch.
Of course, the ultraviolet curve $C$ is the same for the original theory and the lifted one;
hence the coupling constants are also the same.
To understand precisely which theory we obtain, we must consider the behavior of
$\varphi$ near the punctures $z_n$.  For the original $A_1$ theory, around a regular puncture we have locally
\begin{equation}
\varphi \sim \frac{\de z}{z-z_n} \begin{pmatrix} m_n & 0 \\ 0 & -m_n \\ \end{pmatrix} + \cdots
\end{equation}
where $m_n$ is the mass parameter of the defect at the puncture $z_n$.
The level $K$ lift produces a puncture where $\varphi$ has a simple pole, with residue
\begin{equation}
{\rm Diag}\,\{(K-1) m_n, (K-3) m_n, \dots, (3-K) m_n, (1-K) m_n\}.
\end{equation}
As long as $m_n \neq 0$, these eigenvalues are distinct, so we obtain a \emph{full puncture}
in the lifted theory.

We close with some remarks:

\begin{enumerate}

\item So far we have described how $A_1$ theories with only regular singularities, i.e. conformal
$A_1$ theories, lift to conformal $A_{K-1}$ theories.
In a similar way, asymptotically free $A_1$ theories --- which arise when $\varphi$ has
higher-order poles at the punctures --- lift to asymptotically free $A_{K-1}$ theories.
The simplest case of lifting a higher-order pole is discussed in detail in
\S \ref{sec:FlagsFlatSections} below.

\item Our lift gives an embedding of a hyperk\"ahler manifold (the Coulomb branch
 of the $A_1$ theory reduced to three dimensions) into a higher-dimensional hyperk\"ahler
manifold (the Coulomb branch of the lifted theory reduced to three dimensions).
We will refer to the image of this embedding as the \emph{lift locus}.
As the vacuum of the lifted theory approaches the
lift locus, many BPS states become massless.  On general principles the normal bundle
to this locus is a hyperholomorphic vector bundle with
connection induced from the Levi-Civita connection on the
tubular neighborhood \cite{VerbitskyN.S.}.
It should be possible to investigate its geometry
by studying the quantum effects of these states
on the \hk\ metric in the limit that they become massless.

\item Suppose we begin with an $A_1$ theory at a generic point of
its Coulomb branch. Then all the branch points of $\Sigma \to C$
are simple, and hence in a sufficiently small
neighborhood of each branch point we have $\phi_2(z) \approx z (\de z)^2$
for an appropriate local coordinate $z$.  This is exactly the form of $\phi_2(z)$ in
the first superconformal Argyres-Douglas theory $AD_1$.  Therefore,
we can study WKB spectral networks of the lifted theory --- at least in
these neighborhoods --- by studying
WKB spectral networks of the level $K$
lifts of $AD_1$.

\item  We will argue later that
understanding the lifts of the $AD_N$ theories is key to
understanding the lift of general $A_1$ theories of class $S$.
The $AD_2$ theory has a spectral network involving two branch points,
and the level $K$ lift of this theory can be understood as
a result of ``gluing together'' two lifts of $AD_1$.
Similarly, the spectral network of $AD_N$ has $N$ branch points,
and the level $K$ lift of $AD_N$ can be viewed as ``gluing together''
$N$ level $K$ lifts of $AD_1$.

\item  The notion of ``gluing'' just mentioned can be made somewhat precise by
studying the Darboux coordinate systems $\{\CY_\gamma\}$
on moduli spaces of flat connections.  In this context the
geometrical gluing corresponds to Fock and Goncharov's \ti{amalgamation} procedure
on cluster algebras. See \S \ref{sec:Amalgamation} for further discussion.

\item The consistency of wall-crossing for the $AD_N$ theories is
guaranteed by the pentagon identity
 \cite{Gaiotto:2009hg}, which first appears in the $AD_3$ theory.
It follows that the level $K$ lifts of the
$AD_1$, $AD_2$, and $AD_3$ theories are of fundamental importance.
Most of this paper is devoted to understanding these cases.
\S\S \ref{sec:LevelKAD1}-\ref{sec:Vary-Theta} focuses on the level $K$ lift
of the $AD_1$ theory, \S \ref{sec:Amalgamation} focuses on the lift of the
$AD_2$ theory, and \S \ref{sec:SpinLift-WCF} briefly discusses the lifts of the
wall-crossing identities.

\item A lifting construction similar to ours appears in \cite{MR1419455}, embedding the centered $2$-monopole moduli space into the
centered $K$-monopole moduli space.

\end{enumerate}

\section{Coulomb branch and mass deformations of the level \texorpdfstring{$K$}{K} lift of \texorpdfstring{$AD_1$}{AD1}}\label{sec:LiftN1AD}

The spectral curve of the superconformal point of the $AD_N$ theory is
\begin{equation}\label{eq:N1eq}
\lambda^2 = z^N (\de z)^2.
\end{equation}
In this section we explain the Coulomb branch and mass deformations
of the level $K$ lift of this theory.
Readers uninterested in the details should at least note
the rank of the charge lattice \eqref{eq:RankGamma}
and its rewriting as \eqref{eq:WriteCoords}.
(An analysis similar to what we do here for some related theories has been
carried out in \cite{Xie:2012hs}.)

In organizing the perturbations it is very useful to introduce
$R$-charge (following a strategy first used in \cite{Argyres:1995xn}.)
The $1$-form $\lambda$ has mass dimension $1$ and
$R$-charge $1$.  Therefore,
from \eqref{eq:N1eq}, $z$ has $R$-charge and dimension $2/(N+2)$.
The perturbations of the lifted equations \eqref{eq:rep-k-ev}
and \eqref{eq:rep-k-odd} should not change the behavior
$\lambda \sim  \kappa z^{N/2} \de z$ at $z\to \infty$.  (Here $\kappa$ is
a nonzero constant. There is an exceptional case where the growth
is subleading to $z^{N/2}$, dealt with below.)  This puts
strong constraints on what kinds of perturbations
we can consider.

Let $P_0^{(K)}$ be the polynomials \eqref{eq:rep-k-ev}
for $K$ even and \eqref{eq:rep-k-odd} for $K$ odd, with
$\phi_2 = - z^N (\de z)^2$.
These have a Newton polygon with a line as a boundary:
namely, writing the equation as
\begin{equation}
P_0^{(K)} = \lambda^K + \sum_{p=2}^{K} \sum_{a\geq 0} c_{p,a}^{(0)} \lambda^{K-p} z^a (\de z)^{p} = 0,
\end{equation}
the nonzero terms have $2a = Np$.  Requiring that $\lambda \sim \kappa z^{N/2} \de z$ as $z \to \infty$
means we should only allow perturbations within the
Newton polygon:
that is, writing the spectral curve in the form
\begin{equation}
P_0^{(K)} + \sum_{p=0}^K \lambda^{K-p} \delta \phi_p = 0,
\end{equation}
we can only add perturbations of the form
\begin{equation}\label{eq:gen-phi}
\delta \phi_p = \sum_{a=0}^{\left[ \frac{Np}{2} \right]} \delta c_{p,a} z^a (\de z)^p.
\end{equation}
Let us verify that --- as guaranteed by Newton --- the
asymptotic growth of the roots $\lambda$ for $z\to \infty$
is not changed by the perturbations \eqref{eq:gen-phi}.  Let $P^{(K)}$ be the general polynomial obtained
by such perturbations:
\begin{equation}
P^{(k)} =  \lambda^K + \sum_{p=2}^K \sum_{a=0}^{\left[\frac{Np}{2}\right]} c_{p,a} z^a (\de z)^p \lambda^{K-p},
\end{equation}
with $c_{p,a} = c_{p,a}^{(0)} + \delta c_{p,a}$.
Let $\lambda_0 = \kappa z^{N/2} \de z$ be one of the roots of the
unperturbed equation.  If the $\delta c_{p,a}$ are small then we can
compute the perturbed $\lambda = \lambda_0 + \delta \lambda$ from
\begin{equation}
\frac{\p P_0^{(K)} }{\p \lambda} \delta \lambda + \sum_{p=2}^K \sum_{a=0}^{\left[\frac{Np}{2}\right]} \delta c_{p,a} z^a (\de z)^p \lambda_0^{K-p} = 0.
\end{equation}
Using this, we can easily confirm that, if $\lambda_0 = \kappa z^{N/2} \de z$ with $\kappa\not=0$,
then $\delta \lambda$ also grows at most like $z^{N/2} \de z$ as $z \to \infty$.\footnote{There
is one special case we should address.  If $K$ is odd, then one of the roots is $\lambda_0=0$. In this case
$\frac{\p P_0^{(K)} }{\p \lambda}\vert_{\lambda_0} \neq 0$ because there is a
constant term in $P_0^{(K)}$.  Likewise, the only terms which can contribute to $\delta\lambda$
are those with $p=K$. The most dangerous of these is the one with $a = \left[\frac{NK}{2} \right]$.
One easily checks that this give $\delta \lambda \sim z^{\left[N/2\right]} \de z$
as $z \to \infty$, so no root grows faster than $z^{N/2} \de z$, although the perturbation
of this special root might grow more slowly than $z^{N/2} \de z$.}

A useful interpretation of this constraint on the allowed perturbations is obtained by
considering the dimension of $c_{p,a}$. Recall that
$\phi_p$ has mass dimension $[\phi_p] = p$ (e.g. because it is a
Casimir made from $p$ powers of $\varphi$, and in 5D SYM $\varphi$ has mass dimension $1$),
and therefore $c_{p,a}$ has mass dimension
\begin{equation}
[c_{p,a}] = p - \frac{2}{N+2} (a+p) = \frac{Np-2a}{N+2} := D[p,a].
\end{equation}
Thus \eqref{eq:gen-phi} is the general perturbation whose coefficients have nonnegative
mass dimension,
\begin{equation}
 D[p,a] \ge 0.
\end{equation}

Now let us analyze normalizable and non-normalizable deformations. We consider a polynomial
$P^{(K)}$ of the above type and perturb it to
\begin{equation}
P^{(K)} + \epsilon_{p,a} z^a (\de z)^p \lambda^{K-p}.
\end{equation}
By the same analysis we just performed, the change in the
asymptotic behavior of the roots is
\begin{equation}
\frac{\p P^{(K)}}{\p \lambda} \delta \lambda + \epsilon_{p,a} z^a (\de z)^p \lambda^{K-p}=0.
\end{equation}
If $\lambda \sim z^{N/2} \de z$ at infinity then we find the
asymptotic behavior
\begin{equation}\label{eq:del-lam-est}
\delta \lambda \sim const \cdot z^{a + \frac{N}{2} - \frac{Np}{2}} \de z,
\end{equation}
where the constant depends on which root we perturb around.
Note that the power of $z$ can be written as $\frac{N}{2}  -  \frac{N+2}{2} D[p,a]$.
The perturbation is considered to be \emph{normalizable} if
\begin{equation}\label{eq:norm-cond}
\int_{\Sigma} \vert \delta \lambda \vert^2 < \infty.
\end{equation}
The only source of divergence is the integration on the sheets covering
the region $\vert z\vert \to \infty$. Therefore the potential
divergence of \eqref{eq:norm-cond} is controlled by the potential
divergence of the integral
\begin{equation}
\int^{\infty} \frac{\de \vert z \vert}{\vert z \vert^{(N+2) D[p,a] - (N+1)}}
\end{equation}
from the region at infinity.
and therefore we conclude that:\footnote{Note that if we use
the root growing like $z^{[N/2]} \de z$
and $N$ is odd then the estimate \eqref{eq:del-lam-est} is changed.  However,
a parameter is non-normalizable if for \emph{some} root
the integral \eqref{eq:norm-cond} is divergent.}

\begin{enumerate}

\item If $D[p,a]>1$ the perturbation is \emph{normalizable}.

\item If $D[p,a]=1$ the perturbation leads to a log-divergent integral.
These are called \emph{mass perturbations}.

\item If $D[p,a]<1$ the perturbation is \emph{non-normalizable}.

\end{enumerate}

Now we can count the allowed perturbations.  The number of pairs $(p,a)$
with $p=3,\dots, K$ and $a=0,1,\dots, \left[ \frac{Np}{2} \right] $
with $D[p,a]>1$ gives the number of normalizable deformations. These are
local coordinates in the normal bundle to the lift locus
in the Coulomb branch of the lifted $A_{K-1}$ theory.
Similarly, the pairs with $D[p,a]=1$ give the mass perturbations.

The mass perturbations are the ones with $a = \half N(p-1) - 1$
for $p=2, \dots, K$.  Since $a$ must be integral, in the case when
$N$ is odd we only have mass parameters for odd values of $p$.  Thus
the number of mass parameters is:
\begin{equation}\label{eq:MassDefs}
{\rm Mass}(N,K) =
\begin{cases}
K-1 & N= 0\ \mod\ 2, \\
\frac{(K-2)}{2} & N=1\ \mod\ 2,\,K = 0\ \mod\ 2, \\
\frac{(K-1)}{2} & N=1\ \mod\ 2,\,K = 1\ \mod\ 2.
\end{cases}
\end{equation}
Similarly, counting the number of normalizable parameters we find
\begin{equation}\label{eq:NormDefs}
{\rm Norm}(N,K) =
\begin{cases}
\frac{1}{4} N K (K-1) - ( K-1) & N = 0\ \mod\ 2, \\
\frac{1}{4}(N-1)K(K-1) + \frac{(K-2)^2}{4} & N =1\ \mod\ 2,\,K = 0\ \mod\ 2, \\
\frac{1}{4}(N-1)K(K-1) + \frac{(K-1)(K-3)}{4} & N =1\ \mod\ 2,\,K = 1\ \mod\ 2.
\end{cases}
\end{equation}

In the $A_{K-1}$ theory there is an IR charge lattice $\Gamma$.
$\Gamma$ should have an electric and magnetic charge generator for each independent
normalizable deformation, and a flavor charge generator for each independent
mass deformation.  We therefore expect that the rank of $\Gamma$ will be
\begin{equation}
{\rm rank}(\Gamma) = 2\,{\rm Norm}(N,K) + {\rm Mass}(N,K).
\end{equation}
Using \eqref{eq:MassDefs} and \eqref{eq:NormDefs} this works out to
\begin{equation}\label{eq:RankGamma}
{\rm rank}(\Gamma) = \left(\frac{1}{2} NK-1\right)(K-1),
\end{equation}
in all cases.

Now, the $AD_N$ theory is associated with a Hitchin system on $\IC\IP^1$, of rank $K=2$, with
an irregular singularity at $z = \infty$ supporting $N+2$
Stokes sectors, as described in \cite{Gaiotto:2009hg}.  In particular, each
$\CS$-wall of a WKB spectral network in this theory is asymptotic to one of the Stokes rays
as $z \to \infty$.  As we have explained, the level $K$ lift does not change
the number of Stokes sectors at infinity:  the $\CS$-walls of the lifted network
asymptote to the same Stokes rays as those of the original network.
We now check that the result \eqref{eq:RankGamma} is indeed as expected for a
rank $K$ Hitchin system on the complex plane with $N+2$ Stokes sectors at infinity.

Viewing the Hitchin system as a moduli space of flat connections,
the moduli can be encoded in a set of
$K\times K$ Stokes matrices $S_n$, $n=1, \dots, N+2$, cyclically ordered.  The $S_n$
alternate between upper and lower triangular, with $1$ on the diagonal.  We must discuss the
cases $N$ even and odd separately.

For $N$ even the product of the
Stokes matrices is the formal monodromy at infinity,
which we diagonalize:
\begin{equation}\label{eq:even-mod}
\prod_{n=1}^{N+2} S_n = {\rm Diag}\{ \mu_1, \dots, \mu_K \}.
\end{equation}
This formal monodromy encodes the mass parameters. Since $\prod_{i=1}^K \mu_i = 1$
there are $(K-1)$ independent mass parameters in agreement with \eqref{eq:MassDefs}.
Having diagonalized the formal monodromy most of the gauge freedom has been fixed.
We may still conjugate by diagonal matrices in $SL(K,\IC)$. Thus we expect a
moduli space with complex dimension
\begin{equation}\label{eq:checkeven}
(N+2)\cdot \frac{K(K-1)}{2} - (K^2-1) - (K-1),
\end{equation}
and the reader can check that this nicely coincides with
$2\,{\rm Norm}(N,K)$ in the case $N$ is even.

Now let us consider the case when $N$ is odd. In this case there is
branching at infinity.  For the $AD_N$ theories  with
$N$ odd the monodromy exchanges the two branches. How should we
lift this permutation? We can again use the homomorphism $\rho$.
The formal monodromy for the $AD_N$ theory with $N$ odd is
\begin{equation}
\begin{pmatrix} 0 & 1 \\  -1 & 0 \\ \end{pmatrix}.
\end{equation}
The level $K$ lift of this group element is an antidiagonal
matrix of alternating $+1$ and $-1$'s:
\begin{equation}\label{eq:W0-DEF}
W_0^{(K)} := e_{K,1} - e_{K-1,2} + e_{K-2,3} \pm \cdots  + (-1)^{K-1} e_{1,K}
\end{equation}
Therefore, the branching at
infinity is given by the permutation
\begin{equation}\label{eq:w0}
w_0^{(K)}:  \{1,2,\dots,K-1,  K\} \mapsto \{ K, K-1, \dots , 2 ,1\}.
\end{equation}
This permutation will play an important role in our analysis
and similarly the matrix $W_0^{(K)}$  will be important in subsequent sections.
When $K$ is understood we drop the superscript and just write $W_0$.
Note that $(W_0^{(K)})^2 = (-1)^{K-1}$.

From this discussion follows that the monodromy constraint
along the lift locus is
\begin{equation}
\prod_{n=1}^{N+2} S_n = W_0^{(K)}.
\end{equation}
When counting deformation parameters it is more useful to enlarge our scope of
theories by turning on mass parameters. The monodromy constraint
for this broader class of theories is then of the form
\begin{equation}\label{eq:N-Odd-Monodromy}
\prod_{n=1}^{N+2} S_n = {\rm AntiDiag}\{ \mu_1, \dots, \mu_K \}.
\end{equation}
The remaining gauge symmetry in \eqref{eq:N-Odd-Monodromy} is conjugation by diagonal $SL(K,\IC)$
transformations.  We can usefully fix it further
by gauging so that the masses are $w_0$-invariant, i.e.
\begin{equation}
 \mu_1 = \mu_K, \ \mu_2 = \mu_{K-1}, \dots
\end{equation}
This gives $(K-2)/2$ mass parameters for $K$ even and
$(K-1)/2$ for $K$ odd, in agreement with \eqref{eq:MassDefs}.
Having chosen a generic set of $w_0$-invariant mass parameters
there are still $(K-2)/2$ independent diagonal gauge degrees of freedom
for $K$ even and $(K-1)/2$ independent diagonal gauge degrees of
freedom for $K$ odd. Thus we expect that the dimension of the moduli space
for $N$ odd is
\begin{equation}\label{eq:checkodd}
2\,{\rm Norm}(N,K) =
(N+2) \cdot \frac{K(K-1)}{2} - (K^2-1) - \begin{cases} \frac{K-2}{2} & K = 0\,\mod\,2 \\ \frac{K-1}{2} & K = 1\,\mod\, 2 \\
\end{cases}
\end{equation}
which indeed agrees with \eqref{eq:NormDefs}.

Finally, let us briefly consider the expected structure of the
``Darboux coordinates'' in this case. These are locally-defined functions $\CY_\gamma$
on the Hitchin moduli spaces, discussed extensively in \cite{Gaiotto:2009hg,Gaiotto:2010be,Gaiotto2012}.
We rewrite the rank of
$\Gamma$, \eqref{eq:RankGamma}, as
\begin{equation}\label{eq:WriteCoords}
{\rm rank}(\Gamma)
= N \frac{(K-1)(K-2)}{2} + (N-1) (K-1).
\end{equation}
So this is the number of independent coordinates we expect.
As we have stressed, for small $\delta c_{p,a}$, the WKB spectral
network of the lifted theory approximates that of the original $A_1$
theory.  In particular, the WKB spectral network of the $AD_N$ theory (considered as
a particular example of an $A_1$ theory)
determines an ideal triangulation of the disk with $N+2$ vertices.  These vertices
represent the asymptotic directions of the $\CS$-walls on the plane.  There are
$N$ triangles in this triangulation, and $N-1$ internal edges.
Now, recall that for $A_1$ theories each internal
edge of the triangulation is associated with one independent coordinate $\CY_{\gamma}$,
and there are no coordinates associated with triangles; so in the original $A_1$ theory
there are a total of $N-1$ coordinates.
The result \eqref{eq:WriteCoords} suggests how this statement extends to the
level $K$ lift:  each triangle will contribute $\frac{(K-1)(K-2)}{2}$
independent ``triangle coordinates,'' and each internal
edge will contribute $(K-1)$ independent ``edge coordinates.''
In the following sections we will verify that this is indeed the case.

\section{Spectral networks for the level \texorpdfstring{$K$}{K} lift of \texorpdfstring{$AD_1$}{AD1}}\label{sec:LevelKAD1}

In this section we make use of $m$-triangles.
This is a concept introduced and used extensively
in the work of Fock and Goncharov on higher Teichm\"uller theory
\cite{MR2233852}.  See Appendix \ref{app:m-triangles} for $m$-triangles, and Appendix
\ref{app:LinearAlgebra} for important background on some associated linear algebra.

\subsection{Definition of minimal spectral networks}\label{subsec:Def-Minimal-SWALL}

Our first order of business is to define \emph{minimal spectral networks}
of the level $K$ lift of the $AD_1$ theory.  Recall that the $AD_1$ theory has spectral curve
\begin{equation}
\lambda^2 = z (\de z)^2,
\end{equation}
or equivalently $\phi_2 = - z (\de z)^2$.
The lifted polynomial is order $K$ in $\lambda$, as are its
perturbations.  Its discriminant therefore is a polynomial
in $z$ of degree $\half K(K-1)$, and hence for generic
perturbations there are $\half K(K-1)$ simple branch points.

In general, the corresponding spectral networks are very complex.
Still, there are a few general things we can say.
Three $\CS$-walls emerge from each branch point. At large
$\abs{z}$ each of them asymptotes to one of the three $\CS$-walls of the
$AD_1$ theory. We will refer to each
collection of $\CS$-walls asymptoting to a single $AD_1$ $\CS$-wall as
a \emph{cable}.  The lift of $AD_1$ thus has three cables, each consisting
of at least $\half K(K-1)$ $\CS$-walls; in general each cable contains
more than $\half K(K-1)$ $\CS$-walls, because there can be secondary
$\CS$-walls emanating from joints.

\insfigscaled{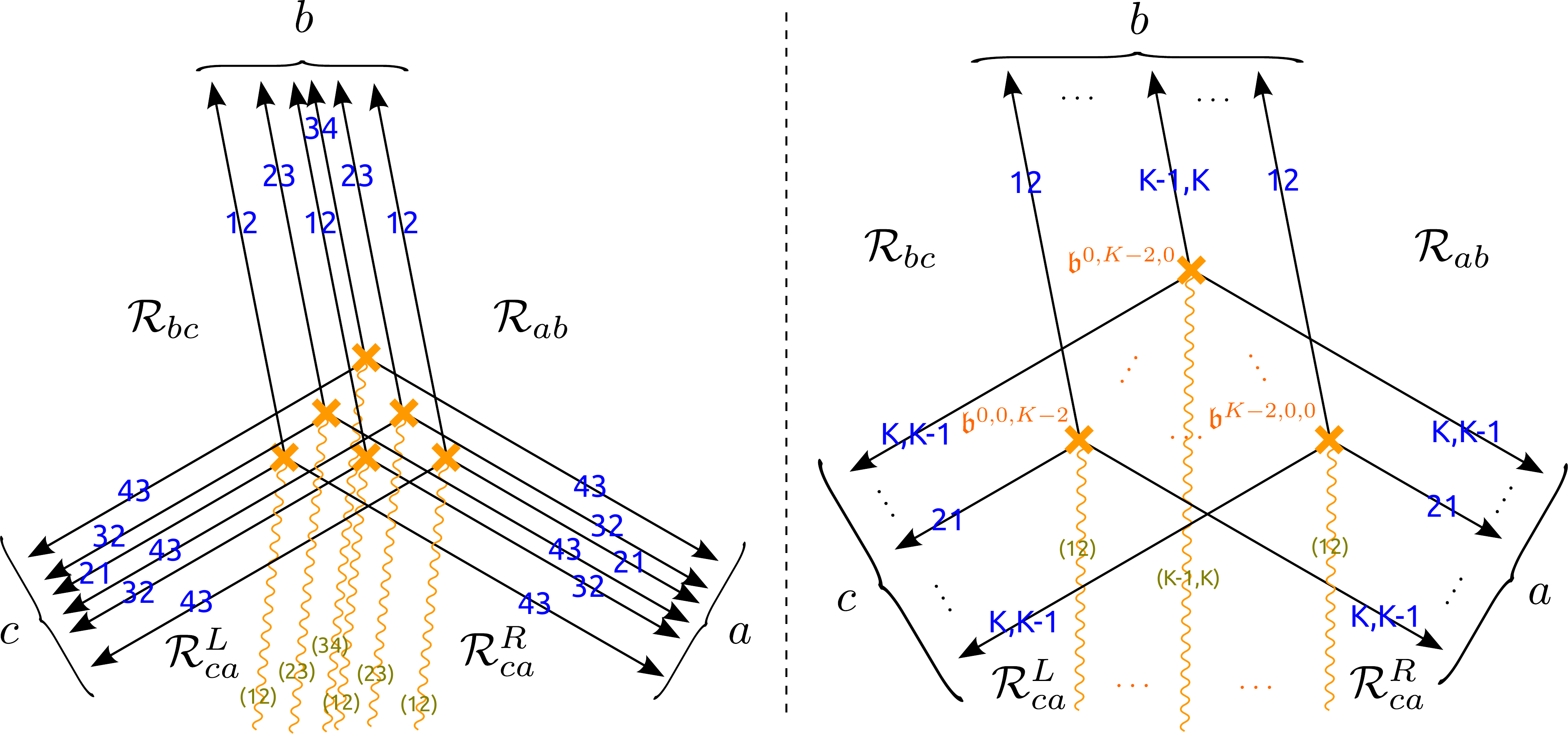}{0.36}{Minimal spectral networks of \emph{Yin} type, for $K=4$ (left)
and general $K$ (right).}

\insfigscaled{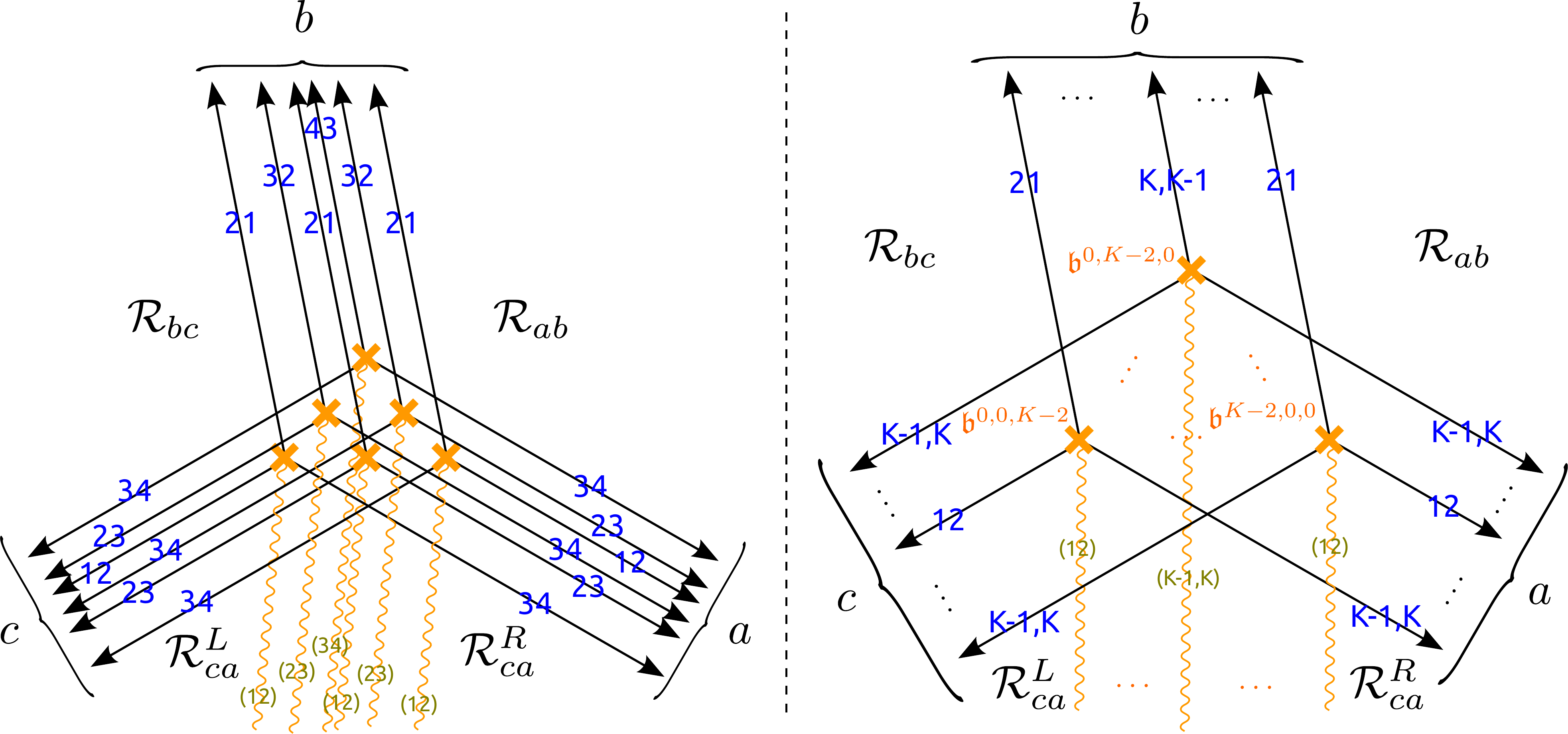}{0.36}{Minimal spectral networks of \emph{Yang} type, for $K=4$ (left)
and general $K$ (right).}

We now define a \emph{minimal level $K$ spectral network}.
In fact we will need to distinguish between two basic types,
which we call minimal spectral networks of \emph{Yin}-type and \emph{Yang}-type.
See Figures \ref{fig:min-sn-yin} and \ref{fig:min-sn-yang}.
A minimal spectral network is a spectral
network of the level $K$ lift of the $AD_1$ theory,
with the following properties:

\begin{enumerate}

\item Each of the three cables consists of precisely $\half K (K-1)$
$\CS$-walls.

\item Each branch point is associated with a point in a $(K-2)$-triangle.
We label the branch points $\fb^{x,y,z}$, where $x,y,z$ are nonnegative
integers with $x+y+z=K-2$.

\item The branch points sit in an
approximate triangular array in the $z$-plane, with $\fb^{0,K-2,0}$ the
topmost vertex, $\fb^{0,0,K-2}$ the bottom left vertex and $\fb^{K-2,0,0}$ the
bottom right vertex. The three cables are labeled $a,b,c$ with $b$ pointing
(roughly) to the north, $a$ to the southeast and $c$ to the southwest. There are
consequently three ``large'' open regions near $z=\infty$. We call the region
between the $a$- and $b$-cables $\CR_{ab}$, lying in the northeast.
The region between the $b$- and $c$-cables is denoted
$\CR_{bc}$, and lies in the northwest.  Finally, $\CR_{ca}$ lies in the south.

\item There exists a trivialization of the branched cover
$\Sigma \to \IC$ (i.e. choice of branch cuts)
such that each branch point is of type $(i,i+1)$ for some $i$.
In this trivialization
$\fb^{x,y,z}$ is a branch point of type $(y+1,y+2)$, where
$0\leq y \leq K-2$.
Thus, the top vertex is of type $(K-1,K)$, the next row has two branch points
of type $(K-2,K-1)$, the next row has three branch points of type $(K-3,K-2)$,
and so on until the bottom row has $K-1$ branch points of type $(1,2)$.
The cuts point down and slightly to the southwest, and
asymptote to a single line, also pointing down and slightly to the southwest.
The cuts thus divide the region $\CR_{ca}$ into western and eastern halves.
Moving from the southwest to the southeast, we encounter cuts of
type
\begin{equation}
\omega_{K-1} \omega_{K-2} \cdots \omega_2 \omega_1,
\end{equation}
where
\begin{equation}
\begin{split}
\omega_{K-1} & = (1,2) (2,3) \cdots (K-1,K), \\
\omega_{K-2} & = (1,2) (2,3) \cdots (K-2,K-1), \\
\vdots &  \\
\omega_{2} & = (1,2)(2,3), \\
\omega_{1} & = (1,2). \\
\end{split}
\end{equation}
Reading these words from left to right describes the cuts encountered
as one moves on the $z$-plane from left to right, i.e. counterclockwise
relative to an origin inside the triangle.
(See the cuts in
Figures \ref{fig:min-sn-yin} and \ref{fig:min-sn-yang}.)

\item The net transformation of sheets across all the cuts is the permutation
$w_0^{(K)}$, which
takes $(1, 2, \dots, K-1, K) \to (K, K-1, \dots, 2, 1)$.  Equivalently it is the
product of $[K/2]$ transpositions:
\begin{equation}
w_0^{(K)} =
\begin{cases} (1,K)(2,K-1)\cdots (\frac{K}{2}, \frac{K}{2}+1) & K = 0\,\mod\,2, \\
(1,K)(2,K-1)\cdots (\frac{K-1}{2}, \frac{K+3}{2}) & K = 1\,\mod\,2. \\
\end{cases}
\end{equation}

\item There are two essentially different kinds of minimal spectral networks,
which we call \emph{Yin}-type and \emph{Yang}-type. In the
\emph{Yin}-type, in each of the three cables there is a unique $\CS$-wall
either of type $K-1,K$ or of type $21$.  In the \emph{Yang}-type,
each cable contains a unique $\CS$-wall of type $K,K-1$ or of
type $12$.  See Figures \ref{fig:min-sn-yin} and \ref{fig:min-sn-yang}.
In the following we describe in detail the structure of the cables for
the \emph{Yin}-type spectral network.  The \emph{Yang}-type is obtained
by ``transposing'' all the $\CS$-walls $ij \to ji$.

\item With the same trivialization of $\Sigma \to C$
used above, $\fb^{x,y,z}$ is associated with an $\CS$-wall in the $b$-cable, of type $K-y-1, K-y$.
One can attach to each wall of type $ij$ an ``$\cS$-factor'' of type $ij$, which concretely means
a matrix whose only nonzero off-diagonal entry is in the $ij$ position.
We abuse notation by representing any such matrix as $S_{ij}$.
(A precise definition of these matrices can be found in
\eqref{eq:SN-BC-2} below.)  Then the product of $\cS$-factors
which one encounters in crossing the $b$-cable is
\begin{equation}\label{eq:b-cable-factors-1}
T_{b-cable} = \beta_1\beta_2 \cdots \beta_{K-2} \beta_{K-1} ,
\end{equation}
where
\begin{equation}\label{eq:b-cable-factors-2}
\begin{split}
\beta_{K-1} & = S_{K-1,K} S_{K-2,K-1} S_{K-3,K-2} \cdots S_{2,3} S_{1,2}, \\
\beta_{K-2} & = S_{K-2,K-1} S_{K-3,K-2}   \cdots S_{2,3}S_{1,2}, \\
\beta_{K-3} & = S_{K-3,K-2} \cdots S_{2,3} S_{1,2}, \\
 \vdots & \\
\beta_{2} & =  S_{2,3}S_{1,2}, \\
\beta_{1} & = S_{1,2}. \\
\end{split}
\end{equation}
Reading this product from right to left gives the $\cS$-factors encountered in moving
from $\CR_{ab}$ to $\CR_{bc}$ at large $\vert z \vert$
from right to left, that is, counterclockwise.

\item The $\cS$-walls for the $a$-cable are
\begin{equation}\label{eq:Ta-cable}
T_{a-cable} = \alpha_{1} \alpha_{2} \cdots \alpha_{K-2} \alpha_{K-1},
\end{equation}
\begin{equation}
\begin{split}
\alpha_{K-1} & = S_{K,K-1}, \\
\alpha_{K-2} & =S_{K,K-1} S_{K-1,K-2} , \\
 \vdots & \\
\alpha_{2} & =S_{K,K-1}S_{K-1,K-2} \cdots  S_{3,2}, \\
\alpha_{1} & = S_{K,K-1} S_{K-1,K-2}\cdots S_{3,2} S_{2,1}. \\
\end{split}
\end{equation}
Reading the product \eqref{eq:Ta-cable} from right to left gives
the $\cS$-factors encountered as one moves counterclockwise from the
region $\CR_{ca}^R$ (to the right of the cuts) to $\CR_{ab}$.

\item The $\cS$-walls for the $c$-cable are
\begin{equation}\label{eq:Tc-cable}
T_{c-cable} =  \gamma_1 \gamma_2 \cdots  \gamma_{K-2} \gamma_{K-1},
\end{equation}
\begin{equation}
\begin{split}
\gamma_{K-1} & =S_{2,1}S_{3,2}  \cdots  S_{K-1,K-2} S_{K,K-1}, \\
\gamma_{K-2} & =  S_{3,2} \cdots S_{K,K-1}, \\
 \vdots & \\
\gamma_{2} & =S_{K-1,K-2} S_{K,K-1}, \\
\gamma_{1} & = S_{K,K-1}. \\
\end{split}
\end{equation}
Here again, reading the product from right to left gives
the pattern of $\cS$-walls encountered as one moves
counterclockwise from region $\CR_{bc}$ to $\CR_{ca}^L$
(to the left of the cuts).

\end{enumerate}

We would like to make a number of remarks about the minimal spectral network.

\begin{enumerate}

\item In a minimal spectral network there are many intersections
of $\CS$-walls.  As explained in \cite{Gaiotto2012}, the rules of spectral
networks demand that intersections of $\CS$-walls in general generate new
$\CS$-walls. However, in the minimal spectral network the intersections are always
such that no new $\CS$-walls are produced.
This is the reason this network is called ``minimal.''

\item  In accounting for the configuration of
$\CS$-wall factors in the $a$- and $c$-cables it is crucial to take into account
the branch cuts.  This is evident e.g. in
Figures \ref{fig:min-sn-yin} and \ref{fig:min-sn-yang}.

\item In Section \ref{subsec:SpecialCurves} we give some evidence, for small
values of $K$, that there are indeed branched covers $\Sigma\to C$ which are
small perturbations of the level $K$ lift, and for which the corresponding WKB spectral network is a
minimal spectral network.  Of course, the minimal level $K$
spectral networks are perfectly legitimate spectral networks in the
sense of \cite{Gaiotto2012}, irrespective of whether they arise
as WKB spectral networks or not.

\item There are $4=2\times 2$ possible definitions we could have used for ``minimal spectral network.''
First, there is a two-fold choice of whether to have $(K-1)$ branch points of type
$(12)$ and one branch point of type $(K-1,K)$ or to have $(K-1)$ branch points
of type $(K-1,K)$ and one of type $(12)$.  This choice is only a convention:
one can reverse it just by relabeling the sheets.
There is another two-fold choice of
whether the cable which is not adjacent to the cuts involves $\CS$-walls of type ${i,i+1}$ or ${i+1,i}$.
This choice is \ti{not} a convention:  reversing it exchanges the \Yin\ and \Yang\ type
minimal spectral networks, which are really different from one another.

\item There are $8=2^3$ possible types of cable products,
such as $T_{a-cable}, T_{b-cable}, T_{c-cable}$ above, which we could encounter.  Let us denote factors
such as $\beta_{K-1}, \dots, \beta_{1}$
in \eqref{eq:b-cable-factors-1} as ``strings'' and let us define their ``length'' to be
the number of $\CS$-factors in $\beta_i$.  Then the $8$ possibilities arise as follows:
First, the strings can be growing or shrinking in length as $i$ increases.  Second, the $\CS$-factors
can be of type $S_{i,i+1}$ or $S_{i+1,i}$.
Third, the string of length $1$ can be of type $(1,2)$ or $(K-1,K)$.

\end{enumerate}

\subsection{Realizing minimal spectral networks as WKB spectral networks}\label{subsec:SpecialCurves}

Much of this paper depends only on the existence of minimal spectral networks
as spectral networks in the sense of \S 9 of \cite{Gaiotto2012}.
Nevertheless it is interesting to ask whether this spectral
network actually arises as a WKB spectral network at some point of the Coulomb branch of the
lifted theory.  This will be important if,
for example, we want to use minimal spectral networks
to study the BPS spectrum.

More precisely, we do not really care whether our networks are strictly minimal:
for all of our purposes in this paper
it will be enough to consider networks which are \ti{equivalent} to a minimal spectral network, in the sense of
\cite{Gaiotto2012}.  We call such networks \ti{essentially minimal}.

We have no general proof that there are such points in the Coulomb branch.  However, we have found
explicitly that they exist for all $K \le 5$.
The WKB spectral networks in question are illustrated in Figure \ref{fig:wkb-minimal-45} for $K=4$, $K=5$.

\insfigscaled{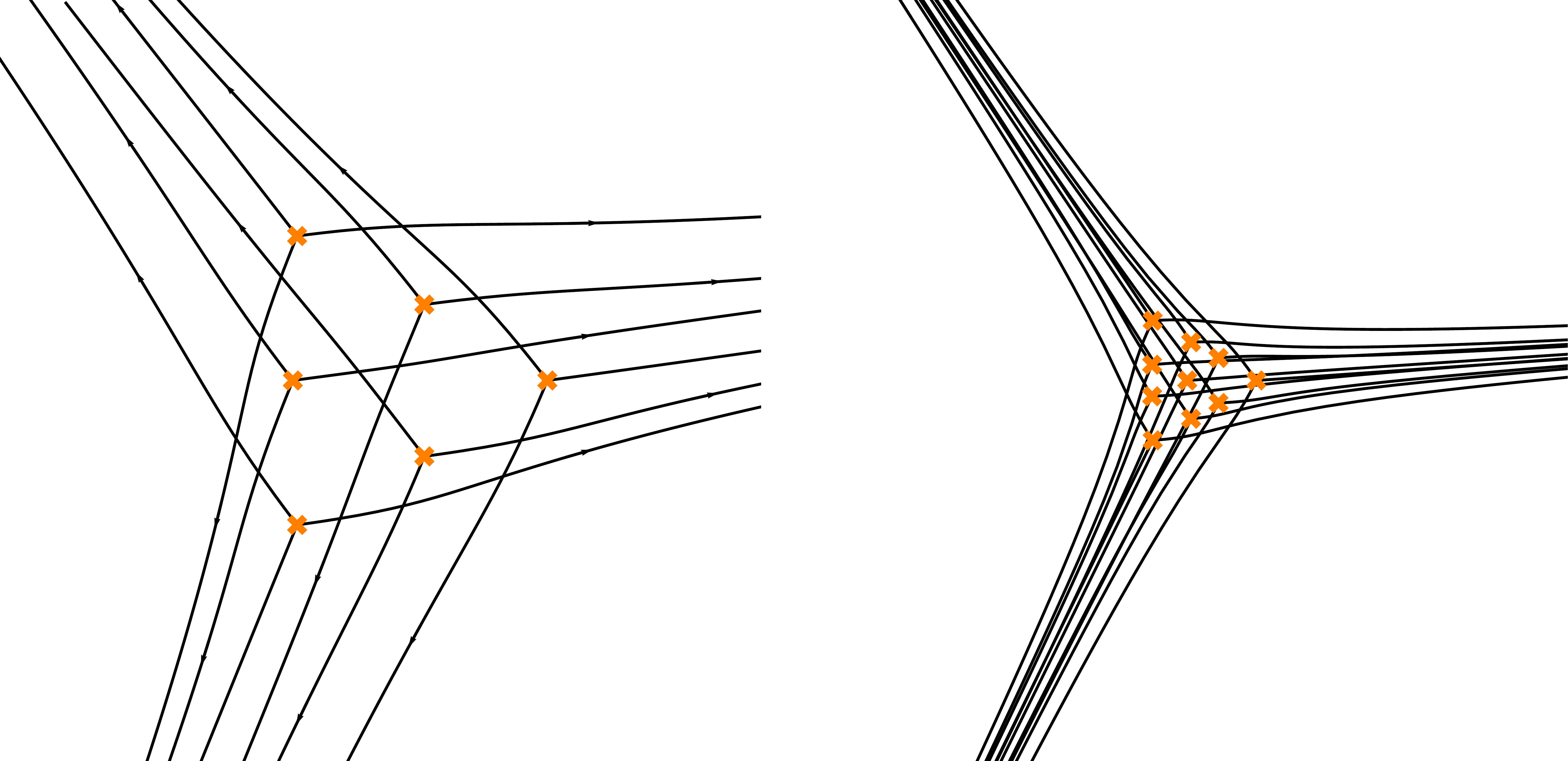}{0.23}{Two essentially minimal WKB spectral networks.
{\bf Left:}  the WKB spectral network for the $K=4$ lift of $AD_1$ theory, with
SW curve $\lambda^4 - 10 z (\de z)^2 \lambda^2 + 4 (\de z)^3 \lambda + 9z^2 (\de z)^4 = 0$,
and $\vartheta = 0.2$.  {\bf Right:}  the WKB spectral network for the $K=5$ lift of $AD_1$ theory, with
SW curve $\lambda^5 - 20 z (\de z)^2 \lambda^3 + (\de z)^3 \lambda^2 + 64z^2 (\de z)^4 \lambda - 2z (\de z)^5 = 0$,
and $\vartheta = 0.1$.}

We have not gotten beyond $K=5$ with this approach;
however, to go further, we can also generalize the lift locus to the more general family of SW curves
\begin{equation}\label{eq:larger-class}
(\lambda^2 + c_1^2 \phi_2) (\lambda^2 + c_2^2 \phi_2) \cdots (\lambda^2 + c_{K/2}^2 \phi_2)
\end{equation}
where $c_i\in \IR$ and $K$ is even, with a similar generalization for $K$ odd.
(Taking $c_i = 2i-1$ we recover the lift locus.)
By considering small perturbations of this family we have realized essentially minimal spectral networks
as WKB spectral networks for $K \le 9$.
Nevertheless, it remains an open problem to find a systematic
procedure for finding minimal or essentially minimal
spectral networks as WKB spectral networks.

\subsection{Cycles on the SW curve associated to a minimal spectral network}\label{subsec:cycles-MSN}

We now give a basis for the lattice $\Gamma$ associated to the
SW cover $\Sigma \to C$ associated with a minimal spectral network.

The branch points of the cover are in 1-1 correspondence with the points of a
$(K-2)$-triangle.  Let us consider the upwards-pointing triangles in this
$(K-2)$-triangle.  These are in 1-1 correspondence with the points of a
$(K-3)$-triangle. If $(x,y,z)$ is a point in the $(K-3)$-triangle,
the three branch points at the vertices of the corresponding shaded triangle are
of the form $\fb^{x,y+1,z}$, $\fb^{x+1,y,z}$, and $\fb^{x,y,z+1}$
where $x+y+z = K-3$. Setting $i = K-1-y$, the branch point at the top vertex is of type $(i-1,i)$,
while the branch points at the bottom left and right
vertices are of type $(i,i+1)$.  Note that the monodromy of the cover
$\Sigma\to C$ around a small path $\wp$ looping counterclockwise
once around these three branch points is the transposition $(i-1,i+1)$.
Therefore, the closed path $\wp$ lifts to a \emph{closed} path $\wp^{(i)}$
on the $i^{th}$ sheet of $\Sigma$.  We denote the homology class
of this path $\gamma^{\fv}$ where $\fv=(x,y,z)$.
There are $\half (K-2)(K-1)$ such cycles, which
can be shown to be linearly independent.\footnote{One proof makes use of the result
\eqref{eq:MainResultOne} below and the fact that the $r^\fv$
are an independent set of coordinates for the moduli space
of three flags in a $K$-dimensional vector space. It would
be nice to have a more direct geometrical argument that the
cycles are linearly independent on the covering surface $\Sigma$.}
Thus, given equation \eqref{eq:WriteCoords} (taking $N=1$),
they form a basis for $\Gamma$ (at least rationally).  In fact, there
is a nice formula for the intersection matrix of the
basis of cycles $\gamma^\fv$:
\begin{equation}\label{eq:Triangle-Intersections}
\langle \gamma^{\fv_1}, \gamma^{\fv_2} \rangle =
\begin{cases} +1 & [\fv_1,\fv_2] \quad {\rm  oriented\  edge} \\
-1 &  [\fv_1,\fv_2] \quad {\rm  anti\ oriented\ edge } \\
0 & {\rm otherwise} \\
\end{cases}
\end{equation}
where ``edge'' refers to edges of the $(K-3)$-triangle of cycles,
oriented counterclockwise.

We will also need basepoints for the relative homology torsors
$\Gamma(z,z)$ and $\Gamma(z_1,z_2)$. (For definitions of these
torsors see \cite{Gaiotto2012}, \S\S 3.2 and 3.3.) These basepoints will
depend on a choice of a cable. For each
branch point $\fb$ and each $z$ on a (primary) $\CS$-wall emerging
from $\fb$, we let $\gamma^{\fb}_{i,i+1} \in \Gamma(z,z)$ denote the
open homology class on $\Sigma$  which projects to the $\CS$-wall
and begins at $z^{(i)}$ (the preimage of $z$ on the $i^{th}$ sheet),
goes down to the ramification point over $\fb$,
and comes back on the $(i+1)^{th}$ sheet to end at the preimage $z^{(i+1)}$.
Similarly define $\gamma^{\fb}_{i+1,i}\in \Gamma(z,z)$.
The class $\gamma^{\fb}_{i,i+1}$ serves as a convenient
basepoint for $\Gamma_{i,i+1}(z,z)$ as a $\Gamma$-torsor, when $z$ is on or near the $\CS$-wall.
Of course there are three $\CS$-walls emerging from $\fb$; when we need to distinguish them, we
denote the basepoints on, say, the $a$-cable by $\gamma^{\fb,a}_{ij}$.

Now, suppose that $\fb$ and $\fb'$ are neighboring branch points of type $(i, i+1)$ on the
$(K-2)$-triangle of branch points, and suppose the points $z$ and $z'$ are large
and located along a common cable. We continue them to a common
point $z_*$ while staying within the cable, in the region near infinity. Then
\begin{equation}
\gamma_{i,i+1}^{\fb}(z_*) + \gamma_{i+1,i}^{\fb'}(z_*) = \gamma^{\mathfrak{v}},
\end{equation}
where $\mathfrak{v}$ is the lattice point in a $(K-3)$-triangle labeling the triangle in the $(K-2)$-triangle
determined by the pair $\fb, \fb'$.

\section{Flags, flat sections, and cluster coordinates
associated with a minimal spectral network}\label{sec:FlagsFlatSections}

\subsection{Flat connections and Hitchin solutions}

Fix some $\vartheta$ and a point of the Coulomb branch of the level $K$ lift
of the $AD_1$ theory near the lift locus,
and assume that the corresponding WKB spectral network $\wnet = \wnet_\vartheta$
is an essentially minimal spectral network.

Recall that given a solution $(A,\varphi)$ of the
Hitchin equations with gauge group $SU(K)$
we can form the flat $SL(K,\IC)$ connection\footnote{Strictly speaking, we should be considering
\ti{twisted} flat connections; we defer this annoying detail to \S\ref{sec:twistings}.}
$\nabla = \de + \CA$, where
\begin{equation}\label{eq:flat-conn}
\CA = \frac{R}{\zeta} \varphi + A + R \zeta \bar \varphi
\end{equation}
on the complex vector bundle $E \to C$.  Here we use the complex structure of $C$
to split $\phi = \varphi + \bar\varphi$ into its $(1,0)$ and $(0,1)$ parts.  We call the
solutions $s$ of the ordinary differential equation $\nabla s = 0$ on $C$
``flat sections'' of $(E,\nabla)$.

The flat connection $\nabla$ has a singularity
only at infinity.  Therefore, there is a $K$-dimensional vector space $\CE$ of
global flat sections.
In what follows we will introduce some special bases of $\CE$ which are attached
to regions in the complement of the spectral network $\wnet$.  The change-of-basis matrix
relating the bases attached to two neighboring regions is particularly simple; we think of this matrix as
associated to ``crossing the $\CS$-wall'' separating the regions.
The change-of-basis matrix obtained by crossing a whole cable of $\CS$-walls can be
interpreted as a Stokes matrix associated to the irregular singularity of $\nabla$ at $z = \infty$.

\textbf{Remark}:  For most of what we will do in the rest of this section, it is not strictly
necessary to start with a solution of Hitchin's equations and corresponding WKB spectral network.
Instead one could just start with a minimal spectral network.\footnote{Here we would use
the definition of ``spectral network'' in \S 9.1 of \cite{Gaiotto2012};
to match that definition precisely, one should cut out a small neighborhood of infinity,
so that we view the spectral network as drawn on the disc, and then perturb the $\CS$-walls slightly
so that each $\CS$-wall ends on one of three
marked points on the boundary (one marked point for each cable).}
This spectral network induces a coordinate system on the moduli space of flat connections on the disc, decorated
by flags associated to three marked points on the boundary.
Much of the following discussion could be carried out in this more abstract context.

\subsection{The flags \texorpdfstring{$A_\bullet, B_\bullet, C_\bullet$}{A.,B.,C.}}\label{subsec:ABC-FLAGS}

From the asymptotics of $\nabla$ as $z \to \infty$ we can define three
flags $A_{\bullet}$, $B_{\bullet}$, $C_{\bullet}$ in $\CE$, associated with
the three cables $a$, $b$, $c$, as follows.

\insfig{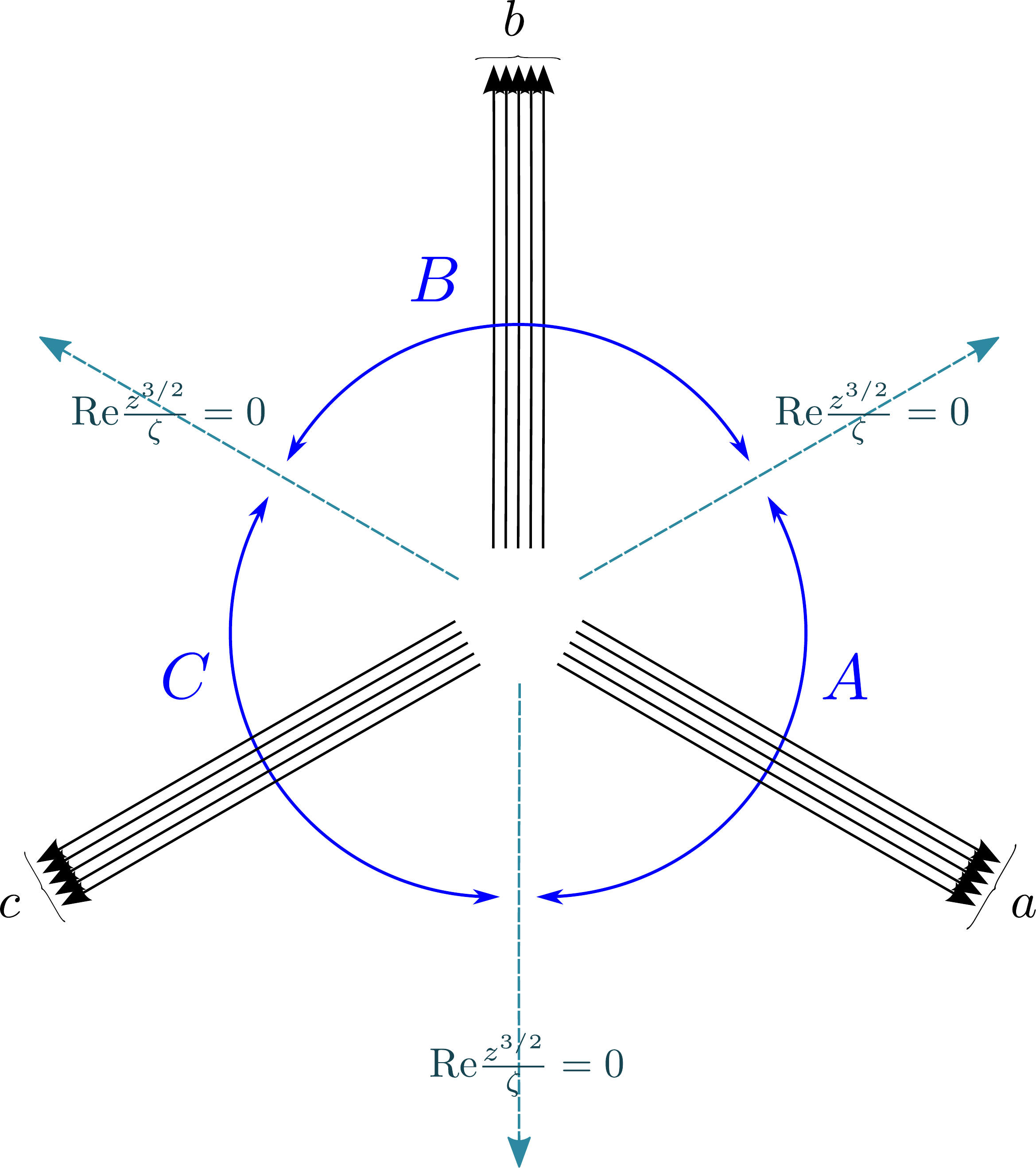}{Regions defined by the $z\to \infty$ asymptotics
of the Higgs field, in the level $K$ lift of the $AD_1$ Hitchin system.
In this figure $\zeta$ has argument $- \frac{\pi}{4}$.
If we vary the argument of $\zeta$ by $\alpha$
then the regions rotate by an angle $\frac{2}{3} \alpha$.
The $a$, $b$, $c$ cables of a minimal WKB spectral network are also depicted.
The position of the cables depends on $\vartheta$; the situation shown here is $\vartheta = \arg \zeta$.
If we vary $\vartheta$ by $\alpha$ then the cables rotate by $\frac{2}{3}\alpha$.}

Having trivialized the cover $\Sigma \to C$ on the complement of the cuts
(which we will fix more precisely below),
we have an unambiguous labeling of the sheets $\lambda_1, \dots, \lambda_K$.
We consider $z$ to be
in the region near $\infty$, so that we may approximate the roots of the
level $K$ spectral curve simply by
\begin{equation} \label{eq:lambda-approx}
\lambda_i \approx (K+1-2i) \sqrt{z} \, \de z \qquad   i=1, \dots, K
\end{equation}
and hence, for any $z_0$ and $\abs{z} \gg \abs{z_0}$,
we may approximate
\begin{equation}
\int_{z_0}^z \lambda_i \approx \frac{2}{3}(K+1-2i) z^{3/2} + const.
\end{equation}
In the region at large $\abs{z}$ there are three rays
where $\Re \, \zeta^{-1} z^{3/2} = 0$, and in the
three regions separated by these lines the $z\to\infty$ asymptotics are
different. We define the $B$-region to be the one in which
\begin{equation}\label{eq:B-growth}
{\rm Re} \int_{z_0}^z \frac{1}{\zeta} \lambda_1 \ll
{\rm Re} \int_{z_0}^z \frac{1}{\zeta} \lambda_2 \ll \cdots \ll {\rm Re} \int_{z_0}^z \frac{1}{\zeta} \lambda_K.
\end{equation}
Then the $A$-region is the region clockwise from the $B$-region and has
\begin{equation}\label{eq:A-growth}
{\rm Re} \int_{z_0}^z \frac{1}{\zeta} \lambda_1 \gg {\rm Re} \int_{z_0}^z \frac{1}{\zeta} \lambda_2
\gg \cdots \gg {\rm Re} \int_{z_0}^z \frac{1}{\zeta} \lambda_K,
\end{equation}
while the $C$-region is the region counterclockwise from the $B$-region and also has
\begin{equation}\label{eq:C-growth}
{\rm Re} \int_{z_0}^z \frac{1}{\zeta} \lambda_1 \gg {\rm Re} \int_{z_0}^z \frac{1}{\zeta} \lambda_2
\gg \cdots \gg {\rm Re} \int_{z_0}^z \frac{1}{\zeta} \lambda_K.
\end{equation}
We have taken all of the branch cuts to be asymptotic to the ray separating the $A$- and $C$- regions.
Crossing all of these cuts implements the permutation $w_0^{(K)}$ on the sheets;
this is important for understanding how the three systems
of inequalities \eqref{eq:B-growth}, \eqref{eq:A-growth}, \eqref{eq:C-growth}
can be consistent.

Let us now relate these regions to those defined by the spectral
network.  According to \cite{Gaiotto2012}, an $\CS$-wall
of type $ij$ emerging from a branch point $\fb$ in a WKB spectral
network is defined by the equation
\begin{equation}\label{eq:S-walls}
\int_{\fb}^z e^{-\I \vartheta} (\lambda_i-\lambda_j) > 0,
\end{equation}
and is naturally oriented outwards, so that the LHS of
\eqref{eq:S-walls} increases to $+\infty$ as $z \to \infty$ along the $\CS$-wall.
Comparing with \eqref{eq:lambda-approx} we see that the $a,b,c$-cables
lie in the $A,B,C$-regions, as shown in Figure
\ref{fig:asymp}, so long as
\begin{equation}
\arg \zeta-\frac{\pi}{2} < \vartheta < \arg\zeta + \frac{\pi}{2}.
\end{equation}

We now define three filtrations $A_{\bullet}$, $B_{\bullet}$ and $C_{\bullet}$
of the vector space $\CE$ of flat sections.
For $z \to \infty$ along a path asymptoting to the $b$-cable,
the condition\footnote{The meaning of ``$< \infty$'' in this equation is that the limit exists.}
\begin{equation}\label{eq:b-filtration-first}
\lim_{z\to\infty} \exp \left[ 2\,{\rm Re} \int_{z_0}^z \frac{R}{\zeta} \lambda_n \right] s(z) < \infty
\end{equation}
defines a $(K+1-n)$-dimensional subspace $B_{K+1-n} \subset \CE$.
For example, suppose we take $n=K$; $\lambda_K$
gives the largest real part for the integral
in the $B$-region at large $\vert z\vert$, and this integral
overwhelms all but the smallest flat section $s(z)$; the condition \eqref{eq:b-filtration-first} with $n = K$
thus singles out a line spanned by $s(z)$.  For $n=K-1$ there is a 2-dimensional space spanned by the smallest
and next-smallest flat sections, and so forth.  In a similar way we can define the other two filtrations $A_\bullet$, $C_\bullet$.

Thus, in summary:
\begin{equation}\label{eq:a-filtration}
\lim_{z\to \infty } \exp\,\left[ 2\,{\rm Re} \int_{z_0}^z \frac{R}{\zeta}\lambda_n \right]  s(z)  < \infty \Leftrightarrow s \in A_{n}
\end{equation}
for $z \to \infty$ along the $a$-cable, where $A_n$ is $n$-dimensional,
\begin{equation}\label{eq:b-filtration}
\lim_{z\to \infty } \exp\,\left[ 2\,{\rm Re}  \int_{z_0}^z \frac{R}{\zeta}\lambda_n \right]  s(z) < \infty \Leftrightarrow s \in B_{K+1-n}
\end{equation}
for $z \to \infty$ along the $b$-cable, where $B_{K+1-n}$ is $(K+1-n)$-dimensional, and
\begin{equation}\label{eq:c-filtration}
\lim_{z\to \infty } \exp\,\left[ 2\,{\rm Re} \int_{z_0}^z\frac{R}{\zeta} \lambda_n \right]  s(z) < \infty \Leftrightarrow s \in C_{n}
\end{equation}
for $z \to \infty$ along the $c$-cable, where $C_n$ is $n$-dimensional.

For generic $\nabla$, the three flags $A_\bullet$, $B_\bullet$, $C_\bullet$ are
in general position. We can therefore apply the linear algebra of Appendix \ref{app:LinearAlgebra}.
In particular, we define $A^n := A_{K-n}$, etc., and a set of
$\half (K+1)K$ lines
\begin{equation}\label{eq:Lx,y,z-def}
\fL^{x,y,z} := A^x \cap B^y \cap C^z, \qquad x+y+z = K-1,
\end{equation}
where $(x,y,z)$ is a lattice point in a $(K-1)$-triangle.
Similarly, we define $\half K (K-1)$ planes
\begin{equation}\label{eq:Px,y,z-def}
\fP^{x,y,z} := A^x \cap B^y \cap C^z, \qquad x+y+z = K-2,
\end{equation}
associated to points in a $(K-2)$-triangle, and
$\half (K-1)(K-2)$ spaces
\begin{equation}\label{eq:Vx,y,z-def}
\fV^{x,y,z} := A^x \cap B^y \cap C^z, \qquad x+y+z = K-3,
\end{equation}
associated to points in a $(K-3)$-triangle. These will play an
important role in the following discussion.

\subsection{Twistings} \label{sec:twistings}

Now let us confront a detail we have hidden up until now (this section might reasonably be skipped on a first reading.)
The construction of \cite{Gaiotto2012} really
involves \ti{twisted} connections over $C$ and $\Sigma$, not ordinary ones.  What this means is that they
are really connections over the bundles of tangent directions $\tilde{C}$ and $\tilde{\Sigma}$,
with holonomy $-1$ around each fiber.  Because of this holonomy it is strictly speaking nonsense
to speak of the space of global flat sections of $\nabla$ --- there are no global flat sections!

There are various ways of dealing with
this twisting, so as to reduce to ordinary connections;
each choice leads to annoying minus signs
appearing in a slightly different place.  For our limited purposes in this paper,
the following prescription will suffice.
Fix any global nonvanishing vector field $\nu$
on the plane; this gives a global section $\hat{C}$ of $\tilde{C} \to C$.
Pulling back $\nabla$ from $\hat{C}$ to $C$, we obtain a flat connection (not a twisted flat
connection) over $C$; by abuse of notation we will also call this flat connection $\nabla$.
After so doing, we may define $\CE$ to be the vector space of global $\nabla$-flat sections over
$C$.

Strictly speaking, different choices of nonvanishing vector field $\nu$ give different vector spaces $\CE(\nu)$.
A homotopy from $\nu$ to $\nu'$ gives an isomorphism $\CE(\nu) \simeq \CE(\nu')$, and any $\nu$ and $\nu'$
are indeed homotopic, so we do have isomorphisms between the different $\CE(\nu)$; however, any $\nu$, $\nu'$ can
be related by distinct homotopies, and so
we do not get \ti{canonical} isomorphisms between the different $\CE(\nu)$.  (One
can easily see this by considering a homotopy from $\nu$ to itself which winds once around the circle;
this corresponds to the automorphism of $\CE(\nu)$ given by $s \mapsto -s$.)  Speaking loosely, we may say
that the space $\CE$ is determined only ``up to multiplication by $-1$.''  On the other hand, the spaces
$\End \CE(\nu)$ are all canonically isomorphic (since conjugation by $-1$ gives the identity automorphism),
so $\End \CE$ is really canonically defined.  Most of our considerations in this paper
ultimately involve only $\End \CE$, not $\CE$ itself; for this reason the choice
of $\nu$ will not play much role in what follows.

The vector field $\nu$ also induces a section $\hat\Sigma$ of $\tilde{\Sigma} \to \Sigma$,
defined on the complement of the ramification locus $R$, and with 1 unit of winding around
each point of $R$.  Thus, given a twisted flat connection $\nabla^\ab$ over $\Sigma$,
pulling back from $\hat\Sigma$ gives a flat connection (not a twisted flat connection)
over $\Sigma \setminus R$.  This flat connection has holonomy $-1$
around each point of $R$.  By abuse of notation we will also call this connection
$\nabla^\ab$.

\subsection{Line decompositions from the nonabelianization map}\label{subsec:LineDecompNonab}

In \cite{Gaiotto:2011tf,Gaiotto2012}, given a spectral network $\CW$, we defined a nonabelianization map $\Psi_\CW$: given a twisted flat connection $\nabla^\ab$
on a complex line bundle $L \to \Sigma$, this map produces a twisted flat connection $\nabla = \Psi_\CW(\nabla^\ab)$ on
a complex rank $K$ bundle $\hat E \to C$.  We conjectured that $\Psi_\CW$ provides
a coordinate system on an open patch of the moduli space of flat rank $K$ connections over $C$
with fixed singularity and flag structure.  We will verify that conjecture here,
in the case of the moduli space corresponding to the level $K$ lift
of the $AD_1$ theory, when we take $\CW$ to be a minimal spectral network.
Indeed, supposing that $\nabla = \Psi_\wnet(\nabla^\ab)$,
we will identify the holonomies of $\nabla^\ab$ around an appropriate set of cycles
with the Fock-Goncharov triangle coordinates of $\nabla$.

Three notes on notation.
First, for simplicity we denote $\hat E$ simply by
$E$.  Second, we use the notation $F(\wp)$ for the $\nabla$-parallel transport along the path $\wp$ (not
for the formal generating function which appeared in \cite{Gaiotto2012}).
Third, our convention for composition of linear transformations is backward
from the usual one:  in particular, linear transformations act on vectors from the right, and
$F(\wp_1) F(\wp_2) = F(\wp_1\wp_2)$.
(See Appendix C of \cite{Gaiotto2012}.)

A minimal spectral network divides the complex plane into
several connected components.  Let us focus on the
$\half K(K-1) + 1$ regions at infinity which abut the
$b$-cable.  Each such region is separated from its neighbor
to the left by an $\CS$-wall emanating from a branch
point $\fb^{x,y,z}$ in the $(K-2)$-triangle of branch points.
We label
these regions as $\CR^{\alpha}$, $\alpha=0, \dots, \half K(K-1)$.
Note that $\CR^0 = \CR_{ab}$ and $\CR^{\half K (K-1)} = \CR_{bc}$.
There are similar regions around the $a$- and $c$-cables.
If we need to distinguish these we write $\CR^{\alpha,a}$ etc.

Suppose we have trivialized the $K$-fold cover $\Sigma \to C$ over $C \setminus {\mathrm{(cuts)}}$ as above.  Then in each region $\CR$ we have an isomorphism
\begin{equation}\label{eq:DirectSumLine}
E \vert_{\CR} \cong \pi_* L = \bigoplus_{i=1}^K L_i,
\end{equation}
where $L_i$ is the line bundle $L \to \Sigma$ restricted to the $i^{th}$ sheet.
Now we can define lines
\begin{equation} \label{eq:def-fL}
\fL_i^{\CR} := \{\psi \in \CE \ \vert \ \psi(z)\in L_i\vert_z \ \forall z\in \CR\}\subset \CE.
\end{equation}
We stress that $\fL^{\CR}_i$ is a line in a fixed $K$-dimensional vector space $\CE$, and is
not to be confused with the fiber of any bundle on $C$.
Thus, to each region $\CR^{\alpha}$ there is an associated decomposition of $\CE$ into a direct sum of lines.

In the interest of clarity we will be somewhat pedantic and
introduce two maps, the restriction map $\rho(z)$ and the extension map
$\varepsilon(z)$. The restriction (or evaluation) map $\rho(z): \CE \to E_z$
simply evaluates a section at $z$. The extension map $\varepsilon(z): E_z \to \CE$
takes a vector in the fiber at $z$ and extends it by parallel transport
with respect to \eqref{eq:flat-conn} to a flat section throughout $C$.\footnote{In the present case,
with $C = \IC\IP^1$ and only a single irregular singular point, there is no monodromy
and $\varepsilon(z)$ can be defined for all $z \in \IC$. In more general situations
one would choose a system of cuts and only define $\varepsilon(z)$ for
$z$ in the complement of the cuts.}
We have trivial identities $\varepsilon(z)\rho(z) = 1$,
$\rho(z) \varepsilon(z)=1$,
and $\varepsilon(z_1)\rho(z_2) = F(\wp(z_1,z_2))$. (Remember that
we write composition of operators from left to right, i.e. subsequent operations
are written to the right.)  Thus,
\begin{equation}\label{eq:Triv-Id}
\rho(z_1) F(\wp(z_1, z_2)) \varepsilon(z_2) = 1
\end{equation}
is the identity transformation of $\CE$.
For all $z\in \CR$ we also have by \eqref{eq:def-fL}
\begin{equation}\label{eq:Rest-Ext}
\rho(z): \fL^{\CR}_i \to L_i\vert_z,  \qquad\quad \varepsilon(z):  L_i\vert_z \to \fL^{\CR}_i.
\end{equation}
 \emph{because} $\nabla^{\ab}$ preserves the direct sum decomposition
\eqref{eq:DirectSumLine}.
Note that \eqref{eq:Rest-Ext} is in general false if $z\notin \CR$.

\subsection{Transformation of line decompositions across \texorpdfstring{$\CS$-walls}{S-walls}}\label{subsec:Trans-S-Wall}

Suppose that two regions $\CR_1,\CR_2$ of $C$ are separated by an $\CS$-wall
of type $ij$.  Let $\wp = \wp(z_1,z_2)$ be a short path from $z_1 \in \CR_1$
to $z_2\in \CR_2$, crossing the $\CS$-wall at $z$.
Assume moreover that there is only one soliton charge $\gamma_{ij}$ supported on the $\CS$-wall, with
$\mu(\gamma_{ij}) \neq 0$; this is the case for any nondegenerate spectral network and in particular for the minimal
spectral networks to which we specialize below.
Then according to \cite{Gaiotto2012} (see particularly (10.4)-(10.6)),
the $\nabla$-parallel transport along $\wp$ is given by
\begin{equation}\label{eq:DefParTrpt}
F(\wp) = \CD(\wp) + \mu(\gamma_{ij}) \CY_{\gamma_{ij}(z_1,z_2)}.
\end{equation}
Here as in \cite{Gaiotto2012} we let $\CY_{a}$ denote the $\nabla^\ab$-parallel transport along a path $a$,
$\CD(\wp) = \sum_{i=1}^K \CY_{\wp^{(i)}}$, and
$\gamma_{ij}(z_1,z_2)$ denotes the ``detour'' path which travels along $\wp^{(i)}$ from $z_1^{(i)}$ to $z^{(i)}$,
then follows the chain $\gamma_{ij}$ from $z^{(i)}$ to $z^{(j)}$, and finally travels along $\wp^{(j)}$
from $z^{(j)}$ to $z_2^{(j)}$.\footnote{To be careful about signs
we should remember that, as explained in \S\ref{sec:twistings}, the sign of the parallel transport $\CY_{\gamma_{ij}}$ depends on how we lift $\gamma_{ij}$ from $H_1(\Sigma, \IZ)$
to $H_1(\Sigma \setminus R, \IZ)$.  From the discussion in \cite{Gaiotto2012} it follows that
there is a canonical way of defining the combination $\mu(\gamma_{ij}) \CY_{\gamma_{ij}(z_1, z_2)}$,
and this is what we are using in \eqref{eq:DefParTrpt}.}

The first term $\CD(\wp)$ in \eqref{eq:DefParTrpt}
preserves the direct sum decomposition $E \simeq \oplus L_i$ globally on $C \setminus {\mathrm{(cuts)}}$.
In contrast, the second term of \eqref{eq:DefParTrpt} involves the non-diagonal
$\CY_{\gamma_{ij}(z_1,z_2)} \in \Hom(L_i\vert_{z_1}, L_j\vert_{z_2})$
(extended to ${\rm Hom}(E\vert_{z_1}, E\vert_{z_2})$ by defining it
to be zero on the other summands.)

In this situation we have the following three properties:

\begin{enumerate}

\item $\fL_s^{\CR_1} = \fL_s^{\CR_2}$ for $s\not= i$.

\item $\fL_i^{\CR_1} \not= \fL_i^{\CR_2}$.

\item \emph{The plane spanned by $\fL_i^{\CR_1}$ and $\fL_i^{\CR_2}$
contains $\fL_j^{\CR_1}=\fL_j^{\CR_2}$.}

\end{enumerate}

Let us give a careful proof of the third, absolutely crucial, statement.
We apply \eqref{eq:Triv-Id} and the definition \eqref{eq:DefParTrpt} of the parallel transport
to an element $s \in \fL_i^{\CR_1}$, to conclude
\begin{equation}\label{eq:ThreeLinesPT}
s = s(z_1) \CD(\wp)\varepsilon(z_2) + \mu(\gamma_{ij}) s(z_1) \CY_{\gamma_{ij}(z_1,z_2)} \varepsilon(z_2).
\end{equation}
Now, $s(z_1) \CD(\wp)\in L_i\vert_{z_2}$, so the extension
$s(z_1) \CD(\wp)\varepsilon(z_2) \in \fL_i^{\CR_2}$.
Similarly, $s(z_1)  \CY_{\gamma_{ij}(z_1,z_2)}\in L_j\vert_{z_2}$, so $s(z_1) \CY_{\gamma_{ij}(z_1,z_2)} \varepsilon(z_2) \in
\fL_j^{\CR_2}$. Thus, \eqref{eq:ThreeLinesPT} is an
equation of the form
\begin{equation}\label{eq:Linear}
 s_i^{\CR_1} = s_i^{\CR_2} + s_j^{\CR_2},
\end{equation}
where $s_i^{\CR_1} \in \fL_i^{\CR_1}$, $s_i^{\CR_2}\in \fL_i^{\CR_2}$, and $s_j^{\CR_2} \in \fL_j^{\CR_2} = \fL_j^{\CR_1}$.
Thus the three lines sit in a common plane.

\subsection{The plane in \texorpdfstring{$\CE$}{E} associated to a branch point}\label{subsec:Plane-And-BP}

\insfigscaled{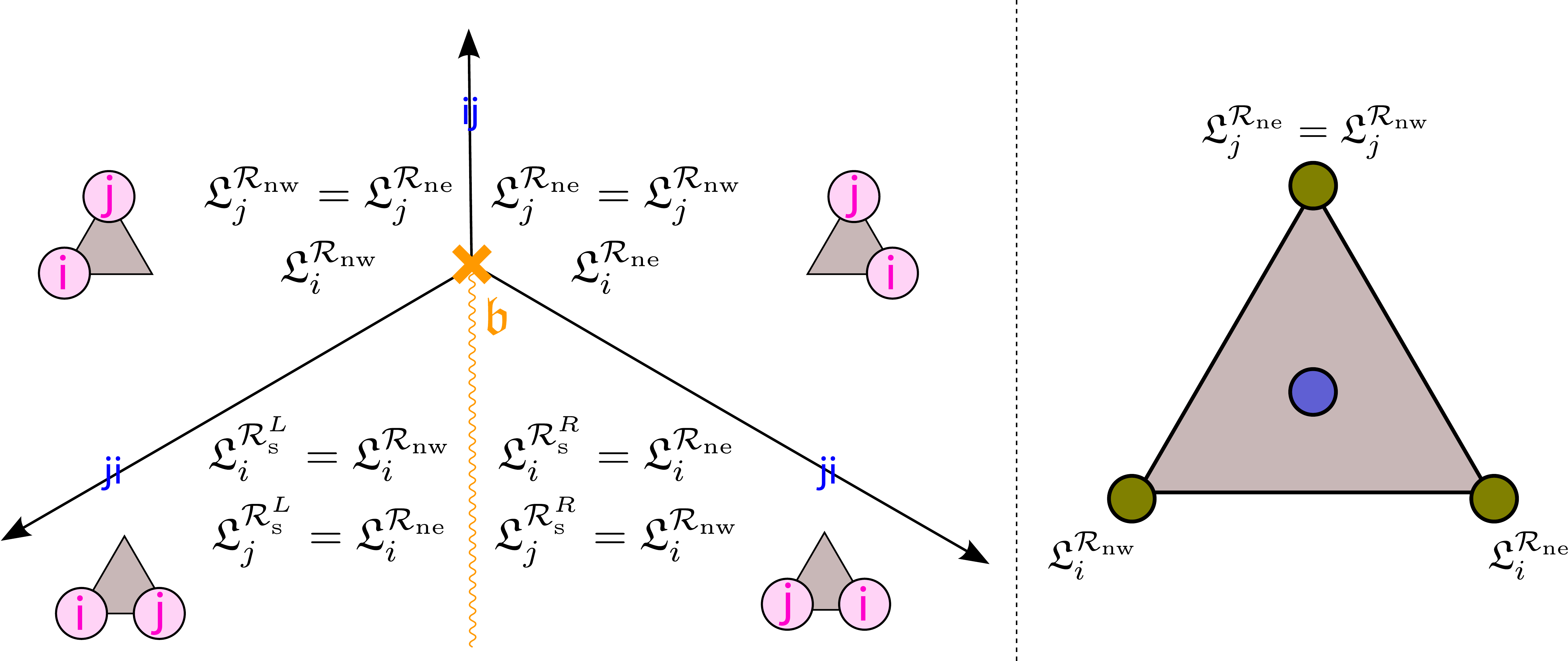}{0.26}{{\bf Left:} Associated to each branch point in any spectral network
there is a distinguished plane in the vector space of local flat sections,
containing several distinguished lines.
The relation between these lines is as indicated here.
{\bf Right:} A $1$-triangle representing the plane determined by
the branch point $\fb$, with the three distinguished lines at its vertices.}

For later use in \S \ref{sec:OpenProblems} we note a corollary
of the previous discussion.  We apply the transformation rules across $\CS$-walls
to the neighborhood of a single branch point of type $(ij)$, with $\CS$-walls of type
$ij$ (north) and $ji$ (southeast and southwest) and a branch cut pointing south.
The branch point determines four regions $\CR_{\mathrm{ne}}, \CR_{\mathrm{nw}}, \CR_{\mathrm{s}}^L, \CR_{\mathrm{s}}^R$
as indicated in Figure \ref{fig:local-plane}.
Our discussion above shows, first, that for $s \not=i,j$ the lines
$\fL_s^{\CR} \subset \CE$ are the same in all four regions.
It shows also that of the eight possible
lines $\fL_i^\CR, \fL_j^\CR$ there are in fact only three lines,
and they are coplanar:
\begin{equation}
\begin{split}
\fL_i^{\CR_{\mathrm{ne}}} & = \fL_i^{\CR_{\mathrm{s}}^R} = \fL_j^{\CR_{\mathrm{s}}^L},\\
\fL_i^{\CR_{\mathrm{nw}}} & = \fL_i^{\CR_{\mathrm{s}}^L} = \fL_j^{\CR_{\mathrm{s}}^R}, \\
\fL_j^{\CR_{\mathrm{ne}}} & = \fL_j^{\CR_{\mathrm{nw}}}.
\end{split}
\end{equation}
To prove this, note that from the three rules above we immediately
get $\fL_j^{\CR_{\mathrm{ne}}} = \fL_j^{\CR_{\mathrm{nw}}}$, $\fL_i^{\CR_{\mathrm{ne}}} = \fL_i^{\CR_{\mathrm{s}}^R}$,
and $\fL_i^{\CR_{\mathrm{nw}}} = \fL_i^{\CR_{\mathrm{s}}^L}$. However, crossing the cut must
exchange $ \fL_i^{\CR_{\mathrm{s}}^L}$ for $\fL_j^{\CR_{\mathrm{s}}^R}$ and $\fL_i^{\CR_{\mathrm{s}}^R}$
for $\fL_j^{\CR_{\mathrm{s}}^L}$.  This is enough to determine all of the lines.

We thus obtain the key result: each branch point
of type $(ij)$ is associated with a \emph{plane} in
the space of local flat sections, containing three lines of types $\fL_i^\CR$ and $\fL_j^{\CR}$.
A $1$-triangle representing this plane, with the three lines at its vertices,
is shown on the right side of Figure \ref{fig:local-plane}.

\subsection{Snakes and the planes}\label{subsec:TriangleLineNonab}

We now apply the discussion of \S \ref{subsec:Trans-S-Wall} to a minimal spectral network of \Yin\ type.
From the various line decompositions which appear in the various regions
$\CR$ we will extract a canonical $(K-1)$-triangle of lines in the vector space $\CE$.
This $(K-1)$-triangle will then be identified with one associated to
the flags $A_\bullet, B_\bullet, C_\bullet$  by Fock and Goncharov in \cite{MR2233852},
thus making the crucial link between spectral networks and their work.

Consider a minimal spectral network of \Yin\ type, with $\CS$-walls
arranged so that when traversing the $b$-cable, moving east to west,
the $\CS$-walls one encounters are (reading the product from right to left):
\begin{equation}\label{eq:New-b-CABLE-DECOMP}
S_{12}\cdot(S_{23} S_{12} ) \cdots (S_{K-2,K-1} \cdots  S_{23} S_{12})\cdot   (S_{K-1,K} \cdots  S_{23} S_{12}).
\end{equation}

Let us describe how the lines $\fL_i^{\CR^\alpha}$ are related to each other.
When we cross the first $\CS$-wall of type $12$ from $\CR^{ab}$ into
region $\CR^{1}$, we find $\fL^{\CR^1}_s = \fL^{\CR_{ab}}_s$ for $s=2,\dots, K$,
but $\fL^{\CR^{1}}_1 \not= \fL^{\CR_{ab}}_1$.  Rather $\fL^{\CR^1}_1, \fL^{\CR_{ab}}_1, \fL^{\CR_{ab}}_{2}$
are all coplanar. Moving further across a wall of type $23$ from $\CR^1$ into $\CR^2$, we have a similar story:
$\fL_s^{\CR^1} = \fL_s^{\CR^2}$ for $s\not=2$, but $\fL_2^{\CR^1}\not=\fL_2^{\CR^2}$ and
$\fL_2^{\CR^1}, \fL_2^{\CR^2}$ and $\fL_3^{\CR^1}$ are all coplanar.  We can continue in this fashion,
reading off the $\CS$-wall types from \eqref{eq:New-b-CABLE-DECOMP}.

Note that near the $b$-cable the line $\fL^{\CR_{ab}}_1$ is spanned by the \emph{largest} section
as $z\to \infty$, while $\fL^{\CR_{ab}}_2 $ is spanned by the next largest. In the next region to the left the largest section has changed by adding a \emph{smaller} section.
This is the standard pattern in Stokes theory:  our basis of sections jumps when
we cross a Stokes line, but the asymptotics of this basis does not jump.

Now we assign the lines $\fL^{\CR^\alpha}_i$ to vertices of a $(K-1)$-triangle,
as follows.
We begin by assigning the lines $\fL_i^{\CR_{ab}}$, $i=1,\dots, K$,
to points on the right side of the $(K-1)$-triangle:  the
vertex $(K-1,0,0)$ corresponds to $\fL_1^{\CR_{ab}}$, and we go
up the right side of the triangle from there, assigning $\fL_i^{\CR_{ab}}$
to $(K-i,i-1,0)$.  Thus the line decomposition in $\CR_{ab}$
corresponds to that associated to the snake from the $B$ vertex to the $A$ vertex.
We next fill in the triangle, assigning lines $\fL^{\CR^\alpha}_i$ to
vertices, so that the change of line decompositions as we step to the left
from $\CR^\alpha$ to $\CR^{\alpha+1}$ corresponds to a step in the canonical path of
snakes.  We can do this precisely because the moves in the canonical
path of snakes match the ordering of $\CS$-wall types given in \eqref{eq:New-b-CABLE-DECOMP}.

For example, crossing into $\CR^1$ we have seen that the line $\fL^{\CR_{ab}}_1$ is changed to
$\fL^{\CR^1}_1$; hence we assign $\fL^{\CR^1}_1$ to the position $(K-2,0,1)$ in
the $(K-1)$-triangle.  Next we assign $\fL^{\CR^2}_{2}$ to the position $(K-3,1,1)$.
We continue this process using the pattern of $\CS$-walls \eqref{eq:New-b-CABLE-DECOMP}.
See Figure \ref{fig:constructing-triangle}.

The resulting $(K-1)$-triangle of lines in $\CE$ has the following properties:

\begin{enumerate}

\item All shaded up-triangles have vertices corresponding to coplanar lines.

\item The set of lines associated to any region $\CR^{\alpha,b}$ constitutes a snake
from the $B$-vertex to the $B$-side in the $(K-1)$-triangle.

\item The sequence of regions $\CR^{\alpha}$ corresponds to the
canonical path of snakes from $\CS\CN_{BA}$ to $\CS\CN_{BC}$
described below \eqref{eq:CanSN-Path}.  In particular, the change of snake from $\CR^{\alpha}$ to
$\CR^{\alpha+1}$ is given by a Fock-Goncharov elementary move of type I or type II, as described in Appendix \ref{app:LinearAlgebra}.

\end{enumerate}

\insfigscaled{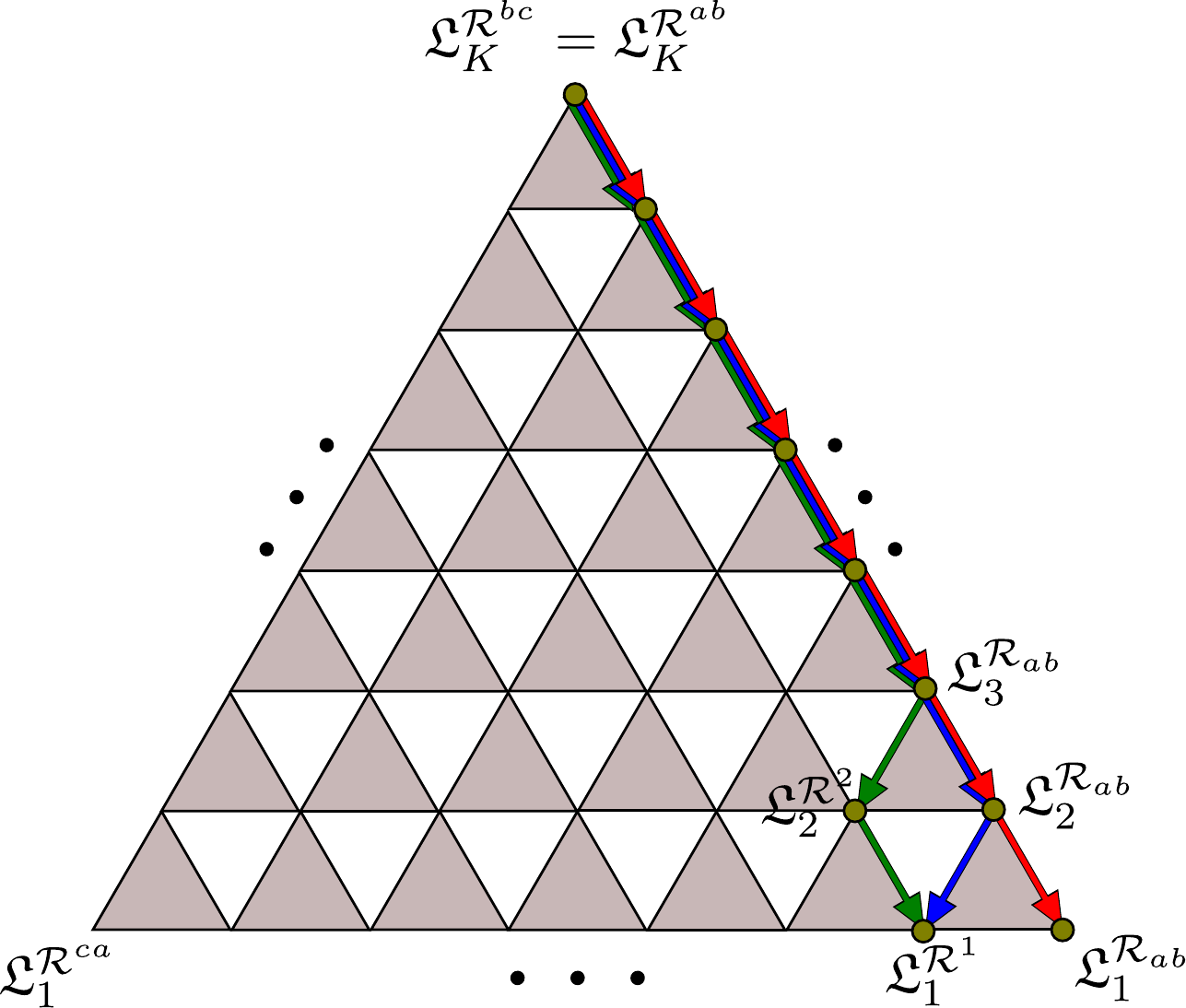}{0.60}{The first two moves in the canonical
path of snakes associated to the $b$-cable,
which we use for construction of a $(K-1)$-triangle of lines.
The red arrows indicate the snake $\CS\CN_{BA}$ which is associated to the region $\CR_{ab}$.
The blue arrows indicate the snake associated to region $\CR^1$, which differs from $\CS\CN_{BA}$
by a type I move.
The green arrows indicate the snake associated to region $\CR^2$.
Note that for crossing the $a$-cable we would use instead the initial snake $\CS\CN_{AB}$,
obtained by reversing all the red arrows.}

Of course, a similar procedure can be applied
to the other two cables.
For example, consider crossing the $a$-cable from $\CR_{ab}$ to $\CR_{ca}^R$.
We first cross an $\CS$-wall
of type ${K,K-1}$ to enter region $\CR^1$.
Thus $\fL^{\CR_{ab}}_s = \fL^{\CR^1}_s$ for $s=1,\dots, K-1$, but
$\fL^{\CR_{ab}}_K \not = \fL^{\CR^1}_K$.  On the $(K-1)$-triangle of lines we have constructed,
this corresponds to a type I move between two snakes running from the $A$ vertex to the $A$ side.
The snake corresponding to $\CR_{ab}$ is the northeast side of the triangle (but
oriented oppositely to the orientation we used when
crossing the $b$-cable.)  Continuing in this way,
following the pattern of $\CS$-walls given in \eqref{eq:Ta-cable}, we obtain
the canonical path of snakes from $\CS\CN_{AB}$ to $\CS\CN_{AC}$.
Similarly, crossing the $c$-cable from $\CR_{bc}$ to $\CR_{ca}^L$, we begin with
a snake from the $C$ vertex to the $B$ vertex (the line decomposition
in $\CR_{bc}$), and then proceed with the canonical path of snakes until we reach
a snake from the $C$ vertex to the $A$ vertex.
Note that the lines in $\CR_{ca}^L$ and $\CR_{ca}^R$ are related across the
cuts by the permutation $w_0^{(K)}$.

One might worry that the triangles of lines constructed by the above procedure
crossing the $a$-, $b$-, and $c$-cables might be different, but in fact
they are all the same.  This follows from the fact that they can all be identified
with the triangle of lines considered by Fock-Goncharov, as we show in the next section.

\subsection{Identifying our triangle with Fock-Goncharov's}\label{subsec:IdentLines}

We now claim that the $(K-1)$-triangle of lines we have just constructed is
identical to the $(K-1)$-triangle
of lines associated by Fock and Goncharov's construction to the triple of
flags $A_\bullet, B_\bullet, C_\bullet$ of \S \ref{subsec:ABC-FLAGS}.

To prove this, first consider the region $\CR_{ab}$.
Applying the WKB approximation to the flat connection $\nabla$ as
$z \to \infty$, we expect that the sections $s_i$ in the line $\fL_i^{\CR_{ab}}$
behave asymptotically as
\begin{equation}
s_i(z) \sim s_i(z_0) \exp\left[-2\Re  \int_{z_0}^z \frac{R\lambda_i}{\zeta} \right]
\end{equation}
throughout the region $\CR_{ab}$. If we take $z\to \infty$ along the
$b$-cable, comparing with the definition \eqref{eq:b-filtration} we
find the limit exists for $n \leq i$ and therefore $s_i$ is in
$B_{K+1-i} = B^{i-1}$. Similarly, taking $z\to \infty$ along the $a$-cable
and comparing with  \eqref{eq:a-filtration} reveals that the limit exists
for $n \geq i$ and hence $s_i$ is also in $A_i = A^{K-i}$.  Therefore,
\begin{equation}
\fL_i^{\CR_{ab}} = A^{K-i}\cap B^{i-1} = \fL^{K-i,i-1,0}.
\end{equation}
Thus we have shown that the lines $\fL_i^{\CR_{ab}}$, which by definition make up
the northeast side of our triangle of lines, agree with the lines $\fL^{K-i,i-1,0}$
which make up the northeast side of Fock-Goncharov's triangle of lines.
Similarly, considering the behavior in $\CR_{bc}$ we see that the lines on the northwest side
of our triangle are given by
\begin{equation}
\fL_i^{\CR_{bc}} = C^{K-i}\cap B^{i-1} = \fL^{0,i-1,K-i}.
\end{equation}

Thus we have found that the northeast and northwest sides
of our $(K-1)$-triangle match with the same two sides of Fock and Goncharov's triangle.
Moreover, as we have already shown, our triangle obeys \ti{coplanarity relations}:
the three vertices of each small shaded up-triangle
correspond to three coplanar lines.  Fock and Goncharov's triangle also obeys these relations.
As we now show, these properties are actually enough to \ti{characterize} Fock and Goncharov's triangle,
and thus prove that the two triangles are the same.

\insfigscaled{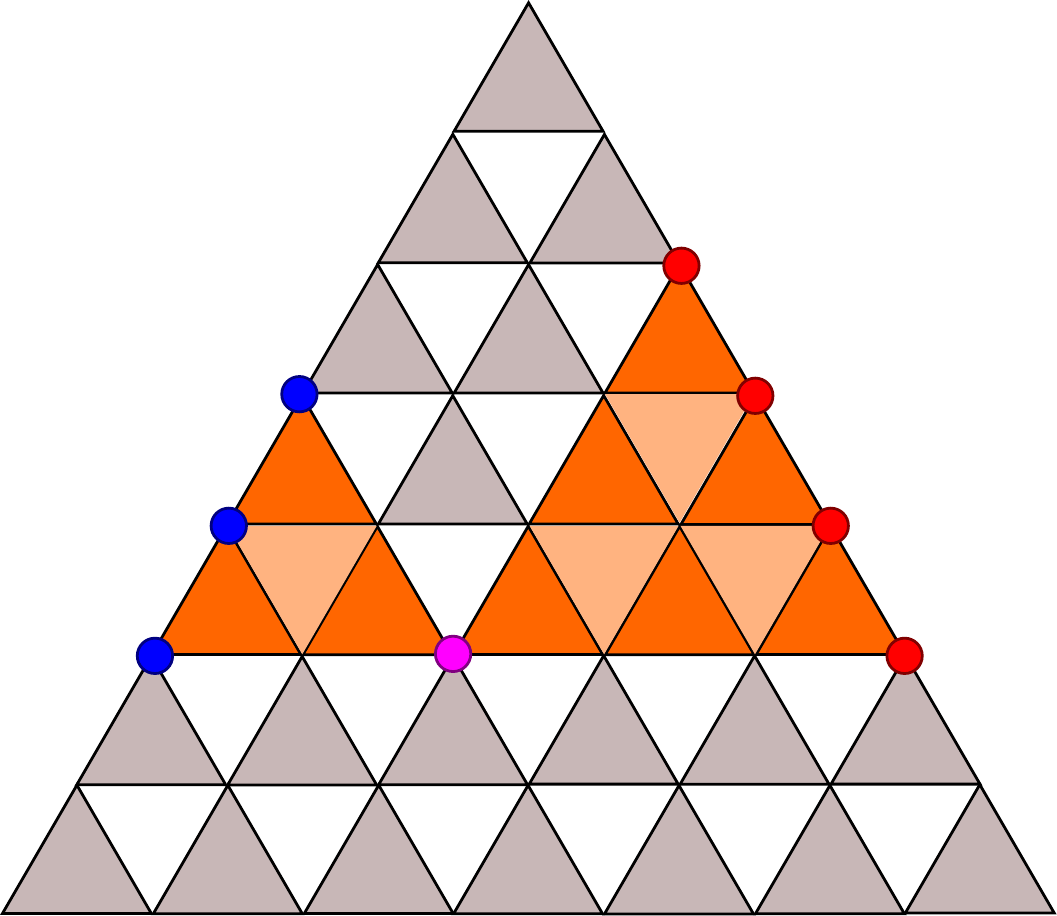}{0.52}{Two subtriangles inside a $(K-1)$-triangle of lines, with a common
vertex $x,y,z$ (purple dot).  The line $\fL^{x,y,z}$ is contained in the direct sum of the lines marked by the blue dots,
and also in the direct sum of the lines marked by the red dots; these two facts together suffice to
determine $\fL^{x,y,z}$.}

Indeed, an easy induction from the coplanarity relations shows that given any upward-pointing
subtriangle inside the $(K-1)$-triangle, the line at any vertex of the subtriangle
is contained in the sum of the lines along the opposite side of the subtriangle.
Now consider a pair of subtriangles, one flush with
the northeast side of the $(K-1)$-triangle and one flush with the northwest side,
with one vertex $x,y,z$ in common, as shown in Figure \ref{fig:radiation-diagram}.
The right subtriangle then implies that the line $\fL^{x,y,z} \subset A^x \cap B^y$,
while the left subtriangle implies that $\fL^{x,y,z} \subset C^z \cap B^y$.
Combining the two we obtain $\fL^{x,y,z} \subset A^x \cap B^y \cap C^z$,
but the latter space is 1-dimensional, so in fact
$\fL^{x,y,z} = A^x \cap B^y \cap C^z$, as desired.

\subsection{Identifying Darboux sections with canonical homs}\label{subsec:IdentHoms}

Now that we have identified our triangle of lines with the Fock-Goncharov
triangle of lines, we invoke some more of
the linear algebra reviewed in Appendix \ref{app:LinearAlgebra}.
In particular, as explained there, given an oriented edge $E$ in
the triangle, with source
and target $s(E)$ and $t(E)$ respectively, and given also an orientation of the
triangle, there is a canonical
hom $x_E \in {\rm Hom}(\fL^{s(E)}, \fL^{t(E)})$.
We are going to identify these canonical homs with our ``Darboux sections,''
by finding chains $\gamma_{ij}(z_1,z_2)$ such that
\begin{equation}
x_E = \rho(z_1) \CY_{\gamma_{ij}(z_1,z_2)} \varepsilon(z_2),
\end{equation}
where $z_1$ and $z_2$ are points on opposite sides of the wall.
To summarize this in a pithier way we can take the limit where $z_1$ and $z_2$ both approach a common point $z$ on the wall,
and define
\begin{equation}
 \hat\CY_{\gamma_{ij}(z)} = \lim_{z_1, z_2 \to z} \rho(z_1) \CY_{\gamma_{ij}(z_1,z_2)} \varepsilon(z_2).
\end{equation}
Then our result becomes simply
\begin{equation}\label{eq:Hom-To-CY}
x_E = \hat\CY_{\gamma_{ij}(z)}.
\end{equation}
Implicit in \eqref{eq:Hom-To-CY} is the fact that the right side does not depend on the chosen point $z$ on the wall.
Indeed, more is true:  we can even push $z$ away from the wall without changing $\hat\CY_{\gamma_{ij}(z)}$,
as long as $z$ does not cross any $\CS$-walls of type
$ki$ or $jk$ for any $k$.  This extra freedom will be important for us below.

Identifying precisely the correct chains $\gamma_{ij}(z)$ takes a bit of care.  We now do this.

\insfigscaled{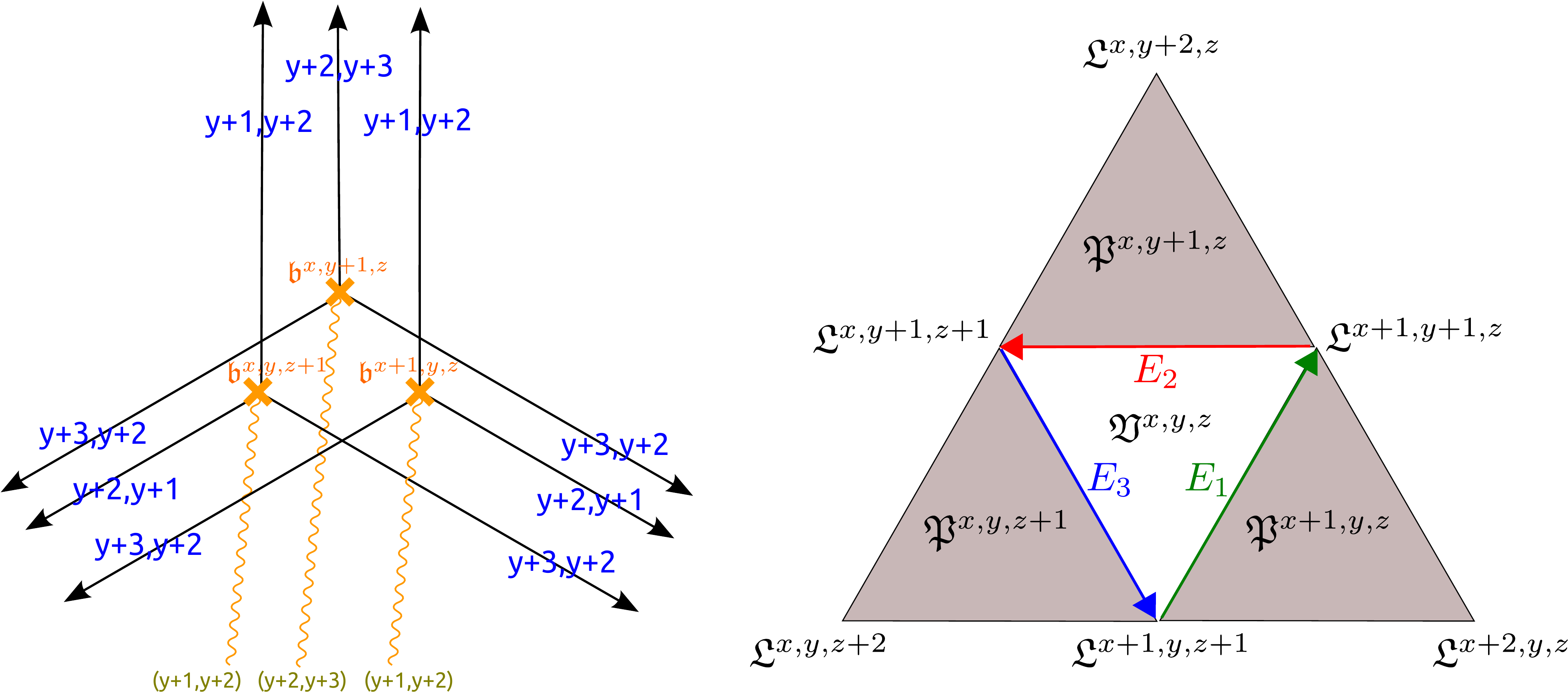}{0.375}{{\bf Left:} In a minimal spectral network there are local clusters of
3 branch points labeled by points $(x,y,z)$ in a $(K-3)$-triangle.  We show here one such cluster,
in a minimal spectral network of \Yin\ type.
We have, roughly, a ``mini'' $AD_2$ theory here.  {\bf Right:}  The associated $2$-triangle of lines,
with three edges $E_i$ marked, for which we will compute the canonical homs $x_{E_i}$ below.}

For later convenience, we will consider three canonical homs at once.
Thus let us consider three adjacent branch points $\fb^{x,y,z+1}, \fb^{x+1,y,z}, \fb^{x,y+1,z}$
with $(x,y,z)\in T(K-3)$.
Associated to this triple is a $2$-subtriangle of the triangle of lines; see Figure \ref{fig:loc-a2-triangle}.

Consider a short path $\wp$ crossing the $b$-cable $\CS$-wall emanating from $\fb^{x+1,y,z}$, say from
$\CR^{\alpha,b}$ to $\CR^{\alpha+1,b}$.  This crossing will be related to
the canonical hom associated with the green internal edge
\begin{equation}
E_1: (x+1,y,z+1) \to (x+1,y+1,z)
\end{equation}
in Figure \ref{fig:loc-a2-triangle}.
Indeed, fix some $s_{y+1}^\alpha \in \fL^{x+2,y,z} = \fL^{\CR^\alpha}_{y+1}$,
and suppose $s^{\alpha+1}_{y+1} \in \fL^{x+1,y,z+1} = \fL^{\CR^{\alpha+1}}_{y+1}$
is related to it by $\nabla^\ab$-parallel transport,
\begin{equation}
s^{\alpha+1}_{y+1} = s_{y+1}^\alpha(z_1) \CD(\wp) \varepsilon(z_2).
\end{equation}
Now we consider the definition \eqref{eq:DefParTrpt} of the $\nabla$-parallel transport along $\wp$.
For this particular $\CS$-wall, the product $\mu(\gamma_{ij}) \CY_{\gamma_{ij}}$ which appears
in \eqref{eq:DefParTrpt} can be described concretely:  one has simply $\mu(\gamma_{ij}) = 1$
and $\gamma_{ij} = \gamma_{y+1,y+2}^{\fb^{x+1,y,z},b}(z_1,z_2)$.
(To be really precise, as noted in \S\ref{sec:twistings}, we have to specify how
the chain $\gamma_{y+1,y+2}^{\fb^{x+1,y,z},b}(z_1,z_2)$ is lifted from $H_1(\Sigma, \IZ)$
to $H_1(\Sigma \setminus R, \IZ)$.
The answer is that $\gamma_{y+1,y+2}^{\fb^{x+1,y,z},b}(z_1,z_2)$ should be represented by
a path which goes clockwise around the ramification
point covering $\fb^{x+1,y,z}$, as in Figure \ref{fig:darboux-hom-chains}.\footnote{This answer can be
checked in laborious fashion beginning with the rules of \cite{Gaiotto2012} and \S\ref{sec:twistings},
but there is also a shortcut.  The rules determining $F(\wp)$ in
\cite{Gaiotto2012} were essentially determined by the
constraint that the connection $\nabla$ should be flat even at the branch points.  In particular
$F(\wp)$ should be the same as $F(\wp')$, where $\wp'$ is
a path which takes a detour around the branch point, thus avoiding this $\CS$-wall.
The term involving $\CY_{\gamma_{y+1,y+2}^{\fb^{x+1,y,z},b}(z_1,z_2)}$ in $F(\wp)$
should thus be identified with a term in $F(\wp')$, which in fact comes from $\CD(\wp')$.
In other words this term should simply be parallel transport along a lift of $\wp'$ to $\Sigma$.
This is enough to fix which way $\CY_{\gamma_{y+1,y+2}^{\fb^{x+1,y,z},b}(z_1,z_2)}$
should wind around the branch point.})
Applying \eqref{eq:DefParTrpt} to $s^\alpha_{y+1}$, we thus obtain
\begin{equation} \label{eq:threept-concrete}
s^{\alpha}_{y+1} = s^{\alpha+1}_{y+1} + s^{\alpha}_{y+1}(z_1) \CY_{\gamma_{y+1,y+2}^{\fb^{x+1,y,z},b}(z_1,z_2)} \varepsilon(z_2).
\end{equation}
Taking the limit $z_1,z_2 \to z_b$ this equation becomes
\begin{equation}
s^{\alpha}_{y+1} = s^{\alpha+1}_{y+1} + s^{\alpha}_{y+1} \hat\CY_{\gamma_{y+1,y+2}^{\fb^{x+1,y,z},b}(z_b)},
\end{equation}
and using the fact that $\hat\CY_{\gamma_{y+1,y+2}^{\fb^{x+1,y,z},b}(z_b)}^2 = 0$ we can also rewrite this as
\begin{equation}
s^{\alpha}_{y+1} = s^{\alpha+1}_{y+1} + s^{\alpha+1}_{y+1} \hat\CY_{\gamma_{y+1,y+2}^{\fb^{x+1,y,z},b}(z_b)}.
\end{equation}
Comparing this with the definition of the canonical hom $x_{E_1}$ in \eqref{eq:canhom3}, we see that
\begin{equation}
x_{E_1} = \hat\CY_{\gamma_{y+1,y+2}^{\fb^{x+1,y,z},b}(z_b)}.
\end{equation}
Similarly, the canonical hom $x_{E_2}$ associated with the red oriented edge
\begin{equation}
E_2: (x+1,y+1,z) \to (x,y+1,z+1)
\end{equation}
arises in crossing the $c$-cable and has
\begin{equation}
x_{E_2} = \hat\CY_{\gamma_{y+2,y+3}^{\fb^{x ,y+1,z},c}(z_c)}.
\end{equation}
Finally, the canonical hom $x_{E_3}$ associated with the blue edge
\begin{equation}
E_3: (x,y+1,z+1) \to (x+1,y,z+1)
\end{equation}
arises in crossing the $a$-cable and has
\begin{equation}
x_{E_3} = \hat\CY_{\gamma_{y+2,y+1}^{\fb^{x ,y,z+1},a}(z_a)}.
\end{equation}
Note that in these equations $z_a,z_b,z_c$ lie on the corresponding
$\CS$-walls and are close to the relevant branch points, in particular
not separated from the branch points by any cuts.

\insfigscaled{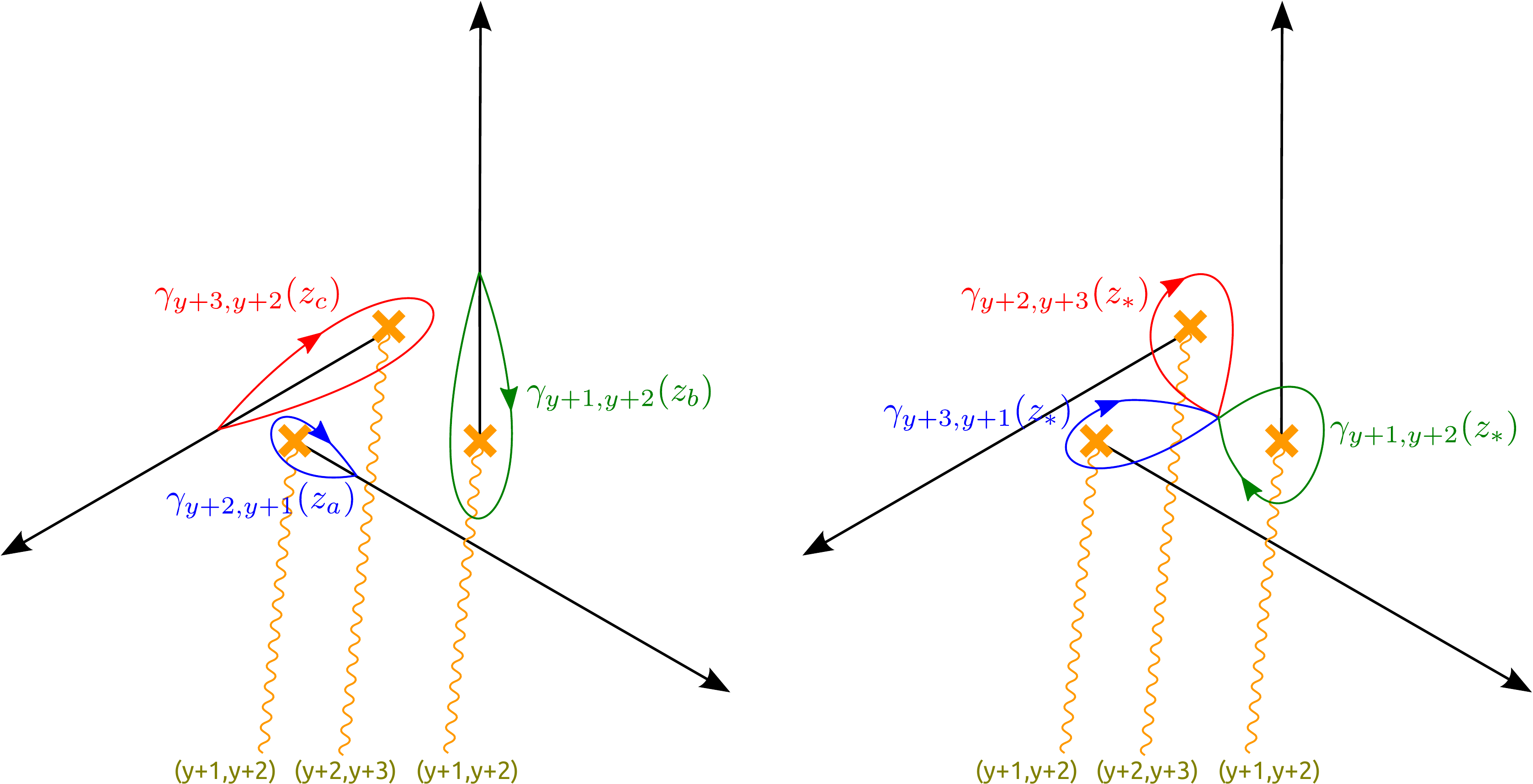}{0.4}{{\bf Left:} The three canonical homs $x_{E_k}$ are identified with
the three operators $\hat \CY_{\gamma_{ij}}$ associated with the three chains $\gamma_{ij}$ on $\Sigma$
shown here.  {\bf Right:}  By moving their basepoints, these three chains
can be deformed so that they combine into a
closed cycle $\gamma^\fv$ on $\Sigma$.}

\subsection{Identifying Fock-Goncharov triangle coordinates with Darboux coordinates}\label{subsec:IdentCoords}

Now consider a point $(x,y,z)$ in a $(K-3)$-triangle. To this point we
can associate a triplet of planes $\fP^{x+1,y,z}$, $\fP^{x,y+1,z}$,
$\fP^{x,y,z+1}$, and a sextuplet of lines as in Figure \ref{fig:loc-a2-triangle}.
On the one hand, we can consider
the composition of canonical homs counterclockwise around the internal
downward pointing triangle of the $(K-1)$-triangle of lines as in
\eqref{eq:ThreeCanHom}.  This gives the Fock-Goncharov coordinate
$r^{x,y,z}$ associated with the triplet of flags in $\CE$.
On the other hand, we have just seen that
these three canonical homs correspond to operators $\hat \CY_{\gamma_{ij}}$, where
$\gamma_{ij}$ runs over the three relative homology classes
\begin{equation}
\gamma_{y+1,y+2}^{\fb^{x+1,y,z},b}(z_b), \quad \gamma_{y+3,y+2}^{\fb^{x,y+1,z},c}(z_c),
\quad \gamma_{y+2,y+1}^{\fb^{x,y,z+1},a}(z_a),
\end{equation}
shown on the left of Figure \ref{fig:darboux-hom-chains}.

As we have commented, these three operators $\hat\CY_{\gamma_{ij}(z)}$ do not change as we
continue $z_a$, $z_b$, $z_c$ to the common point $z_*$
shown on the right of Figure \ref{fig:darboux-hom-chains}.
Note that in doing so we move $z_a$ and $z_c$ across a cut of type $(y+2,y+3)$,
so the indices $ij$ of the chains have to be relabeled appropriately.
These three chains can now be combined into a single closed cycle on $\Sigma$,
\begin{equation}
\gamma^{\fv} = {\rm cl} \left(\gamma_{y+2,y+3}^{\fb^{x,y+1,z},c}(z_*) + \gamma_{y+3,y+1}^{\fb^{x,y,z+1},a}(z_*)
+ \gamma_{y+1,y+2}^{\fb^{x+1,y,z},b}(z_*) \right).
\end{equation}
The composition of the abelian parallel transports around these three open paths gives the abelian parallel transport along the closed cycle $\gamma^\fv$.
Thus we finally arrive at one of the main results of
this paper:
\begin{equation}\label{eq:MainResultOne}
\CY_{\gamma^\fv} = r^{x,y,z}.
\end{equation}

Let us summarize our result.  The minimal spectral network of the level
$K$ lift of the $AD_1$ theory provides a Darboux coordinate system on
the moduli space of
flat $SL(K,\IC)$ connections on the disc with three fixed flags.
What we have shown is that these Darboux coordinates coincide with the
Fock-Goncharov coordinates on the same moduli space.

\subsection{Another viewpoint on the Darboux coordinates}

There is another point of view on the $\CY_{\gamma_{ij}}$ and the
$\CY_\gamma$, used heavily in
\cite{MR2233852} and \cite{Gaiotto:2009hg,Gaiotto:2010be}:  one can construct them directly
as combinations of flat sections determined by flags around singular points.
(For example, in \cite{MR2233852,Gaiotto:2009hg}, the $\CY_\gamma$ in case $K=2$ are defined using cross-ratios
of 4-tuples of distinguished ``small'' flat sections associated to 4 singular points.)
We will not make much use of the generalization
of these formulas to the higher rank case in this paper, but we note here how
they could in principle be obtained.

The essential remark is that we can interpret
\eqref{eq:Linear} as a transformation from a basis for the plane $P$ using
flat sections $s_a^{\CR_1}$, $a=i,j$, associated to region $\CR_1$, to a basis
using flat sections $s_a^{\CR_2}$, $a=i,j$, associated to region $\CR_2$:
\begin{equation}
\begin{split}
s_i^{\CR_2} & = s_i^{\CR_1}-  s_j^{\CR_1},\\
s_j^{\CR_2} & = s_j^{\CR_1}.
\end{split}
\end{equation}
This may be written as
\begin{equation}
s_a^{\CR_2} = s_a^{\CR_1} \CO^{\CR_1 \to \CR_2}
\end{equation}
where
\begin{equation}
\CO^{\CR_1 \to \CR_2} = 1 - \left( \frac{ \cdot \wedge s_j^{\CR_1}}{s_i^{\CR_1} \wedge s_j^{\CR_1}} \right) \otimes s_j^{\CR_1} \cdot
\left( \frac{ s_i^{\CR_2} \wedge s_i^{\CR_1}}{s_i^{\CR_2} \wedge s_j^{\CR_1}} \right).
\end{equation}

Here we have focused on a plane $\fP$ of flat sections in the $K$-dimensional space $\CE$, so these wedge products are
valued in the line $\wedge^2 \fP$, and hence their ratios are scalars.
From this starting point,
one can write an expression for $\CY_{\gamma_{ij}}$ in terms of flat sections.  In turn the $\CY_\gamma$ can be obtained
by composition of appropriate choices of $\CY_{\gamma_{ij}}$.

It can be useful to rewrite the ratios of elements of $\wedge^2 \fP$ in a way that does not
distinguish $\fP$. To do this one chooses a complementary basis of sections $s_t$, $t=1,\dots, K-2,$
and interprets
\begin{equation}
  \frac{ s_i^{\CR_2} \wedge s_i^{\CR_1}}{s_i^{\CR_2} \wedge s_j^{\CR_1}}
=   \frac{ s_i^{\CR_2} \wedge s_i^{\CR_1}\wedge s_1 \wedge \cdots \wedge s_{K-2} }
{s_i^{\CR_2} \wedge s_j^{\CR_1}\wedge s_1 \wedge \cdots \wedge s_{K-2} }
\end{equation}
as a ratio of elements of the line $\wedge^K \CE$. The choice of basis sections
$s_t$ does not matter.

\subsection{Compact regions}\label{subsec:CompactRegions}

\insfigscaled{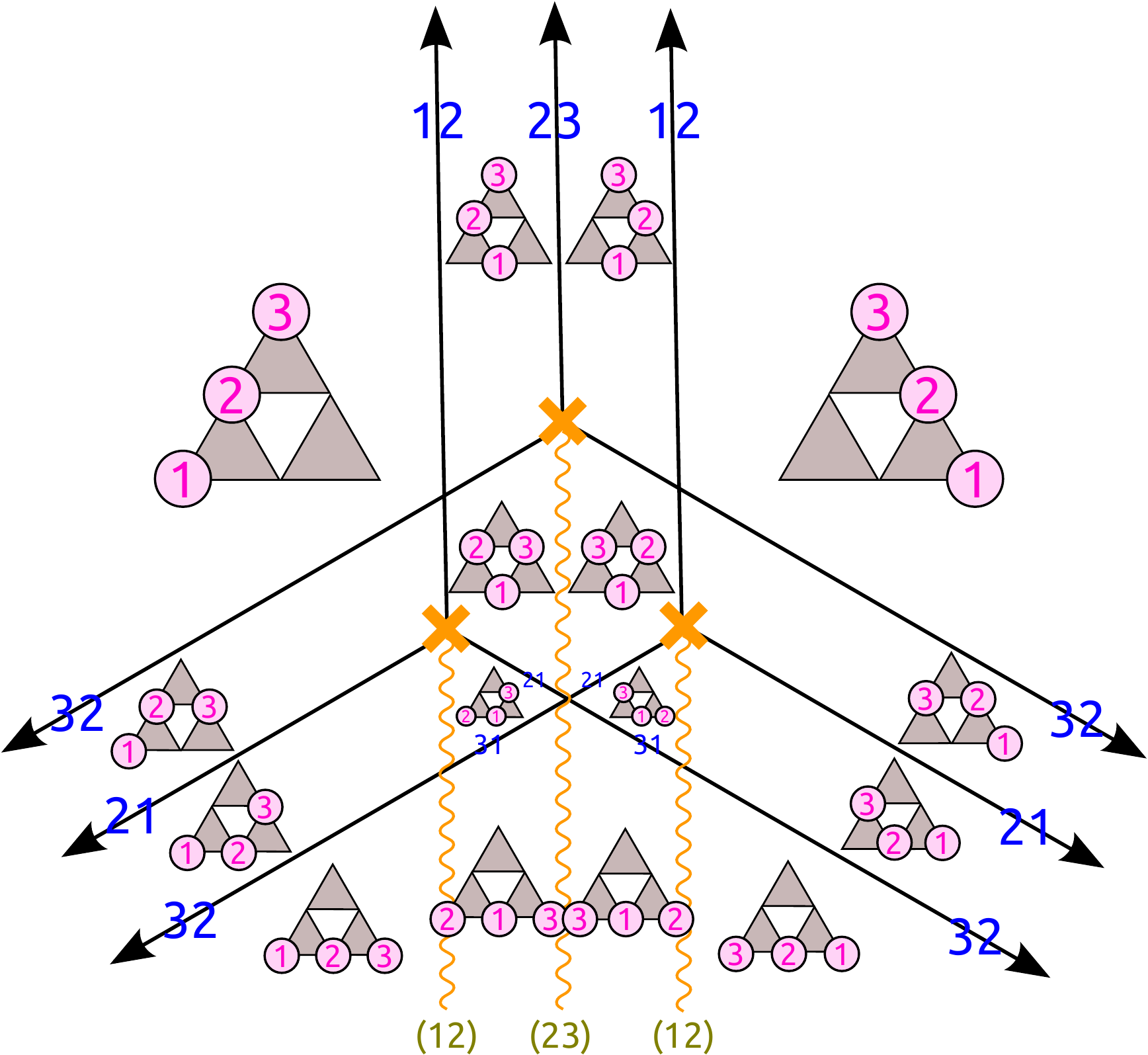}{0.5}{The line decompositions of $\CE$ in the various regions of the complement of a minimal
spectral network, with $K=3$.  In each region we display the canonical $2$-triangle of lines, and mark the three
lines which occur in the line decomposition.  (Decompositions separated by a branch cut are essentially the same, differing
only in the numbering of the lines.)  In all of the asymptotic regions the line decompositions correspond to
snakes, while in the hexagonal interior region the line decompositions do not correspond to any snake.}

Thus far we have focused on the noncompact regions cut out by a minimal
spectral network. For $K>2$, there are of course also compact regions in $C \setminus \wnet$.
The nonabelianization
map again gives a decomposition of $\CE$ into lines associated with these regions,
and the transformations across $\CS$-walls are again governed by the discussion of
Section \S \ref{subsec:Trans-S-Wall}.  Indeed, one can show that the decomposition into lines
associated with an internal region $\CR$ is again given by a
family of $K$ lines each of which is of the type $\fL^{x,y,z}$.
However, these $K$ lines need not lie along a snake.
In Figure \ref{fig:lines-K3} we indicate these decompositions in the case $K=3$.

\subsection{WKB asymptotics}

Let us briefly comment on another important feature of our line decompositions of $\CE$:
as explained in \S 10.9 of \cite{Gaiotto2012},
if our spectral network is a WKB spectral network
$\wnet_{\vartheta}$, these decompositions are naturally related to the WKB asymptotics
of the flat sections as $\zeta \to 0$.  Indeed, in each region of
$C \setminus \wnet_{\vartheta}$, our line decomposition of $\CE$
consists of flat sections with simple asymptotics (schematically
$s_i \sim s_i(z_0)\exp (-  R / \zeta \int_{z_0}^z \lambda_i)$)
as $\zeta \to 0$ in the half-plane $\IH_{\vartheta}$.
When we cross the $\CS$-walls,
the lines of asymptotically larger sections get modified by the addition
of asymptotically smaller sections.  This discussion applies equally well to the
compact and non-compact regions.

\subsection{Formal monodromy}

As we remarked at the beginning of this section,
the change-of-basis matrix obtained by crossing a whole cable of $\CS$-walls can be
interpreted as a Stokes matrix associated to the irregular singularity of $\nabla$ at $z = \infty$.
So we have three such Stokes matrices $T_{a-cable}$, $T_{b-cable}$, $T_{c-cable}$.
Each one is further decomposed into $\CS$-wall matrices according to \eqref{eq:b-cable-factors-1}-\eqref{eq:Tc-cable}.
Some concrete examples of these matrices appear in Appendix \ref{app:snakes-ex}.

In general the product of all the Stokes matrices around an irregular singularity gives the formal monodromy.
In particular, if the mass parameters are all set to $1$, then the formal monodromy is just $W_0$, so
the decomposition of the cables gives a solution to the equation
\eqref{eq:N-Odd-Monodromy}:
\begin{equation}\label{eq:abc-Identity}
T_{c-cable} T_{b-cable} T_{a-cable} = W_0,
\end{equation}
where $W_0$ is the antidiagonal matrix defined in \eqref{eq:W0-DEF}.

\subsection{The \Yang\ case}

So far in this section we have exclusively considered a minimal spectral network of \Yin\ type.
If we consider instead a minimal spectral network of \Yang\ type,
the story is slightly different.
Each region still corresponds to a line decomposition of $\CE$,
but the lines which occur in these decompositions
are not the ones
in the Fock-Goncharov triangle associated to
the three flags $A_\bullet$, $B_\bullet$, $C_\bullet$ in $\CE$.
To describe the decompositions which we do get,
first observe that a line decomposition of $\CE$ induces a
dual line decomposition of $\CE^*$.  Hence in each
region we have a line decomposition of $\CE^*$.
In the \Yang\ case it is these dual
decompositions which turn out to have a simple description:
they are the ones in the Fock-Goncharov triangle associated to the three
flags $\check{A}_\bullet$, $\check{B}_\bullet$, $\check{C}_\bullet$
in $\CE^*$.

In order to check this, it is convenient to use a volume form to identify $\CE^*$
as $\wedge^{K-1} \CE$, and then realize the Fock-Goncharov triangle of lines
in $\CE^*$ as
\begin{equation} \label{eq:dual-lines}
\begin{split}
\hat \fL_K^{\CR} & = \fL_1^{\CR} \wedge \cdots \wedge \fL_{K-1}^{\CR} \\
\hat \fL_{K-1}^{\CR} & = \fL_1^{\CR} \wedge \cdots \wedge \fL_{K-2}^{\CR}\wedge \fL_{K}^{\CR} \\
\hat \fL_{K-2}^{\CR} & = \fL_1^{\CR} \wedge \cdots \wedge \fL_{K-3}^{\CR}
\wedge \fL_{K-1}^{\CR}\wedge \fL_{K}^{\CR}\\
\vdots & \qquad\qquad \vdots \\
\hat \fL_1^{\CR} & = \fL_2^{\CR} \wedge \cdots \wedge \fL_{K }^{\CR}.
\end{split}
\end{equation}
Using this representation one can check directly that the canonical path of snakes in $\CE^*$
is obtained by crossing the $b$-cable of the \Yang network.

\section{Example: \texorpdfstring{$\CS$-walls}{S-walls} and coordinates for the level \texorpdfstring{$3$}{3} lift of \texorpdfstring{$AD_1$}{AD1}}\label{sec:Spin1-N1B}

\insfigscaled{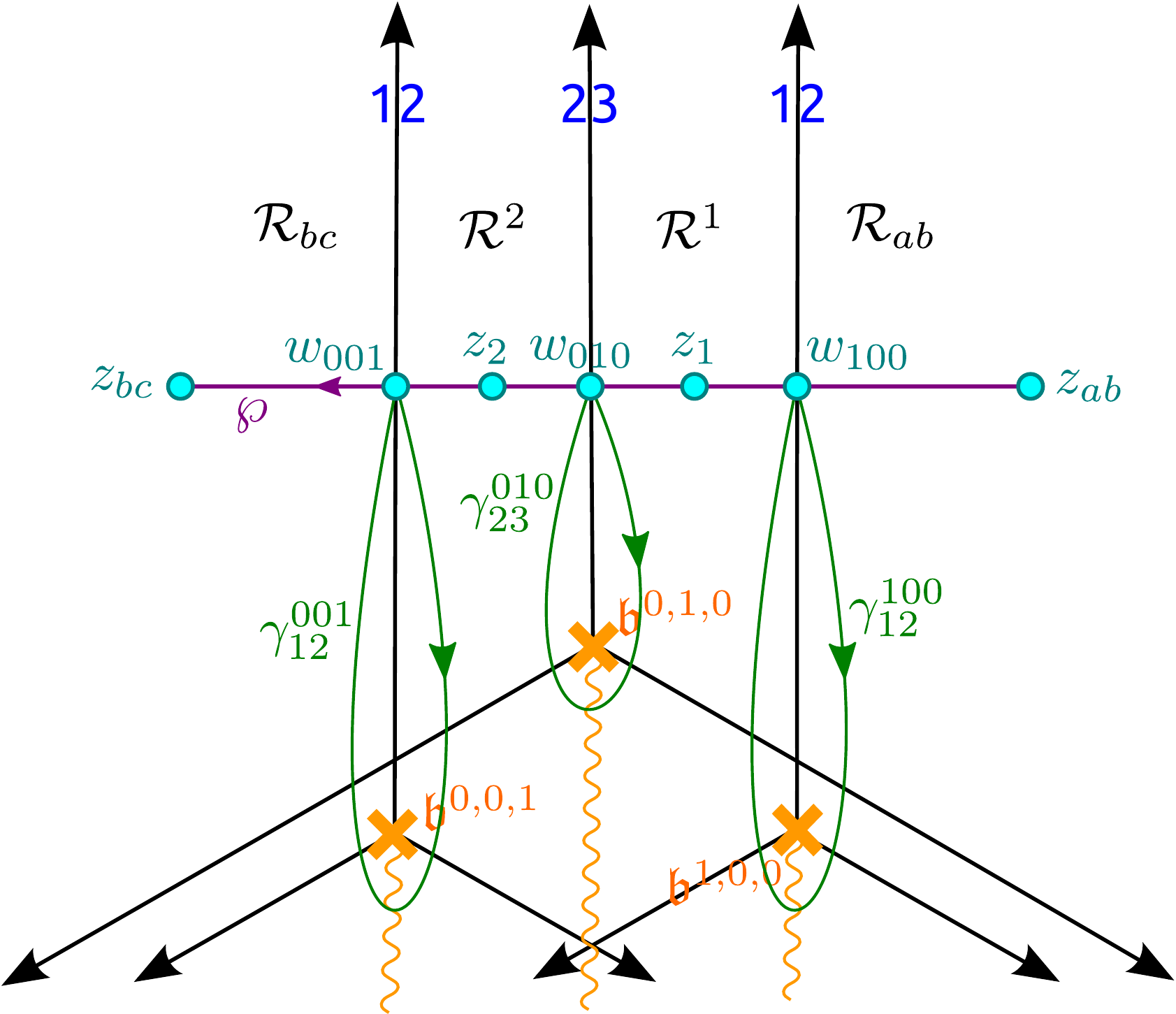}{0.5}{Parallel transport of sections in the $K=3$ lift of $AD_1$.}

\insfigscaled{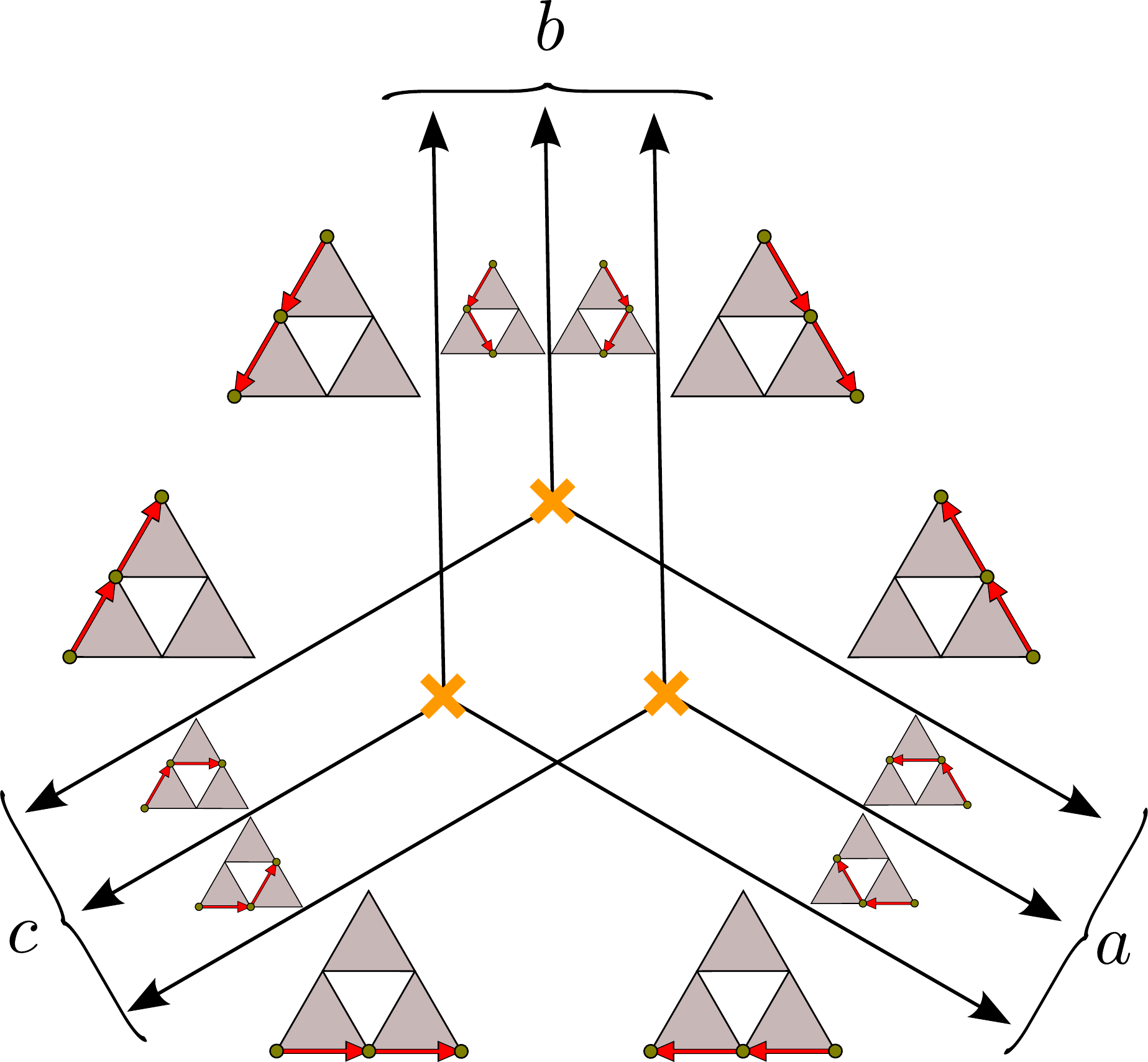}{0.5}{The snakes which are used in our construction of the triangle of lines
for $K=3$. Note that in the regions $\CR_{ab}$, $\CR_{bc}$
and $\CR_{ca}$ we use two different orientations for the snakes, the choice
depending on which cable we choose to cross.
The outer transitions are type I, but the transition across the inner wall in each cable is type II and introduces the
Darboux coordinate $\CY_{\gamma^\fv}$ into the Stokes matrix.}

In this section we illustrate some of the general claims of
Section \ref{sec:FlagsFlatSections}  above  in the case of $K=3$.
We will focus on the changes of basis which one meets in
crossing the $b$-cable, using parallel transport across the
path $\wp$ illustrated in Figure \ref{fig:paths-K3}.
The various snakes which we will use on the $2$-triangle of lines are illustrated in
Figure \ref{fig:snakes-K3}.

We begin by choosing three vectors
\be\label{eq:K3EX-1}
s_i^{\CR_{ab}}(z_{ab}) \in L_i\vert_{z_{ab}}.
\ee
We have made a choice of three scale factors here.
We then extend these three vectors to flat sections $s_i^{\CR_{ab}}\in \CE$.
More formally, we apply the extension map:
$s_i^{\CR_{ab}} = s_i^{\CR_{ab}}(z_{ab}) \varepsilon(z_{ab})$.

Next we define three vectors in $L_i\vert_{z_{1}}$ by
\be\label{eq:K3EX-2}
s_i^{\CR^1}(z_1):= s_i^{\CR_{ab}}(z_{ab}) \CY_{\wp^{(i)}(z_{ab},z_1)} \in L_i\vert_{z_{1}}.
\ee
We stress that \eqref{eq:K3EX-2} is merely a definition of reference sections
in the lines above $z_1$, not parallel transport with respect to the
flat nonabelian connection $\nabla$. Parallel transport with
respect to $\nabla$ of course preserves flat sections:
\be
s_i^{\CR_{ab}}(z_f) = s_i^{\CR_{ab}}(z_i) F(\wp'(z_i,z_f))
\ee
for parallel
transport along \emph{any} path $\wp'(z_i,z_f)$ from $z_i$ to $z_f$.
In particular,
\be\label{eq:K3EX-3}
s_i^{\CR_{ab}}(z_1)= s_i^{\CR_{ab}}(z_{ab}) F(\wp(z_{ab},z_1)).
\ee

Combining \eqref{eq:K3EX-2} and \eqref{eq:K3EX-3} and writing out the
definition of $F(\wp)$, we arrive at
\be\label{eq:K3EX-4}
\begin{split}
s_3^{\CR_{ab}}(z_1) & = s_3^{\CR_{ab}}(z_{ab}) \CY_{\wp^{(3)}(z_{ab},z_1)} = s_3^{\CR^1}(z_1), \\
s_2^{\CR_{ab}}(z_1) & = s_2^{\CR_{ab}}(z_{ab}) \CY_{\wp^{(2)}(z_{ab},z_1)} = s_2^{\CR^1}(z_1), \\
s_1^{\CR_{ab}}(z_1) & = s_1^{\CR_{ab}}(z_{ab}) \CY_{\wp^{(1)}(z_{ab},z_1)}  \\
& + s_1^{\CR_{ab}}(z_{ab}) \CY_{\wp^{(1)}(z_{ab},w^{100} )}
\CY_{\gamma^{100}_{12}(w^{100})} \CY_{\wp^{(2)}(w^{100},z_{1} )}.
\end{split}
\ee
Now we can simplify the last line of \eqref{eq:K3EX-4}, by noting that there must be a
constant $\kappa_1\in \IC^\times$ such that
\be\label{eq:K3EX-5}
s_1^{\CR_{ab}}(z_{ab}) \CY_{\wp^{(1)}(z_{ab},w^{100} )}
\CY_{\gamma^{100}_{12}(w^{100})} \CY_{\wp^{(2)}(w^{100},z_{1} )} = \kappa_1 s_2^{\CR_{ab}}(z_{ab}) \CY_{\wp^{(2)}(z_{ab},z_1)}.
\ee
Applying the extension map $\varepsilon(z_1)$ to the definitions \eqref{eq:K3EX-2}
we define sections $s_i^{\CR^1}\in\CE$. It follows from \eqref{eq:K3EX-4} and \eqref{eq:K3EX-5} that
we have the linear relations:
\be\label{eq:K3EX-5b}
\begin{split}
s_1^{\CR^1} & = s_1^{\CR_{ab}} - \kappa_1 s_2^{\CR_{ab}}, \\
s_2^{\CR^1} & = s_2^{\CR_{ab}}, \\
s_3^{\CR^1} & = s_3^{\CR_{ab}}.
\end{split}
\ee
Now we describe the change of basis from $\CR^1$ to $\CR^2$ in an
entirely analogous way. We begin by defining
\be\label{eq:K3EX-6}
s_i^{\CR^2}(z_2):= s_i^{\CR^1}(z_1) \CY_{\wp^{(i)}(z_1,z_2)}.
\ee
On the other hand, since $s_i^{\CR^1}$ are flat sections we have
\be\label{eq:K3EX-7}
s_i^{\CR^1}(z_2)= s_i^{\CR^1}(z_1) F(\wp(z_1,z_2)).
\ee
Writing this out using the definition of $F$,
we arrive at equations completely analogous to \eqref{eq:K3EX-4}.
In analogy to \eqref{eq:K3EX-5}, we define $\kappa_2\in \IC^\times$ by
\be\label{eq:K3EX-8}
s_2^{\CR^1}(z_1) \CY_{\wp^{(2)}(z_1,w^{010} )}
\CY_{\gamma^{010}_{23}(w^{010})} \CY_{\wp^{(3)}(w^{010},z_{2} )}= \kappa_2 s_3^{\CR^1}(z_1) \CY_{\wp^{(3)}(z_1,z_2)}.
\ee
This leads to the analog of \eqref{eq:K3EX-5b}:
\be\label{eq:K3EX-9}
\begin{split}
s_1^{\CR^2} & = s_1^{\CR^1}, \\
s_2^{\CR^2} & = s_2^{\CR^1} - \kappa_2 s_3^{\CR^1}, \\
s_3^{\CR^2} & = s_3^{\CR^1}.
\end{split}
\ee
Finally, proceeding from $\CR^2$ to $\CR_{bc}$,
we introduce a third constant $\kappa_3\in \IC^\times$ in the equation
\be\label{eq:K3EX-10}
s_1^{\CR^2}(z_2) \CY_{\wp^{(1)}(z_2,w^{001} )}
\CY_{\gamma^{001}_{12}(w^{001})} \CY_{\wp^{(2)}(w^{001},z_{bc})} = \kappa_3 s_2^{\CR^2}(z_2) \CY_{\wp^{(2)}(z_2,z_{bc})},
\ee
leading to the change of basis
\be\label{eq:K3EX-11}
\begin{split}
s_1^{\CR_{bc}} & = s_1^{\CR^2}- \kappa_3 s_2^{\CR^2},    \\
s_2^{\CR_{bc}} & = s_2^{\CR^2},  \\
s_3^{\CR_{bc}} & = s_3^{\CR^2}.
\end{split}
\ee
Combining these linear transformations, we have altogether
\be\label{eq:K3EX-12}
\begin{split}
s_1^{\CR_{bc}} & = s_1^{\CR_{ab}} - (\kappa_1+\kappa_3) s_2^{\CR_{ab}} + \kappa_2 \kappa_3 s_3^{\CR_{ab} }, \\
s_2^{\CR_{bc}} & = s_2^{\CR_{ab}} - \kappa_2 s_3^{\CR_{ab}}, \\
s_3^{\CR_{bc}} & = s_3^{\CR_{ab}}.
\end{split}
\ee

\insfigscaled{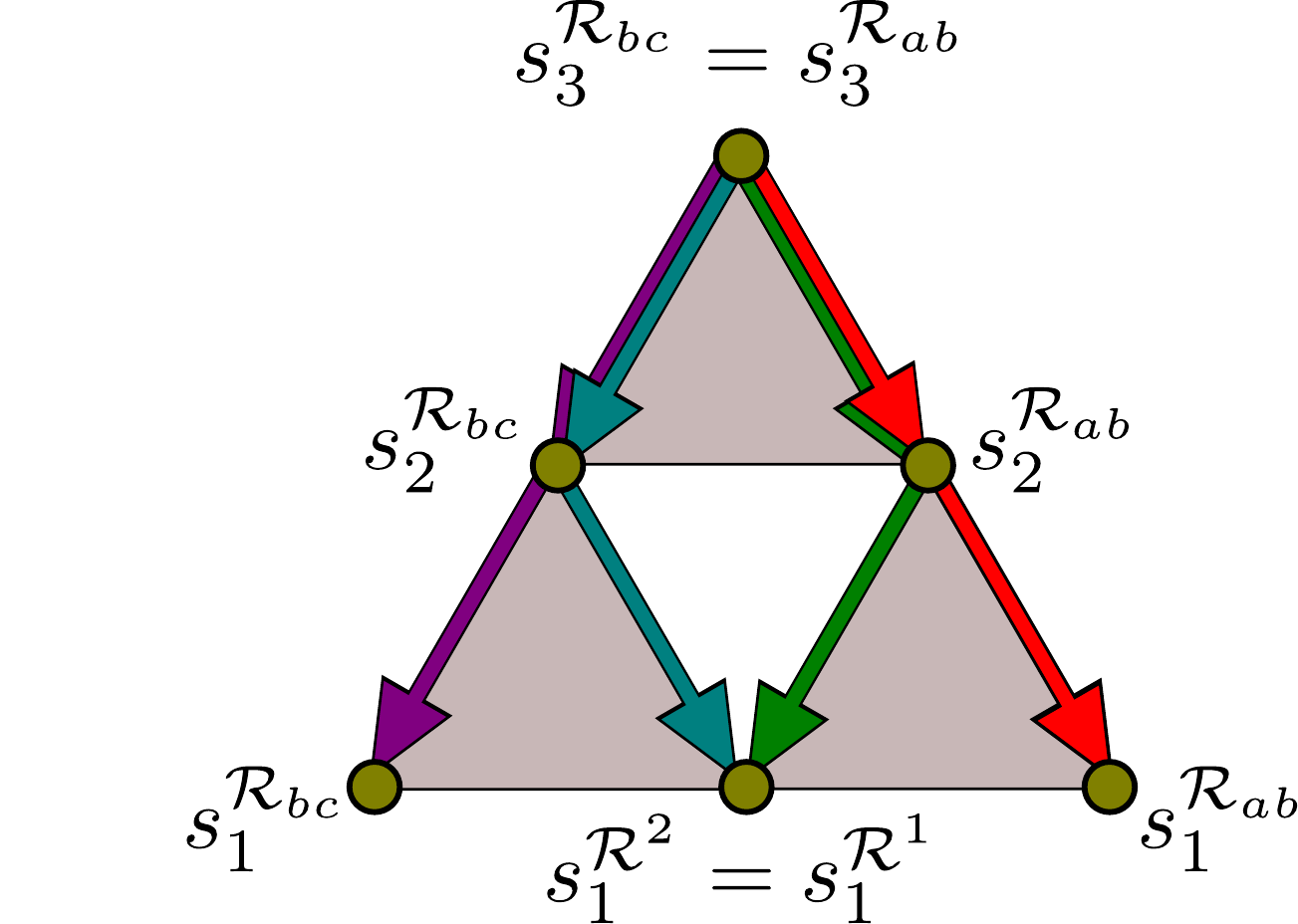}{0.42}{The canonical $2$-triangle of lines in the $K=3$ lift of the $AD_1$ theory.  We also indicate basis sections in the various lines, and the snakes in the canonical path.}

Now, we consider the canonical $2$-triangle of lines, illustrated in
Figure \ref{fig:K3-example-sections}.  We choose normalizations so that the basis in $\CR_{ab}$ and
$\CR_{bc}$ coincide with those determined by the snakes indicated in
Figure \ref{fig:K3-example-sections}.
For $s_1^{\CR_{ab}}$ to be obtained from $s_2^{\CR_{ab}}$ by the canonical hom, a glance at
\eqref{eq:K3EX-5b} shows we must choose $\kappa_1=-1$.
Similarly, for $s_2^{\CR_{ab}}$ to be obtained from $s_3^{\CR_{ab}}$ by the canonical hom,
\eqref{eq:K3EX-12} shows that we must choose $\kappa_2=-1$.

At this point we have fixed the normalization of $s_2^{\CR_{ab}}$ and
$s_3^{\CR_{ab}}$ relative to $s_1^{\CR_{ab}}$,
as expected. The remaining constant $\kappa_3$ is therefore invariant.

We are now in a position to show that the coordinate $\kappa_3$ is related to the
coordinate Fock and Goncharov associate to a triple of flags in a
three-dimensional vector space. See \eqref{eq:tripleratio}. In the present case we
have three flags:
\begin{equation}
A_\bullet :\quad  s_1^{\CR_{ab}} \IC \subset s_1^{\CR_{ab}} \IC\oplus s_2^{\CR_{ab}} \IC \subset \CE,
\end{equation}
\begin{equation}
B_\bullet :\quad  s_3^{\CR_{ab}} \IC \subset s_3^{\CR_{ab}} \IC\oplus s_2^{\CR_{ab}} \IC \subset \CE,
\end{equation}
\begin{equation}
C_\bullet :\quad  s_1^{\CR_{bc}} \IC \subset s_1^{\CR_{bc}} \IC\oplus s_2^{\CR_{bc}} \IC \subset \CE.
\end{equation}
Using the change of basis \eqref{eq:K3EX-12} and the definition \eqref{eq:tripleratio},
or, more elegantly, using equations \eqref{eq:K3EX-5b}, \eqref{eq:K3EX-9}, \eqref{eq:K3EX-11},
and \eqref{eq:TwoHomCom}, we find that the triple ratio for these three flags is
\begin{equation}\label{eq:K3EX-15}
r(A_{\bullet}, B_{\bullet}, C_{\bullet}) = \kappa_3^{-1}.
\end{equation}
On the other hand, comparing \eqref{eq:K3EX-5} with  \eqref{eq:K3EX-10},
then using the definitions \eqref{eq:K3EX-2} and \eqref{eq:K3EX-6},
we can deduce
\begin{multline} \label{eq:K3EX-13}
s_1^{\CR_{ab}}(z_{ab}) \CY_{\wp^{(1)}(z_{ab},w^{001} )}
\CY_{\gamma^{001}_{12}(w^{001})} \CY_{\wp^{(2)}(w^{001},z_{bc} )} =\\ \kappa_3
 s_1^{\CR_{ab}}(z_{ab}) \CY_{\wp^{(1)}(z_{ab},w^{100} )}
\CY_{\gamma^{100}_{12}(w^{100})} \CY_{\wp^{(2)}(w^{100},z_{bc} )}.
\end{multline}
The two sides here are parallel transport of the same section around two different paths,
differing by $\gamma^{000}$; from this we learn
\be\label{eq:K3EX-14}
\kappa_3= \CY_{\gamma^{000}}^{-1}.
\ee
Combining this with equation \eqref{eq:K3EX-15} we confirm our main result,
\be\label{eq:K3EX-16}
r(A_{\bullet}, B_{\bullet}, C_{\bullet})  = \CY_{\gamma^{000}}.
\ee

\section{How to find the BPS spectrum of the level \texorpdfstring{$K$}{K} lift of \texorpdfstring{$AD_1$}{AD1}}\label{sec:Vary-Theta}

In this section we will give a recipe for determining the BPS spectra of
level $K$ lifts of $AD_1$.

\subsection{The spectrum generator} \label{sec:spec-gen}

We first recall that the general formalism of wall-crossing in $\N=2$ theories
\cite{ks1,Gaiotto:2008cd,Cecotti:2009uf,Dimofte:2009tm,Gaiotto:2010be,Manschot:2010qz,Pioline:2011gf}
implies that the BPS spectrum of any $\N=2$ theory can be organized into a single natural object, the
\emph{spectrum generator} $\IS$.
Abstractly speaking, $\IS$ is simply the birational transformation which relates the quantities
$\CY_\gamma$ computed at a phase $\vartheta$ to the $\CY_\gamma$ at phase $\vartheta + \pi$ (we will
see below what this means concretely in theories of class $S$).
$\IS$ is a wall-crossing invariant (up to conjugation), but one can extract from it
all of the BPS degeneracies $\Omega(\gamma)$.  The idea is that after fixing a point of the
Coulomb branch and hence in particular an ordering of the ``BPS phases'' $\arg -Z_\gamma$,
$\IS$ admits a unique factorization into a product of the form
\begin{equation} \label{eq:factorization}
 \IS = \prod_{\gamma: \vartheta \le \arg -Z_\gamma < \vartheta+\pi} \CK_\gamma^{\Omega(\gamma)},
\end{equation}
where the product is taken in order of increasing $\arg -Z_\gamma$,
and $\CK_\gamma$ denotes a transformation of the form\footnote{The term $\sigma(\gamma) \CY_\gamma$
appearing in \eqref{eq:k-def} needs a little explanation:  $\CY_\gamma$ is the holonomy of the twisted
connection $\nabla^\ab$ along the canonical lift of $\gamma$ to $H_1(\tilde\Sigma, \IZ)$, and
$\sigma(\gamma) = \pm 1$ is a tricky sign, conjecturally a canonical quadratic refinement of the $\mod\,2$
intersection pairing in $\Gamma$ (this conjecture is true at least for $K=2$).}
\begin{equation} \label{eq:k-def}
 \CK_\gamma : \CY_{\gamma'} \to \CY_{\gamma'}(1 - \sigma(\gamma) \CY_\gamma)^{\inprod{\gamma',\gamma}}.
\end{equation}
Different chambers of the Coulomb
branch correspond to different orderings of the $\arg Z_\gamma$ and thus to different factorizations
of $\IS$; this is why the $\Omega(\gamma)$ can undergo wall-crossing even while $\IS$ is invariant.
Indeed, because it is more invariant, in some sense $\IS$ is superior to the $\Omega(\gamma)$ as
a way of packaging the BPS particle content of the theory.

It was pointed out in \cite{Gaiotto:2009hg} that the spectrum generator
$\IS$ can sometimes be computed directly without prior knowledge of the
BPS spectrum (hence the name). In particular an algorithm based on
spectral networks (actually, on their dual WKB triangulations) was given
for $A_1$ theories in \cite{Gaiotto:2009hg}.\footnote{A nice application of this description of
$\IS$ appeared recently in \cite{Caetano2012}:  there it was noted that
while the TBA-like integral equations of \cite{Gaiotto:2008cd} depend on knowing the $\Omega(\gamma)$,
there is a variant of those equations which depends directly on $\IS$, and this variant can be easier
to use in the application to moduli of Hitchin systems.  This could be further extended to produce
solutions of Hitchin equations themselves, using the extended 2d-4d integral equations in \cite{Gaiotto:2011tf}
and the extended 2d-4d spectrum generator for $A_1$ theories given in \cite{Longhi2012}.}
Below we will determine $\IS$ for level $K$ lifts of $AD_1$.  We will first consider the case
$K=3$ where we will read off the individual factors $\CK_\gamma^{\Omega(\gamma)}$ directly
by looking at the WKB spectral networks (indeed, it will turn out that in this case there is
just a single factor $\CK_\gamma$).
We will then turn to general $K$ where we will use a more powerful approach,
which gives $\IS$ directly rather than its factorization.

\subsection{Morphisms of WKB spectral networks for the level \texorpdfstring{$3$}{3} lift of \texorpdfstring{$AD_1$}{AD1}}

We begin with the simple case $K=3$.

\insfigscaled{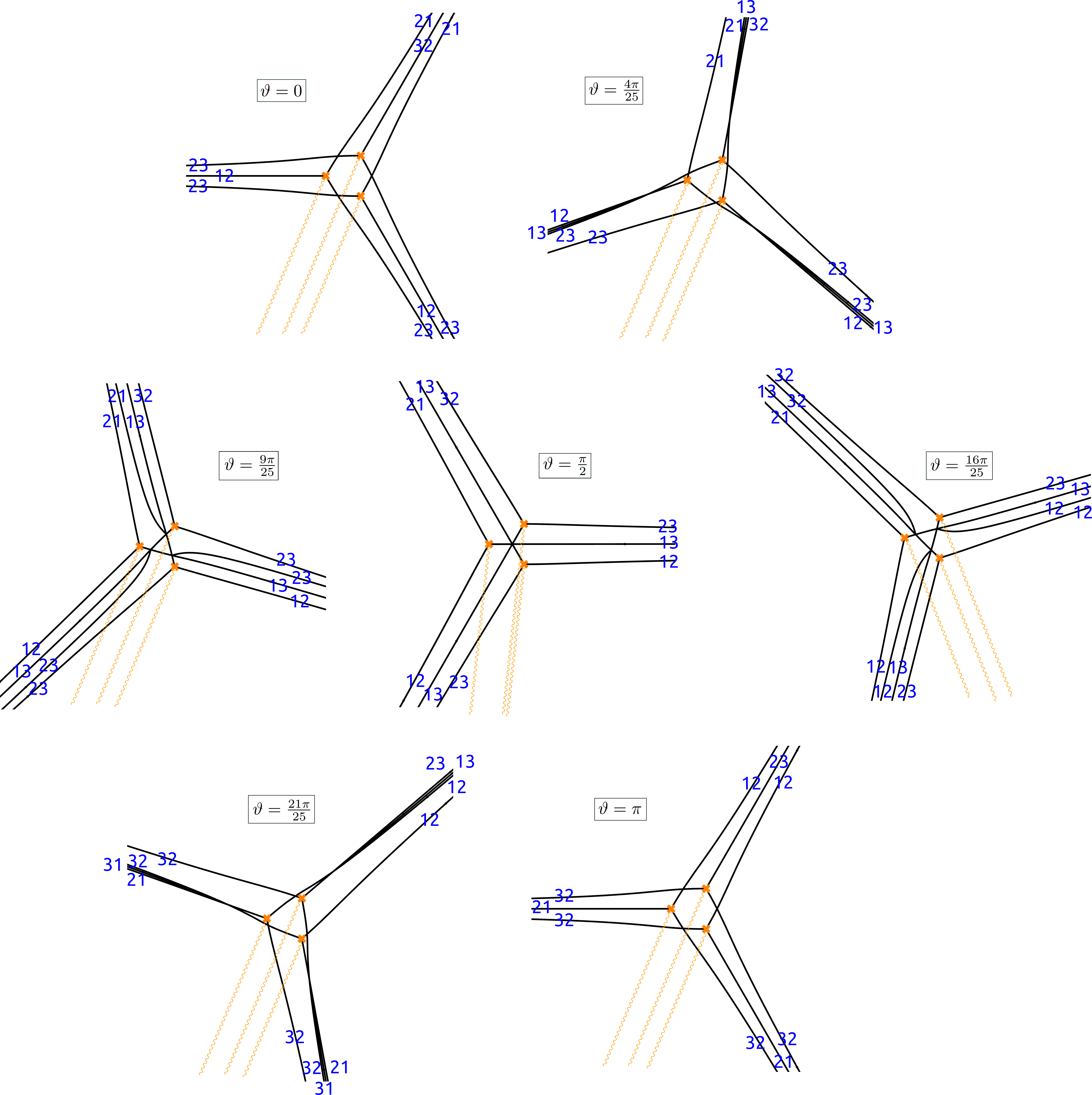}{0.155}{WKB spectral networks $\wnet_\vartheta$
in the $K=3$ lift of the $AD_1$ theory, at the point of the Coulomb branch with
spectral curve $\lambda^3 + (4z \de z^2) \lambda + \de z^3 = 0$
at seven values of $\vartheta$.
Note the three-pronged web which appears in the middle of the figure
at $\vartheta = \pi/2$, and the topology change as we cross this critical phase.}

In Figure \ref{fig:detailed-flip} we depict the variation of the
WKB spectral network $\wnet_\vartheta$ with $\vartheta$
in the $K=3$ lift of the $AD_1$ theory, at the point of the Coulomb branch with
spectral curve
\be
\lambda^3 + (4z \de z^2) \lambda + \de z^3 = 0.
\ee
(Compare the unperturbed equation \eqref{eq:explicit}.)
At $\vartheta = 0$ we have a \Yang\ type
minimal network, while at $\vartheta = \pi$ we have a \Yin\ type minimal network.

On the top line of Figure \ref{fig:detailed-flip} we see the evolution from $\vartheta = 0$
to $\vartheta = \frac{4 \pi}{25}$.  At $\vartheta = 0$ the network is truly minimal,
while at $\vartheta = \frac{4 \pi}{25}$ it is only essentially minimal, i.e. equivalent
to a minimal network, in the sense of \S 9 of \cite{Gaiotto2012}.  For example, in evolving
from $\vartheta = 0$ to $\vartheta = \frac{4 \pi}{25}$, in the cable pointing to the north,
an $\CS$-wall of type 12 and one of type 23 have crossed one another, producing a new joint
from which an $\CS$-wall of type 13 emerges.
For our present purposes, though, the distinction between minimal and essentially minimal is of
little consequence.  What we are really interested in is the phases where $\wnet_\vartheta$ jumps
in a way that is \ti{not} an equivalence.

The latter phenomenon happens in the middle row
of Figure \ref{fig:detailed-flip}.
Focusing on this row, we observe that there is an internal triangle, first visible at
$\vartheta = \frac{9 \pi}{25}$,
which shrinks as $\vartheta$ increases and collapses completely at
the critical phase $\vartheta_c = \frac{\pi}{2}$.  At this critical phase the spectral network degenerates
to include a three-pronged
string web.  This string web lifts
to the cycle $\gamma= \gamma^{000}$ associated to the triangle of branch points
in \S\ref{subsec:cycles-MSN} above.

Such degenerations of spectral networks were studied in \S6 of \cite{Gaiotto2012}.
As explained there, the parallel transport operators $F(\wp)$ jump discontinuously whenever $\vartheta$ crosses a critical phase.
Writing $F(\wp, \vartheta)$ for the parallel transport induced by the
spectral network $\wnet_\vartheta$ and a fixed abelian connection $\nabla^\ab$,
we have
\begin{equation} \label{eq:Fplus-Fminus}
F(\wp, \vartheta^+) = \CK F(\wp, \vartheta^-),
\end{equation}
for a particular transformation $\CK$ which contains the information of the 4d BPS degeneracies
for particles with BPS phase $\vartheta$.
$\CK$ can be determined from the degenerate network $\wnet_\vartheta$ at the critical phase.
The specific type of degenerate network we see at $\vartheta = \frac{\pi}{2}$ in
Figure \ref{fig:detailed-flip} was considered in \S7.3 of \cite{Gaiotto2012},
where we found that $\CK$ is given by
\begin{equation}
\CK = \CK_{\gamma^{000}} : \CY_{\gamma'} \to \CY_{\gamma'} (1 + \CY_{\gamma^{000}})^{\inprod{\gamma',\gamma^{000}}}.
\end{equation}
Comparing this with the general story reviewed in \S\ref{sec:spec-gen}, we see that this
corresponds to the BPS degeneracy $\Omega(\gamma^{000}) = 1$, i.e. to
a single BPS hypermultiplet of charge $\gamma^{000}$ (with $\sigma(\gamma^{000}) = -1$).

This kind of analysis by looking directly at the evolution of the WKB spectral network rapidly becomes
cumbersome as we increase $K$.  In the next section we turn to a more efficient algebraic method.

\subsection{Determining the spectrum generator}

We now describe a scheme which allows one to write the spectrum generator $\IS$ explicitly
for all level $K$ lifts of the $AD_1$ theory.  We also exhibit the answers for $K = 3, 4, 5$.

We consider some phase $\vartheta$ and point of the Coulomb branch
such that the WKB spectral network $\wnet_\vartheta$ is an essentially minimal
spectral network of \Yin type.  For any flat connection $\nabla$ over $C$, there is
a flat abelian connection $\nabla^\ab_\vartheta$ over $\Sigma$ such that
\be
\Psi_{\CW_{\vartheta}}(\nabla^\ab_{\vartheta}) = \nabla.
\ee
As we have described in \S\ref{sec:FlagsFlatSections},
for each $\fv \in T(K-3)$,
the holonomy $\CY_{\gamma^{\fv}}$ of $\nabla^\ab_\vartheta$ is equal to
the coordinate $r^{\fv}$ associated with the flags
$A_{\bullet}, B_{\bullet}, C_{\bullet}$ determined by $\nabla$.
There is another abelian connection $\nabla^\ab_{\vartheta+\pi}$ such that
\be \label{eq:w-gym}
\Psi_{\CW_{\vartheta+\pi}}(\nabla^\ab_{\vartheta+\pi}) = \nabla.
\ee
The spectrum generator $\IS$ is
the birational transformation which takes the holonomies
$\CY_{\gamma}$ of $\nabla^\ab_{\vartheta}$ to the holonomies $\hat \CY_{\gamma}$
of $\nabla^\ab_{\vartheta+\pi}$.

Combining \eqref{eq:w-gym} with (10.16) of \cite{Gaiotto2012}, we obtain
\be\label{eq:sn-dualident}
\Psi_{\wnet_{\vartheta}}((\nabla^\ab_{\vartheta+\pi})^*) = \nabla^*.
\ee
It follows that the holonomies of $(\nabla^\ab_{\vartheta+\pi})^*$ are the coordinates
$\check{r}^\fv$ associated to the flags
determined by $\nabla^*$, i.e. to the
dual flags $\check{A}_\bullet, \check{B}_\bullet, \check{C}_\bullet$.
The desired $\check\cY_\gamma$
are the holonomies of $\nabla^\ab_{\vartheta+\pi}$, which are thus given by $\check\cY_{\gamma^\fv} = 1 / \check{r}^\fv$.

Thus, the problem of determining the spectrum generator $\IS$ is really
a question of linear algebra:  how are the coordinates of a triple of flags related to those
of the dual triple?  We answer this question in Appendix \ref{subsec:rrcheck}:
the formula \eqref{eq:explicit-sg} gives an explicit way to
compute $\check r^\fv$ as functions of the $r^\fv$.

\subsubsection{Example: \texorpdfstring{$K=3$}{K=3}}

As a simple example we first consider $K=3$.  As we work out
in Appendix \ref{app:cable3}, in this case the relation between
$r$ and $\hat{r}$ is particularly simple:
\begin{equation}
 \hat r^{000} = 1 / r^{000}.
\end{equation}
Therefore
\begin{equation}
 \hat \CY_{\gamma^{000}} = \CY_{\gamma^{000}}.
\end{equation}
This is just as expected:  the charge lattice contains only the single charge $\gamma^{000}$,
which is pure flavor, so that the transformation $\cK_{\gamma^{000}}$ induced by the
BPS state of charge $\gamma^{000}$ acts trivially on $\CY_{\gamma^{000}}$.

\subsubsection{Example: \texorpdfstring{$K=4$}{K=4}}

The next example is $K=4$.  In Appendix \ref{app:cable4} we show that in this case
\begin{equation}\label{eq:explicit-rrcheck-K4}
\begin{split}
\hat r^{100} & = \frac{1 + r^{100} + r^{100} r^{010}}{r^{010} \left(1 + r^{001} + r^{001}r^{100}\right)},\\
\hat r^{010} & = \frac{1 + r^{010} + r^{010} r^{001}}{r^{001} \left(1 + r^{100} + r^{100}r^{010}\right)},\\
\hat r^{001} & = \frac{1 + r^{001} + r^{001} r^{100}}{r^{100} \left(1 + r^{010} + r^{010}r^{001}\right)}.
\end{split}
\end{equation}

This exactly matches the result for the $AD_4$ theory discussed in \cite{Gaiotto:2009hg},
equations (11.45)-(11.47).  (This is not a coincidence:  upon writing $\lambda = y\,\de z$ one finds
that the spectral curves in this theory are related to those of the $AD_4$ theory by the exchange
$y \leftrightarrow z$.  In the language of \cite{Cecotti:2010fi}, the theory we are studying here is the $(A_3, A_1)$ theory
while the $AD_4$ theory is the $(A_1, A_3)$ theory.)
We can therefore borrow some results from \cite{Gaiotto:2009hg}.
In particular, as explained in \S 11.3 of \cite{Gaiotto:2009hg}, the factorization
of this spectrum generator shows that there are four BPS states.  These consist of three
states carrying the three charges $\gamma^{\fv}$, together with a fourth state which is a bound state
of two of the original three.
Which two of the three form a bound state depends on which region of the Coulomb
branch we are in.

\insfigscaled{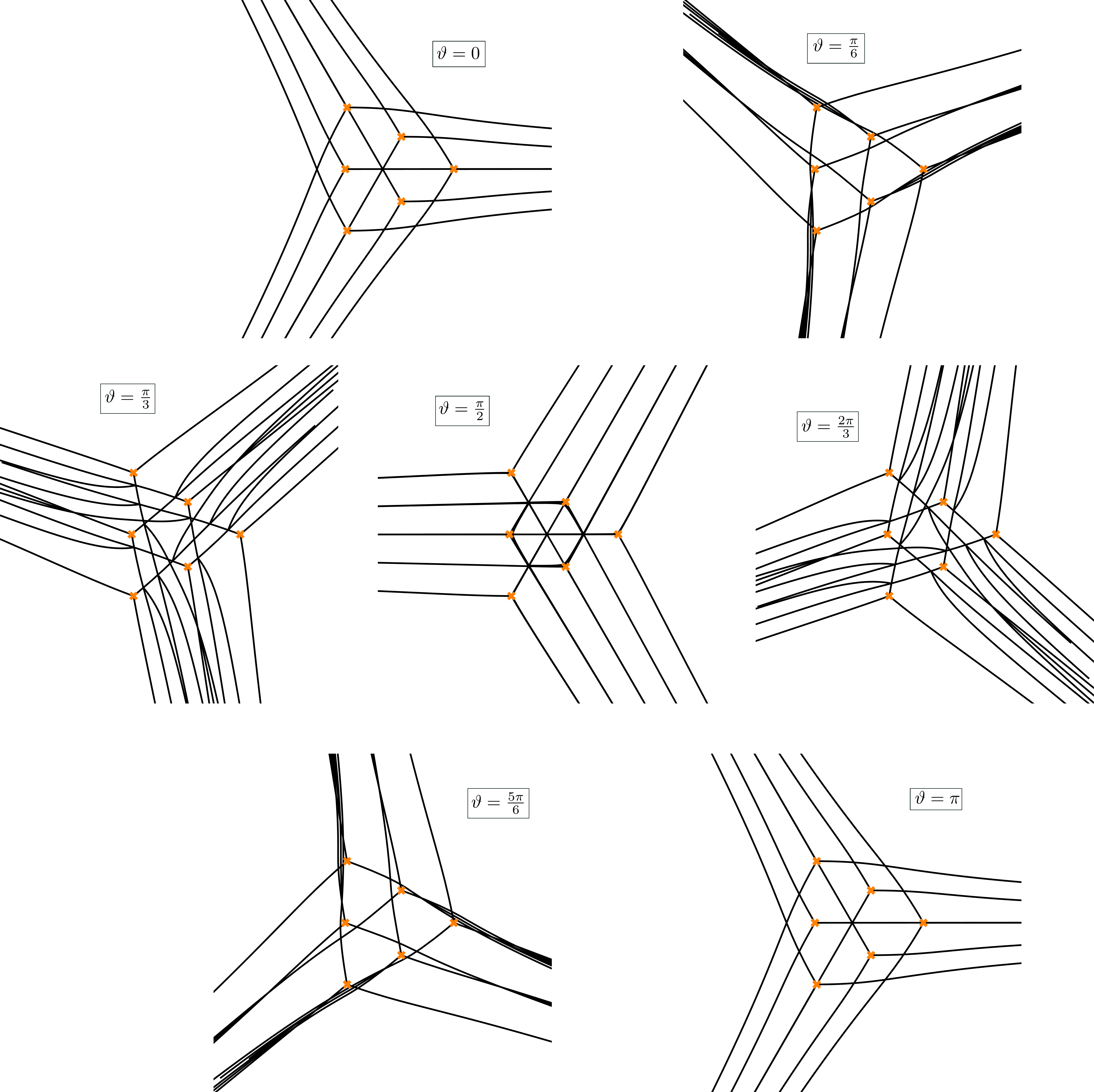}{0.15}{WKB spectral networks for the
spectral curve $\lambda^4 - 10 z (\de z)^2 \lambda^2 + 4 (\de z)^3 \lambda + 9 z^2 (\de z)^4 = 0$,
as $\vartheta$ varies from $0$ to $\pi$.  There is a nontrivial jump at $\vartheta = \frac{\pi}{2}$.}

\insfigscaled{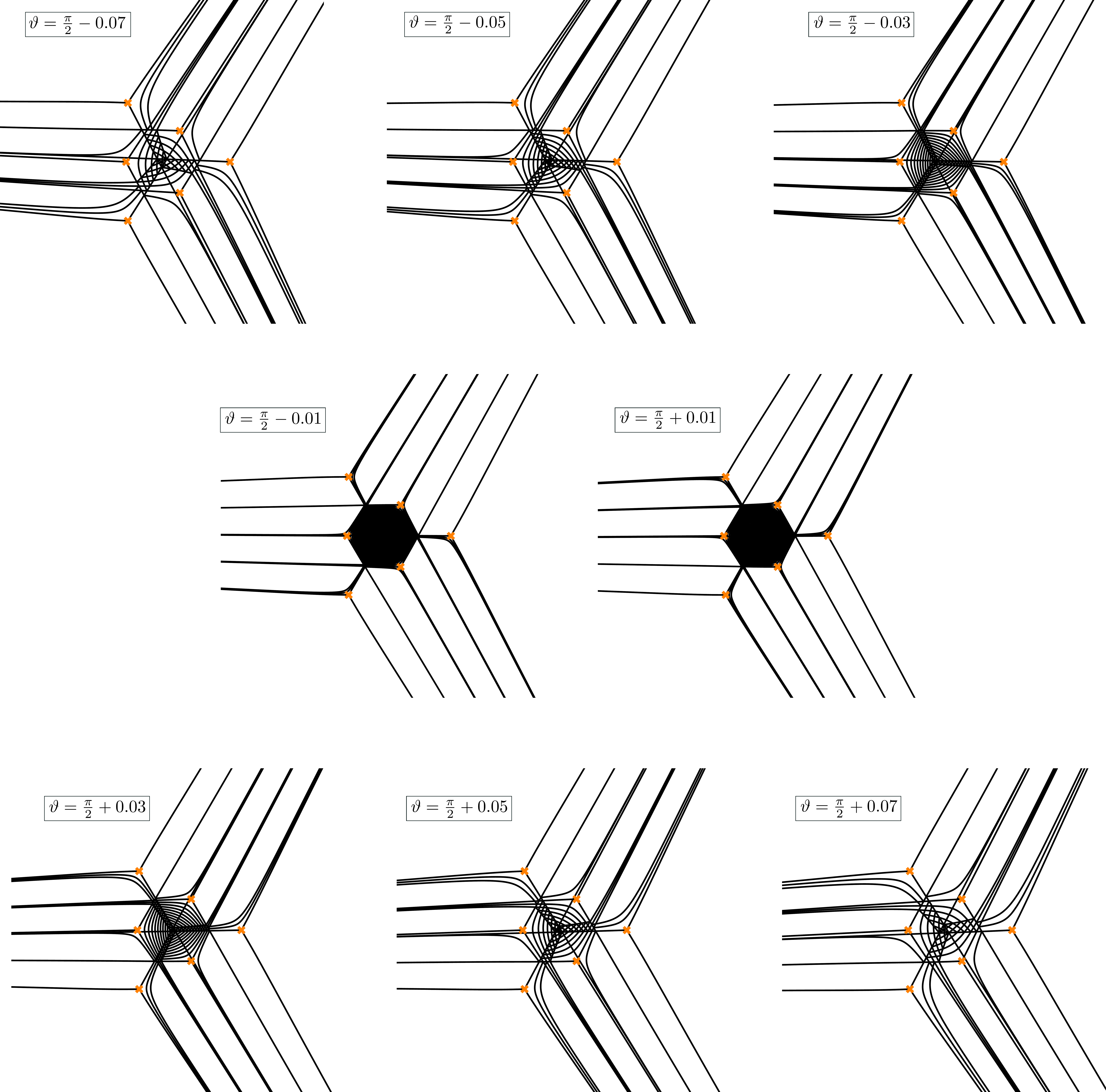}{0.15}{WKB spectral networks for the
spectral curve $\lambda^4 - 10 z (\de z)^2 \lambda^2 + 4 (\de z)^3 \lambda + 9 z^2 (\de z)^4 = 0$,
as $\vartheta$ varies from $\frac{\pi}{2} - 0.07$ to $\frac{\pi}{2} + 0.07$.}

In Figure \ref{fig:detailed-evolution-K4} we show the
variation of the spectral network $\CW_{\vartheta}$ as $\vartheta$ varies from $0$ to $\pi$,
at a point of the Coulomb branch where the spectral curve is
\be
\lambda^4 - 10 z (\de z)^2 \lambda^2 + 4 (\de z)^3 \lambda + 9 z^2 (\de z)^4 = 0.
\ee
(Compare \eqref{eq:explicit} for the unperturbed equation.)
The variation is by equivalences except for a nontrivial jump at $\vartheta = \pi/2$.

In Figure \ref{fig:zoomed-evolution-K4} we zoom in on the behavior near this critical phase.
Note that this behavior looks more complicated than the BPS states we have encountered to this point:
at $\vartheta \to \pi/2$ an infinite number of winding $\CS$-walls coalesce onto the hexagon visible
in Figure \ref{fig:detailed-evolution-K4}.  This behavior is still treatable using the methods of \cite{Gaiotto2012}
(in particular it is somewhat similar to what happens in the example of \S7.2 of that paper), but it is
rather subtle.
Part of the reason for this difficulty is that at the point of the Coulomb branch we have chosen to consider there is
an exact $\IZ_3$ symmetry (the spectral curve $\Sigma$ is invariant under $(z,\lambda) \mapsto (e^{2 \pi \I / 3} z,\lambda)$),
and in particular all three
of the $Z_{\gamma^\fv}$ have the same phase.  Since the $\gamma^\fv$ are mutually non-local charges and all support
BPS states, this means we are sitting on a wall of marginal stability, where we may expect that the picture becomes
more intricate.
Perturbing slightly to move off the wall, we should expect to
find the four states predicted in the previous paragraph.  Indeed, we have confirmed experimentally
that this is the case.

\subsubsection{Example: \texorpdfstring{$K=5$}{K=5}}

Just for the record, we also give the spectrum generator for $K=5$.
The new coordinates at the vertices of the $2$-triangle of spaces
are given by a fairly simple formula:
\be\label{eq:explicit-rrcheck-K5}
\begin{split}
\hat r^{200} & = \frac{ (
 1 + r^{110} + r^{020} r^{110} + r^{101} r^{110} + r^{020} r^{101} r^{110} +
  r^{011} r^{020} r^{101} r^{110})}{
 r^{020} (1 + r^{011} + r^{002} r^{011} + r^{011} r^{110} + r^{002} r^{011} r^{110} +
    r^{002} r^{011} r^{101} r^{110})},\\
    \hat r^{020} & = \frac{ (
 1 + r^{011} + r^{002} r^{011} + r^{011} r^{110} + r^{002} r^{011} r^{110} +
  r^{002} r^{011} r^{101} r^{110})}{
 r^{002} (1 + r^{101} + r^{011} r^{101} + r^{101} r^{200} + r^{011} r^{101} r^{200} +
    r^{011} r^{101} r^{110} r^{200})},\\
    \hat r^{002} &= \frac{ (
 1 + r^{101} + r^{011} r^{101} + r^{101} r^{200} + r^{011} r^{101} r^{200} +
  r^{011} r^{101} r^{110} r^{200})}{r^{200}(1 + r^{110} + r^{020} r^{110} + r^{101} r^{110} +
    r^{020} r^{101} r^{110} + r^{011} r^{020} r^{101} r^{110}) },
\end{split}
\ee
while the transformation laws for the intermediate coordinates are a bit more intricate:
\be
\begin{split}
\hat r^{110} & = \frac{ A^{110} B^{110} }{r^{011} C^{110} D^{110} }, \\
A^{110} & =  (1 + r^{101} + r^{011} r^{101} + r^{101} r^{200} +
    r^{011} r^{101} r^{200} + r^{011} r^{101} r^{110} r^{200}),\\
    B^{110} & = (1 + r^{020} + r^{011} r^{020} +
    r^{002} r^{011} r^{020}),\\
C^{110} &= (1 + r^{110} + r^{020} r^{110} + r^{101} r^{110} + r^{020} r^{101} r^{110} +
    r^{011} r^{020} r^{101} r^{110}),\\
D^{110} & = (1 + r^{002} + r^{002} r^{101} + r^{002} r^{101} r^{200}),
\end{split}
\ee
\be
\begin{split}
\hat r^{101} & = \frac{ A^{101} B^{101} }{r^{110} C^{101} D^{101}   }, \\
A^{101} &= (1 + r^{011} + r^{002} r^{011} +
    r^{011} r^{110} + r^{002} r^{011} r^{110} + r^{002} r^{011} r^{101} r^{110}),\\
B^{101} &= (1 + r^{200} +
    r^{110} r^{200} + r^{020} r^{110} r^{200}),\\
C^{101} &=   (1 + r^{101} + r^{011} r^{101} + r^{101} r^{200} +
    r^{011} r^{101} r^{200} + r^{011} r^{101} r^{110} r^{200}),\\
D^{101} &= (1 + r^{020} + r^{011} r^{020} +
    r^{002} r^{011} r^{020}),
\end{split}
\ee
\be
\begin{split}
\hat r^{011} &= \frac{ A^{011} B^{011} }{r^{101}  C^{011} D^{011} }, \\
A^{011} &=     (1 + r^{110} +
    r^{020} r^{110} + r^{101} r^{110} + r^{020} r^{101} r^{110} + r^{011} r^{020} r^{101} r^{110}),\\
B^{011} &=(1 +
     r^{002} + r^{002} r^{101} + r^{002} r^{101} r^{200}),\\
C^{011} &= (1 + r^{011} + r^{002} r^{011} + r^{011} r^{110} + r^{002} r^{011} r^{110} +
    r^{002} r^{011} r^{101} r^{110}),\\
D^{011} &= (1 + r^{200} + r^{110} r^{200} +
    r^{020} r^{110} r^{200}).
\end{split}
\ee
It appears likely that these formulas can be written in terms of trees on the triangle of spaces,
and thereby generalized to arbitrary $K$. We leave a detailed discussion of that
for another occasion.

\section{Gluing of triangles, amalgamation, and identifying Fock-Goncharov edge coordinates with Darboux coordinates}\label{sec:Amalgamation}

\insfigscaled{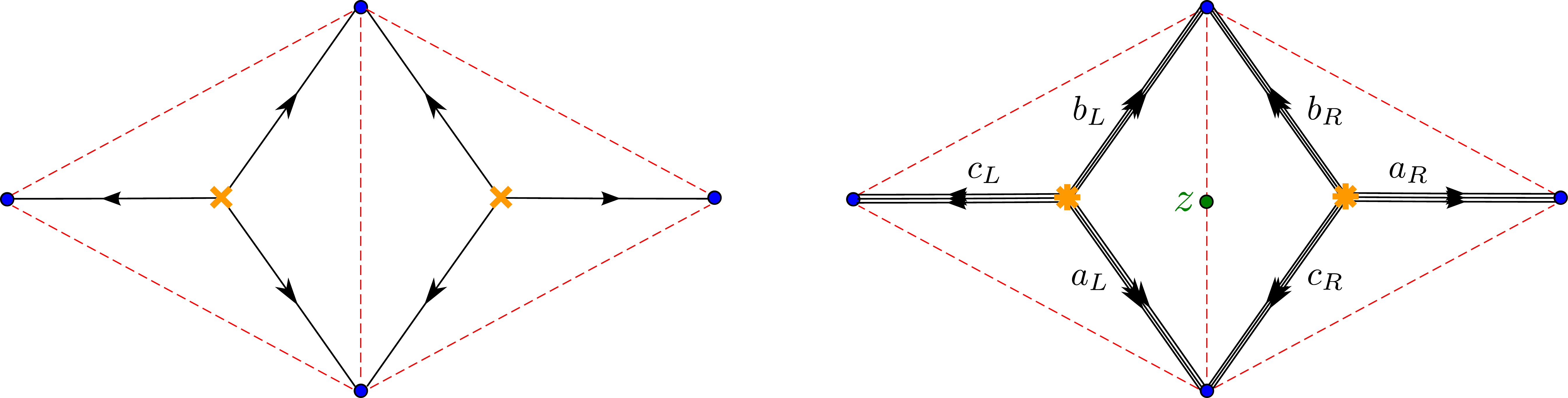}{0.265}{{\bf Left:}  An idealized picture of
a quadrilateral formed by two adjacent triangles in a WKB triangulation
$T_\WKB(\vartheta)$ of $C$.
The $\CS$-walls of the corresponding WKB spectral network are shown in black.
The edges of the WKB triangulation are shown in dashed red;
they run between singular points, shown in blue.
{\bf Right:}  An idealized picture of the level $K$ lift of
the picture at left, at a point of the Coulomb branch
near the lift locus.  Each triangle now contains a spectral network
for the level $K$ lift of the $AD_1$ theory. Each branch point has been fattened into a cluster
of branch points, and the $\CS$-walls of the original theory
have been thickened into cables.  We also mark a ``pinning point'' $z$ for later use.}

Let us now consider level $K$ lifts of more general theories
$S[\fg,C,D]$ with $\fg=A_1$. Choosing $\vartheta$ and a point in
the Coulomb branch, we obtain a WKB spectral network, which is roughly dual to
a decorated ideal triangulation $T_{\rm WKB}(\vartheta)$ of $C$.
(This triangulation is described at length in \cite{Gaiotto:2009hg}, and its relation to the
$K=2$ WKB spectral network is explained in \cite{Gaiotto2012}.)
A local picture of this triangulation is shown on the left side of Figure \ref{fig:local-triangulation};\footnote{In
Figure \ref{fig:local-triangulation} we have shown only the simplest situation; in general one also has to consider
degenerate triangles, obtained as quotients of Figure \ref{fig:local-triangulation}
where some of the singular points are identified.}
this picture can be considered as part of a larger triangulation, or as a triangulation associated to the theory $AD_2$.

Now consider making a small perturbation away from the lift locus of the level $K$ lifted theory.
Each branch point $\fb$ then splits into $\half K (K-1)$ branch points $\fb^{x,y,z}$
with $(x,y,z) \in T(K-2)$.  Each of the $\CS$-walls
shown on the left of Figure \ref{fig:local-triangulation} splits into a cable of $\CS$-walls in the perturbed lift.
Topologically speaking, each triangle contains a spectral network
for the lifted $AD_1$ theory.  The idealized picture is shown on the right side
of Figure \ref{fig:local-triangulation}, while  Figure \ref{fig:fattening} is an
actual realization of this picture by WKB spectral
networks, in the case $K=3$.

In general, the WKB spectral networks which appear in the various triangles of the lifted theory
need not be minimal.
We will assume, however, that there exist regions of the Coulomb branch of the
lifted theory for which the WKB spectral network in each triangle is essentially
minimal.  (It is not strictly necessary to use WKB spectral networks at all:  we could
work with more general spectral networks as defined in \S 9 of \cite{Gaiotto2012}.
In this case we could simply declare that we are going to study the case where the
network is minimal in each triangle.
The only disadvantage of doing this is that then we cannot relate our results
directly to BPS states at a particular point of the Coulomb branch of the lifted theory.)

\insfigscaled{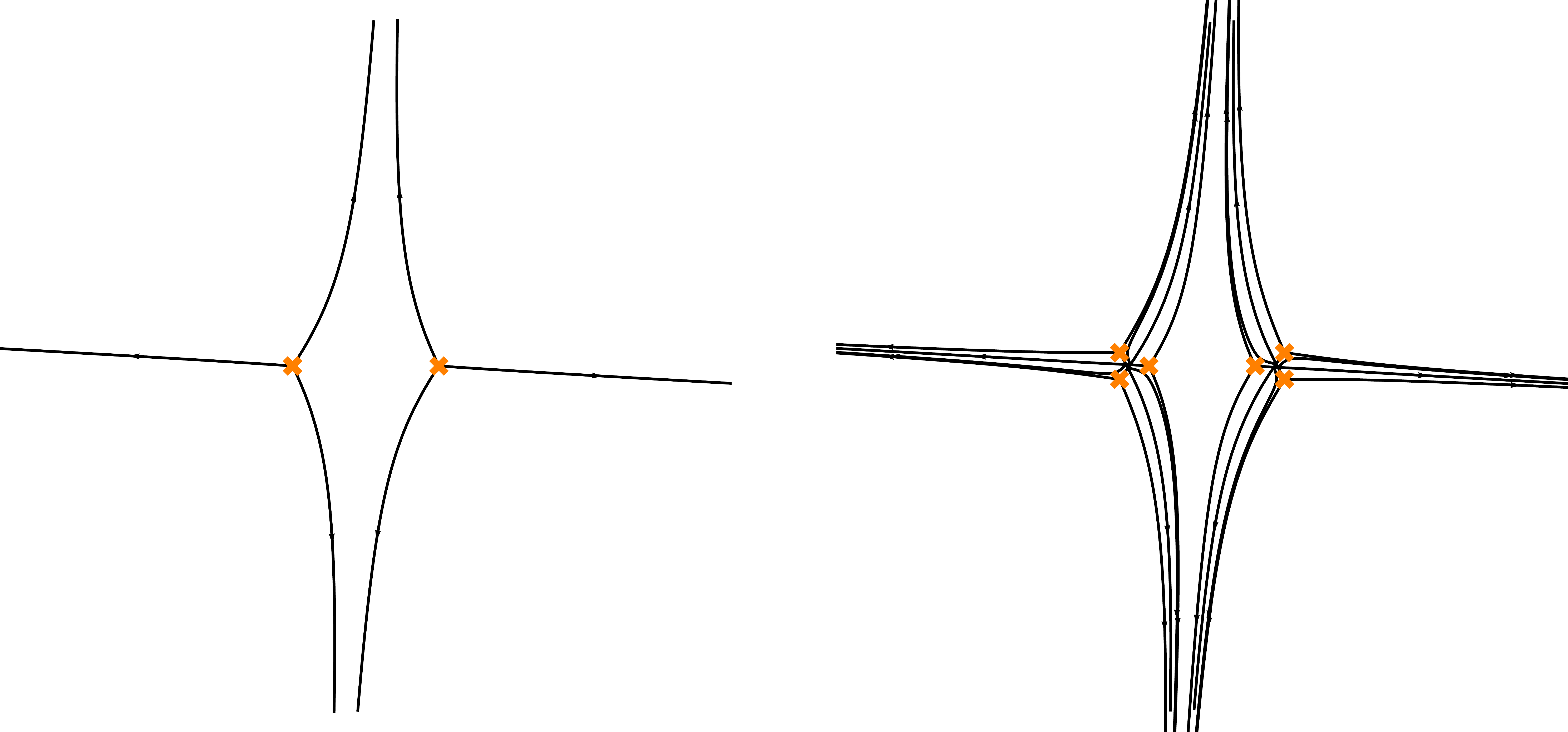}{0.245}{{\bf Left:} a WKB spectral network $\wnet_\vartheta$ in the $AD_2$ theory, at the point of the Coulomb branch with $\phi_2 = 4(z^2-1) \de z^2$, with $\vartheta = \frac{7 \pi}{15}$.  {\bf Right:} a WKB spectral
network $\wnet_\vartheta$ in the level $K=3$ lift of the $AD_2$ theory,
at the point of the Coulomb branch with $\phi_2 = 4(z^2-1) \de z^2$, $\phi_3 = \de z^3$,
with $\vartheta = \frac{7 \pi}{15}$.  This
is a small perturbation of the lift of the original theory on the left.
Each branch point on the left has split into 3 branch points on the right,
and the $\CS$-walls on the left have thickened into cables on the right.}

The spectral network now provides Darboux
coordinates on the moduli space of flat connections associated to the lifted theory.  Our object is to relate these coordinates
to the Fock-Goncharov coordinates of \cite{MR2233852}.

Some of the Darboux coordinates are easy to identify with Fock-Goncharov coordinates:  indeed, in each triangle
we have a collection of cycles $\gamma^\fv$, and the Darboux coordinates $\CY_{\gamma^\fv}$
will be identified with corresponding Fock-Goncharov ``triangle coordinates'' just as in \S\ref{subsec:IdentCoords}.
What is new here is that, when we have more than one triangle, we also get
coordinates associated with the internal edges of the
triangulation.

To understand these ``edge coordinates,'' let us return to the local picture in Figure \ref{fig:local-triangulation}.
The spectral network in each of the two triangles shown will be either of \Yin\ or \Yang\ type.
There are therefore four cases we should consider, but
it suffices to consider just the cases (\Yin,\Yin) and (\Yin,\Yang).

\insfigscaled{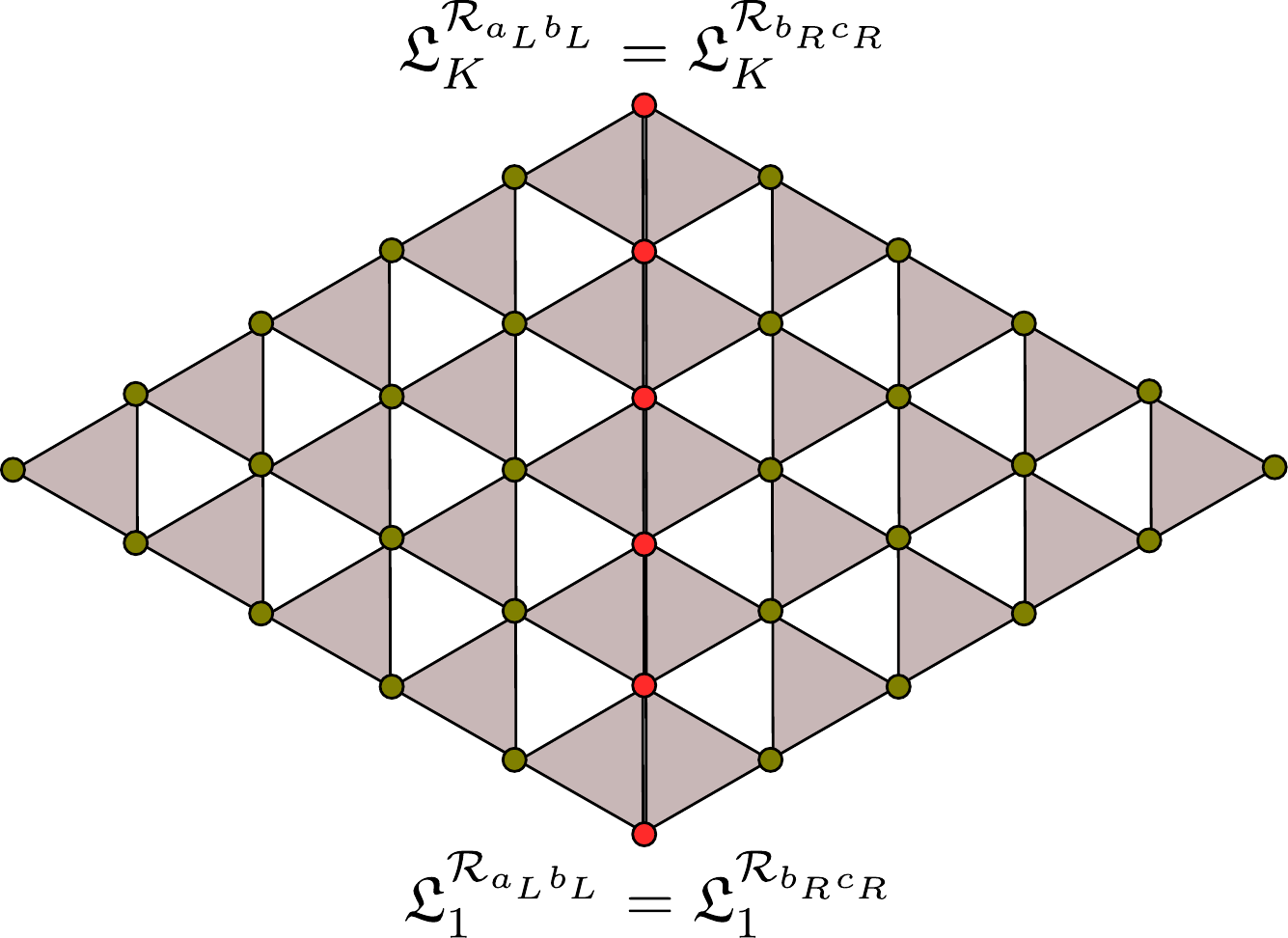}{0.6}{Each of the two minimal spectral networks in Figure
\ref{fig:local-triangulation} is associated to a
triangle of lines.  Since there is a common region $\CR_{a_L b_L} = \CR_{b_R c_R}$,
the two triangles of lines can be considered to share a common edge,
as depicted here.  The lines along the common edge are highlighted.}

First consider the (\Yin,\Yin) type lift. Because there is a common region
$\CR_{a_L b_L} = \CR_{b_R c_R}$ there is a common edge in the corresponding
triangles of lines, as shown in Figure \ref{fig:two-triangles-glued}. There are $(K-1)$ edge segments
along this common edge, which we call
\be
E_j^L: \fL_{K+1-j}^{\CR_{a_L b_L}} \to  \fL_{K-j}^{\CR_{a_L b_L}}, \qquad E_j^R: \fL_{j}^{\CR_{b_R c_R}} \to \fL_{j+1}^{\CR_{b_R c_R}}.
\ee
Here $j=1, \dots, K-1$, and we have labeled the edges so that $E_{K-j}^L$ and $E_j^R$
are the same edge, but oppositely oriented.  Therefore, if we compose the
canonical homs attached to these edges (defined in \eqref{eq:canhom3}),
we get a scalar:
\be\label{eq:edge-coordinate}
r_{E_{K-j}^L} = r_{E_j^R} := x_{E_j^R} x_{E_{K-j}^L}.
\ee
These scalars are the Fock-Goncharov edge coordinates.

Next, we relate the Fock-Goncharov coordinates $r_{E_{K-j}^L}$ to Darboux
coordinates $\CY_{\gamma}$ for appropriate $\gamma$.
Choose a ``pinning point'' $z$ as indicated in Figure \ref{fig:local-triangulation}.
As in \S\ref{subsec:IdentHoms}, the canonical homs can then be identified with the Darboux sections for chains
$\gamma_{ij}^R(z), \gamma_{ji}^L(z)\in \Gamma(z,z)$:
\begin{equation} \label{eq:x-ident}
x_{E_j^R}  = \CY_{\gamma_{ij}^R(z)}, \qquad
x_{E_{K-j}^L} = \CY_{\gamma_{ji}^L(z)}.
\end{equation}
Here $\gamma_{ij}^R(z)$ is an open chain which begins on sheet $i$ above
$z$, circles the branch point $\fb_R^{0,K-1-j,j-1}$ in the triangle $t(\fb_R)$ and
returns to sheet $j$ above $z$.
There is a similar description for  $\gamma_{ji}^L(z)$,
which meets the branch point $\fb_L^{j-1,K-1-j,0}$.
Concatenating these chains gives a closed cycle $\gamma = {\rm cl}(\gamma_{ij}^R(z) + \gamma_{ji}^L(z))$.
Now using \eqref{eq:edge-coordinate} and \eqref{eq:x-ident} we obtain
\begin{equation}
 r_{E_{K-j}^L} = r_{E_j^R} = \CY_\gamma.
\end{equation}
In other words, we have identified the $(K-1)$ Darboux coordinates $\CY_\gamma$ associated with these closed
curves with the $(K-1)$ Fock-Goncharov edge coordinates.

If we have two neighboring triangles of type (\Yin,\Yang), then along the common edge of the
triangle of lines we pair the lines $\fL_j^{\CR_{a_L b_L}}$ with the
dual lines $\hat \fL_j^{\CR_{b_R c_R}}$.  Under this dual pairing
$\fL_i^{\CR_{a_L b_L}}$ is orthogonal to $\hat \fL_j^{\CR_{b_R c_R}}$
for $i\not=j$ and has a nondegenerate pairing for $i=j$. The analog
of \eqref{eq:edge-coordinate} is that we choose arbitrary nonzero vectors
$v_j \in  \fL_j^{\CR_{a_L b_L}}$ and $\hat v_j \in  \hat \fL_j^{\CR_{a_L b_L}}$
and consider
\be\label{eq:yinyang-edge}
r = \frac{ x_{E}(v_j) \cdot \hat x_{E}(\hat v_j)}{v_j \cdot \hat v_j}.
\ee
As before, $r$ is the composition of Darboux sections for open paths $\CY_{\gamma_{ij}(z)}$,
and the ratio in \eqref{eq:yinyang-edge} is $\CY_\gamma$, for $\gamma$ a
cycle encircling an $ji$ branch point $\fb_L^{j-1,K-1-j,0}$ in the left triangle and an $ij$ branch
point $\fb_R^{0,j-1,K-1-j}$ in the right triangle.

The above concatenation procedures have a nice algebraic interpretation.
In \cite{amalg} Fock and Goncharov introduced a notion
of \emph{amalgamation} of cluster algebras and cluster varieties.  The procedure
we have outlined above appears to fit naturally into that framework, as follows.
Each triangle corresponds to a seed.
This seed includes some
``unfrozen'' or ``active'' variables corresponding to the Darboux coordinates
$\CY_{\gamma^{\fv}}$; it also includes
``frozen variables,'' corresponding to the Darboux sections $\CY_{\gamma_{ij}(z)}$
associated with the $(K-1)$ edge segments on each edge of the triangle.
One can now amalgamate the seeds corresponding to the various triangles of some
triangulation, to make a seed in a bigger cluster algebra.
This bigger algebra has unfrozen variables which are built as products of the frozen
variables from the smaller seeds:  this corresponds to
concatenating the open paths $\gamma_{ij}(z)$ along common edges of the triangles
to make closed paths, as we have described above.

In this section we have not taken careful account of the twistings discussed in \S \ref{sec:twistings}.
We leave a proper discussion of that detail to the future.

\section{BPS spectra of level \texorpdfstring{$K$}{K} lifts} \label{sec:GeneralPicture}

In this section we give a description of some (perhaps all)
of the BPS states of the level $K$ lift of a general $A_1$ theory $S[A_1, C, D]$,
in special regions of the Coulomb branch near the lift locus.

So, fix an $A_1$ theory $S[A_1, C, D]$, and fix a point $u$ of its Coulomb branch.
We assume that at $u$ the $A_1$ theory has
only BPS hypermultiplets, not vectormultiplets.
(For example, for gauge theories with gauge group $SU(2)$ and $N_f\leq 4$,
this condition is satisfied when $u$ lies in a strong coupling region.)
Let $\vartheta_\alpha$ denote the phases of the central charges of
these BPS hypermultiplets, with $\alpha$ running over the hypermultiplets.
As we vary $\vartheta$, the WKB spectral network $\wnet_\vartheta(u)$
undergoes a flip at each critical phase $\vartheta_\alpha$.

Now, we are going to consider the level $K$ lift of our original $A_1$
theory.  In the Coulomb branch of the lifted theory, there is a point of the lift locus
corresponding to $u$.  At this point there are a lot of massless BPS charges, so rather than working exactly
at this point, we will
study the lifted theory at a point $u'$ slightly perturbed from the lift of $u$.  At $u'$, for generic $\vartheta$,
the relation between the WKB spectral networks $\wnet_\vartheta(u')$ and the original $\wnet_\vartheta(u)$
is as indicated in Figure \ref{fig:local-triangulation}.
In particular, $\wnet_\vartheta(u')$ is naturally divided into subnetworks
embedding into triangles on $C$, and each of these subnetworks looks like a
spectral network for the level $K$ lift of $AD_1$.
Now, we make an important assumption:  we assume that for most phases
$\vartheta$ the restriction of
$\wnet_\vartheta(u')$ to each triangle is essentially minimal, and that all the triangle states in each triangle
occur in a narrow range of phases.  The minimality is then violated only in this narrow range; for phases
on one side of this critical range the triangle is of \Yin\ type, for phases on the other side
it is \Yang.

Under this assumption, we now describe three different
types of BPS states in the lifted theory, which we call \emph{triangle states},  \emph{lifted dyons},
and \emph{lifted flavor states}.

We first consider the triangle states. As we have discussed,
each branch point $\fb$ of the $A_1$ theory splits into $\half K (K-1)$
branch points $\fb^{x,y,z}$, $(x,y,z) \in T(K-2)$, in the lifted theory.
The triangle states are the BPS states associated with the level $K$ lift
of the ``local $AD_1$'' theory associated with $\fb$.
As we have seen in examples in \S \ref{sec:Vary-Theta}, among the triangle states
there are $\half (K-1)(K-2)$ ``elementary'' hypermultiplets
carrying charges $\gamma^\fv$, with $\fv \in T(K-3)$, and in general
there are also additional bound states of these hypermultiplets.
Also as discussed in \S \ref{sec:Vary-Theta},
the triangle states for any given triangle can be enumerated by considering the spectrum generator $\IS$
determined by \eqref{eq:explicit-sg}, and decomposing $\IS$
into an ordered product of elementary $\CK$-transformations as in \eqref{eq:factorization}.
This decomposition depends on the phases of the central charges $Z_{\gamma^\fv}$, hence depends on
$u'$.
We have such a collection of triangle states
for each of the triangles of the $A_1$ theory.

Next we consider the lifted dyons.  Consider a BPS hypermultiplet of the $A_1$ theory,
with central charge of phase $\vartheta_\alpha$.  At the phase $\vartheta_\alpha$,
the WKB spectral network $\wnet_\vartheta(u)$ of the $A_1$ theory undergoes a flip.  This flip involves two
triangles which we call $T_L$, $T_R$, containing two branch points
which we call $\fb_L$, $\fb_R$.  At phases $\vartheta$ near $\vartheta_\alpha$,
restricting the network $\wnet_\vartheta(u')$ of the lifted theory to these two triangles
gives two essentially minimal spectral networks.
When the phase $\vartheta$ crosses $\vartheta_\alpha$, in the original $A_1$ theory,
one of the $\CS$-walls emerging from $\fb_L$ sweeps across one of the $\CS$-walls emerging from $\fb_R$.
In the lifted theory, this means that one of the cables emerging from the left triangle $T_L$ sweeps
across one of the cables emerging from the right triangle $T_R$.
In this process, an $\CS$-wall of type $ij$ in the left cable can collide head-on with an $\CS$-wall of type $ji$
in the right cable; each time this happens we get a BPS hypermultiplet in the lifted theory, which we call
a lifted gauge state.

The precise counting of these lifted dyons
depends on the type of minimal spectral networks we have in the two triangles $T_L$, $T_R$ at phases near
$\vartheta_\alpha$.
There are 4 cases:  (\Yin,\Yin), (\Yin,\Yang), (\Yang,\Yin), or (\Yang,\Yang) types.

In the (\Yin,\Yin) or (\Yang,\Yang) case,
each of the $(K-i)$ $\CS$-walls coming from $T_L$ of type $(i,i+1)$
can meet each of the $i$ $\CS$-walls of type $(i+1,i)$ coming from $T_R$. Thus we obtain
a total of
\begin{equation}
\sum_{i=1}^{K-1} i(K-i) = \frac{1}{6} K (K^2-1)
\end{equation}
lifted gauge hypermultiplets.  All of these hypermultiplets have phase close to $\vartheta_\alpha$.

On the other hand, in the (\Yin,\Yang) or (\Yang,\Yin) case,
each of the $i$ $\CS$-walls from $T_L$ of type $(i+1,i)$
can meet each of the $i$ S-walls of type $(i,i+1)$ from $T_R$. Thus there will
be a total of
\begin{equation}
\sum_{1}^{K-1} i^2 = \frac{1}{6} K (K-1) (2K-1)
\end{equation}
lifted gauge hypermultiplets, again all with
phases close to $\vartheta_\alpha$.
In Figure \ref{fig:K3-lift-evolution} we show an example:  the single BPS hypermultiplet in the $AD_2$ theory
lifts to $5$ BPS hypermultiplets in the $K=3$ lift of $AD_2$.

\insfigscaled{K3-lift-evolution}{0.1055}{{\bf Top:}  Evolution of $\wnet_\vartheta$ in the $AD_2$ theory,
with $\phi_2 = 4(z^2-1) \de z^2$, for $\vartheta = \frac{- 7 \pi}{150}, 0, \frac{7 \pi}{150}$.
$\vartheta = 0$ is a critical phase at which $\wnet_\vartheta$ jumps; this frame is boxed (green).
In this frame we see a single $\CS$-wall connecting two branch points, representing a single BPS hypermultiplet.
{\bf Bottom:}  Evolution of $\wnet_\vartheta$ in the level $K=3$
lift of the $AD_2$ theory, with $\phi_2 = 4(z^2-1) \de z^2$ and $\phi_3 = (\frac{1}{2} + \frac{\I}{100}) \de z^3$,
for phases $\vartheta$ running from $\frac{- 8 \pi}{150}$ to $\frac{7 \pi}{150}$ in steps of $\frac{\pi}{150}$.
The two minimal spectral networks shown are in a (\Yin,\Yang) configuration.
There are $3$ critical phases at which $\wnet_\vartheta$ jumps; the frames closest to these phases
are boxed.  At $2$ of these phases (blue boxes) we see a single $\CS$-wall which very nearly connects two branch points,
while at $1$ of them (red box) there are three such $\CS$-walls at once.
Each of these saddle connections corresponds to a BPS hypermultiplet.
Altogether we have $5$ such hypermultiplets:  these are the ``lifted dyons'' associated to the single
BPS hypermultiplet of the $AD_2$ theory.  An animated version of this figure, along with
its \Yin\ counterpart, can be found at \cite{spectral-network-movies}.}

Finally, we come to the \emph{lifted flavor states}.  In an $A_1$ theory, we say a ``pure flavor state''
arises when at a critical phase we have a closed $\CS$-wall,
beginning and ending on the same branch point $\fb$,
and encircling a single singular point.\footnote{These pure flavor states have a special status as they
play no role in the four-dimensional Kontsevich-Soibelman wall-crossing-formula,
since their charges are in the annihilator of the antisymmetric product on $\Gamma$.
Nevertheless, since we can construct them using strings in the $(2,0)$ theory, or
membranes in M-theory, they are full-fledged members of the BPS spectrum with all the
rights, privileges, and responsibilities pertaining thereunto. For example, they \emph{are}
visible in the 2d4d wall-crossing formula \cite{Gaiotto:2011tf}.}
In the level $K$ lifted theory, when we sweep through the critical phase, two cables emerging from
the same triangle will sweep across one another.  In the process we will encounter
$\CS$-walls which begin and end at the same branch point; these correspond to pure flavor states
of the lifted theory.  More interestingly, we will also encounter $\CS$-walls which connect distinct
branch points from the same cluster; these correspond to states of the lifted theory carrying nontrivial gauge
charge.  The number of such hypermultiplets is
\be
\sum_{i=1}^{K-1} i^2 - i = \frac{1}{3} K(K-1)(K-2)
\ee
for each original flavor state. The reason is that there are $i$ branch points of type $(K-i,K-i+1)$,
and there are $i(i-1)$ \emph{ordered} pairs of distinct branch points. We must use ordered
pairs because the strings in the string web are oriented.

As an example, we apply this to the trinion theories $\CT(K)$ of \cite{Gaiotto:2009we}.
The theory $\CT(2)$ is an $A_1$ theory, with $C = \IC\IP^1$ and 3 defects.
In this theory the generic WKB triangulation involves two triangles, each containing a single branch point.
This theory has four charged hypermultiplets (associated with paths connecting the two distinct branch points)\footnote{This
is the expected number, because we have half-hypermultiplets in the
$(2,2,2)$ of the flavor $su(2)\oplus su(2)\oplus su(2)$ symmetry; alternatively this prediction can be
verified directly, e.g. by plotting the spectral network by hand.}
and three pure flavor particles (associated with paths connecting branch points to themselves after looping
around a defect).
The level $K$ lift of $\CT(2)$ is the theory $\CT(K)$.  Thus in theory $\CT(K)$ we will have BPS states
of all the three types described above:  triangle states for each of the two triangles,
lifted dyons for each of the four hypermultiplets, and lifted flavor states for each of the
three flavor states.

In particular, let us consider the case $K=3$.  In this case each triangle contains a single
elementary triangle state, and there is no possibility of any bound triangle states.  Thus we get a total
of $2$ triangle states.  Suppose that each of the $4$ hypermultiplets arises from a quadrilateral
of (\Yin,\Yin) type; in this case each lifts to $4$ more BPS states.
Finally, each of the $3$ flavor states lifts to $2$ BPS states with gauge charge.
Thus we conclude that if our assumptions are all satisfied,
the theory $\CT(3)$ has at least $2 + 4 \times 4 + 3 \times 2 = 24$ BPS states in some region near
the lift locus.  It is interesting to note that this $24$ matches the total number of BPS states
found in some region of the Coulomb branch in
\cite{Alim:2011ae,Alim2011a}; so if the region studied there is the same as the region we are studying,
we would expect that the $24$ states just described exhaust the whole BPS spectrum there. In principle
this could be settled by looking at the spectral networks $\wnet_{\vartheta}$ as $\vartheta$
evolves through an interval of length $\pi$.

\section{Lifting wall-crossing identities}\label{sec:SpinLift-WCF}

Combining the above picture of BPS states of the lifted
theory with the Kontsevich-Soibelman formula \cite{ks1},
which is known to govern the wall-crossing behavior of the BPS spectrum
in $\N=2$ theories \cite{Gaiotto:2008cd,Cecotti:2009uf,Dimofte:2009tm,Gaiotto:2010be,Manschot:2010qz,Pioline:2011gf},
leads to a number of identities on the Kontsevich-Soibelman symplectomorphisms $\CK_{\gamma}$.
In this section we briefly sketch how a few of these identities arise.

\subsection{Conjugating (\Yin,\Yin) to (\Yin,\Yang)}

\insfigscaled{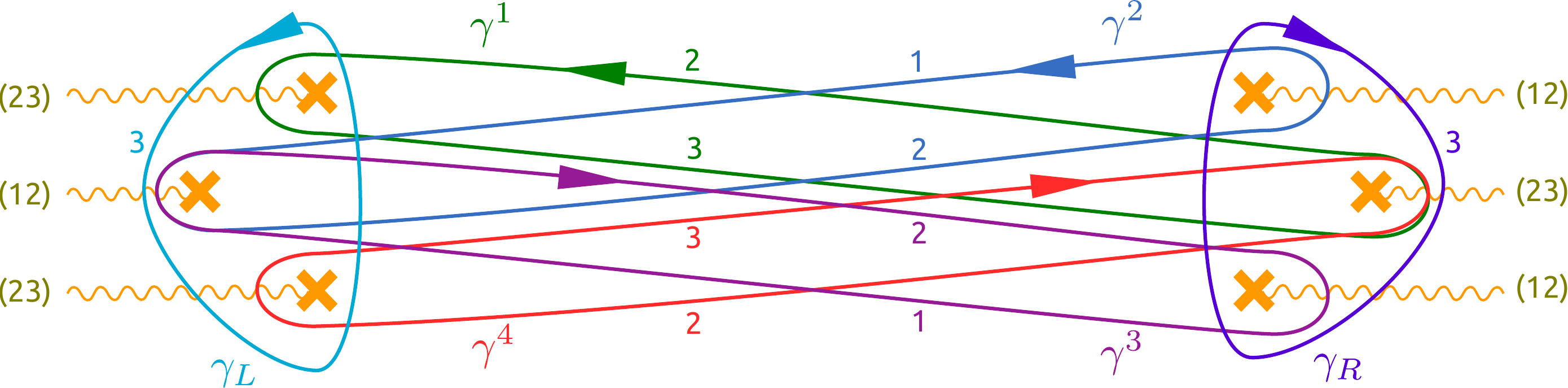}{0.53}{Cycles representing charges of BPS states
which appear in a (\Yin,\Yin) configuration with $K=3$ (4 lifted dyons and 2 triangle states).
The small numbers next to the paths
indicate which sheet of $\Sigma$ they lie on.}

Consider a (\Yin,\Yin) quadrilateral in a level
$K=3$ lift.  The charges of the associated BPS states are
shown in Figure \ref{fig:yin-yin-cycles}.
Each triangle has a single associated BPS triangle state;
we label their charges $\gamma_{L}$ and $\gamma_{R}$.
There are also $4$ charges $\gamma^{i}$ for the $4$ lifted dyons.

Now, imagine varying Coulomb branch parameters so that the
central charge $Z_{\gamma_{L}}$ sweeps past the central
charges $Z_{\gamma^{i}}$.  After so doing we should reach
a (\Yang,\Yin) quadrilateral.  This is nicely reflected
in the behavior of the KS transformations.  Recall the basic
pentagon identity \cite{ks1},
\be\label{eq:pentagon-id}
\CK_{\gamma_2} \CK_{\gamma_2 + \gamma_1} \CK_{\gamma_1} = \CK_{\gamma_1} \CK_{\gamma_2} \qquad \text{ for } \langle \gamma_2,\gamma_1\rangle = +1.
\ee
Using this identity, together with $\langle \gamma_{L} , \gamma^{1}\rangle = +1$
and $\gamma^{4} = \gamma^{1} + \gamma_{L}$, we can write
\be
\CK_{\gamma_{L} } \CK_{\gamma^{4}} \CK_{\gamma^{1}} \CK_{\gamma^{2}} \CK_{\gamma^{3}} =
 \CK_{\gamma^{1}} \CK_{\gamma_{L} }\CK_{\gamma^{2}} \CK_{\gamma^{3}},
\ee
and then using $\langle \gamma_{L}, \gamma^{2}\rangle = \langle \gamma_{L}, \gamma^{3}\rangle = -1$,
we obtain
\be
\CK_{\gamma_{L} } \CK_{\gamma^{4} } \CK_{\gamma^{1}} \CK_{\gamma^{2}} \CK_{\gamma^{3}}
=
\CK_{\gamma^{1}} \CK_{\gamma^{2}} \CK_{\gamma^{2}+\gamma_{L} }   \CK_{\gamma^{3}}
\CK_{\gamma^{3} + \gamma_{L} } \CK_{\gamma_{L} }.
\ee
Thus, with an appropriate ordering of the 4 BPS states of the (\Yin,\Yin) quadrilateral,
conjugation by $\CK_{\gamma_{L}}$ can bring the product of 4 symplectomorphisms representing
these 4 states to the product of 5 symplectomorphisms representing the 5 BPS states of
the (\Yang,\Yin) quadrilateral.

\subsection{Level 3 lift of the pentagon identity}

Consistency of our picture of the BPS spectrum of the lifted theory implies in particular
that there should be higher-$K$ lifts of the basic
``pentagon'' and ``juggle'' identities which guarantee consistency of the spectra of $A_1$ theories.
In this section we consider the $K=3$ lift of the pentagon identity \eqref{eq:pentagon-id}.  The generalizations
to higher $K$ and to lifts of the juggle identity are left as open problems.

We consider three triangles in the $K=3$ lift, or equivalently we consider the $K=3$ lift
of the $AD_3$ theory.  Recall that in the $AD_3$ theory there is a wall of marginal stability:
on one side of the wall there are $2$ BPS hypermultiplets, on the other side there are $3$,
and the consistency of the spectrum follows from the pentagon identity \eqref{eq:pentagon-id}.
We would like to see how the consistency is maintained in the $K=3$ lift.

\insfigscaled{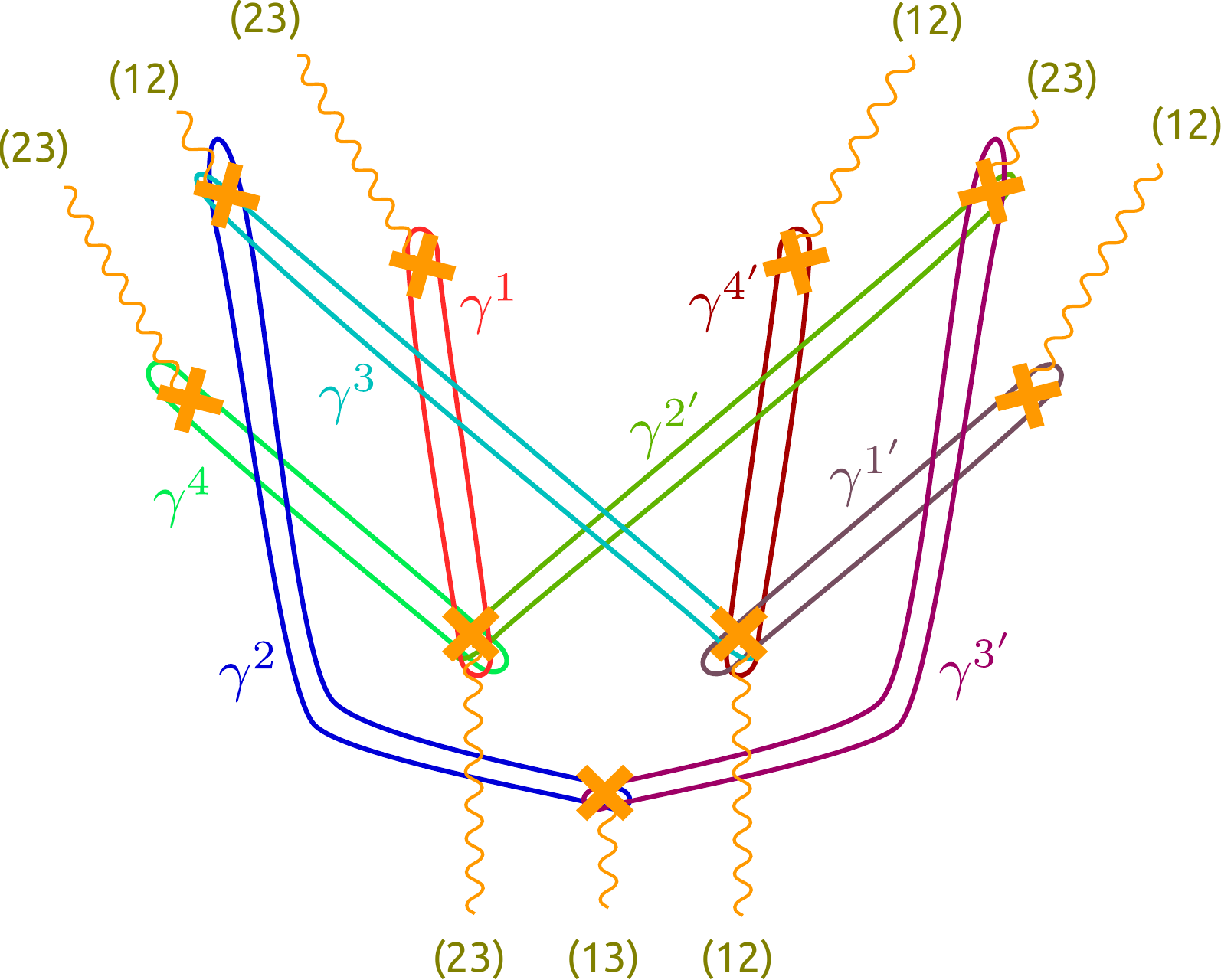}{0.55}{Cycles used in the $K=3$ lift of the pentagon identity.  These cycles
represent the charges of $8$ hypermultiplets obtained by lifting $2$ hypermultiplets of the $AD_3$ theory.}

Consider for simplicity the case where all of the quadrilaterals are of (\Yin,\Yin) type.
Then in the $K=3$ lift, each of the hypermultiplets of $AD_3$ gets lifted to 4 hypermultiplets.
The charges of the lifts of $2$ hypermultiplets of $AD_3$ are shown in Figure
\ref{fig:yin-yin-yin-cycles}.  One can show (just using the
pentagon identity, the intersection numbers, and linear relations among the various cycles)
that one has
\begin{multline}
\left( \CK_{\gamma^{1'}}\CK_{\gamma^{2'}}\CK_{\gamma^{3'}}\CK_{\gamma^{4'}}\right)\left(\CK_{\gamma^{1}} \CK_{\gamma^{2}} \CK_{\gamma^{3}} \CK_{\gamma^{4}} \right)
= \\ \left(\CK_{\gamma^{1}} \CK_{\gamma^{2}} \CK_{\gamma^{3}} \CK_{\gamma^{4}} \right) \left( \CK_{\gamma^1+\gamma^{2'}} \CK_{\gamma^3+\gamma^{1'}} \CK_{\gamma^4+\gamma^{2'}} \CK_{\gamma^{3}+\gamma^{4'}} \right)
\left( \CK_{\gamma^{1'}}\CK_{\gamma^{2'}}\CK_{\gamma^{3'}}\CK_{\gamma^{4'}}\right).
\end{multline}
This is the needed lift of the pentagon identity, which ensures consistency of the spectrum
in the lifted theory.

\insfig{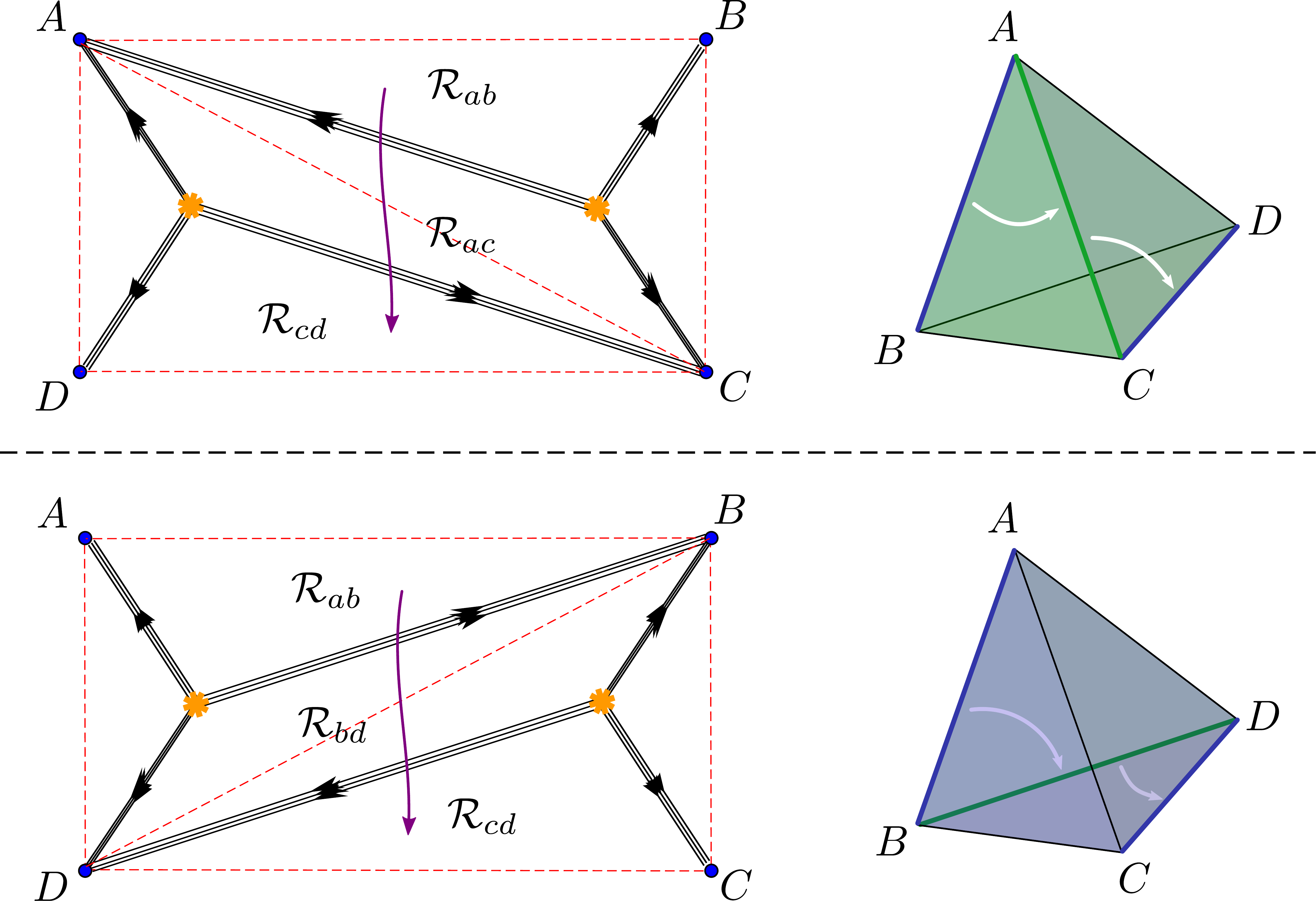}{{\bf Top left:}  A schematic picture of a spectral network in
the level $K$ lift of $AD_3$, and a path connecting three asymptotic regions,
$\CR_{ab} \to \CR_{ac} \to \CR_{cd}$.  {\bf Top right:}  The canonical
tetrahedron of lines in $\CE$, with the snakes $\CS\CN_{AB}$, $\CS\CN_{AC}$, $\CS\CN_{CD}$
marked; our path corresponds to a path of snakes interpolating between these three,
involving snakes sweeping across the faces $ABC$ and $ACD$ (the front faces, shaded green).
{\bf Bottom left:}  Another spectral network in the same theory, related by a flip to the
one at top left.  The region $\CR_{ac}$ has disappeared and been replaced by $\CR_{bd}$, so
our path now connects $\CR_{ab} \to \CR_{bd} \to \CR_{cd}$.
{\bf Bottom right:}  The canonical
tetrahedron of lines in $\CE$, with the snakes $\CS\CN_{AB}$, $\CS\CN_{BD}$, $\CS\CN_{CD}$
marked; our path corresponds to a path of snakes interpolating between these three,
involving snakes sweeping across the faces $ABD$ and $BCD$ (the rear faces, shaded green).
The two paths of snakes on the right are related by a kind of discrete homotopy across the interior of the
tetrahedron; the steps in this homotopy (not pictured here) correspond to the elementary jumps of the spectral network,
whose composition makes up the flip.}

\subsection{Flips and Darboux coordinates}

Finally, let us briefly consider one more natural question.

We have seen that each spectral network of an $A_1$ theory,
i.e. each ideal triangulation of $C$, can be naturally lifted to a spectral network of the level $K$ lifted theory.
When the lifted network has $\Yin$ type in each triangle,
the coordinates $\CY_\gamma$ in the lifted theory match those defined by Fock-Goncharov in \cite{MR2233852}.
We may now ask:  how do these coordinates transform when we flip the underlying ideal triangulation?

This transformation was worked out in \S 10, p. 153, of \cite{MR2233852}:  when we flip the triangulation,
the coordinates transform by a certain sequence of elementary symplectomorphisms.
In the context of the lifted spectral network, these elementary symplectomorphisms have
a natural interpretation.  As we have discussed in \S\ref{sec:GeneralPicture},
the flip decomposes into a sequence of more elementary transformations of the
spectral networks, each one corresponding to a single ``lifted gauge state.''
Indeed, there is a 1-1 correspondence between the lifted dyons we found in the
$(\Yin, \Yin)$ case and the elementary symplectomorphisms which appeared in \cite{MR2233852}.
Moreover, using the results of \cite{Gaiotto2012}, we know already that these lifted gauge
states indeed induce symplectomorphisms of the right form (corresponding to
single ``mutations'' in the language of \cite{MR2233852}).  This is a useful
consistency check of our story.

We believe there is more to say here.
Indeed, the analysis of this transformation in \cite{MR2233852} involves a beautiful generalization
of the combinatorics of $m$-triangles to ``$m$-simplices'' of higher dimension, which should also
have an interpretation in terms of spectral networks.
To be concrete, consider a spectral network in the level $K$ lift of the $AD_2$ theory, with 4 cables of lines
asymptoting to 4 singular points, and 4 associated flags $A_\bullet, B_\bullet, C_\bullet, D_\bullet$.
See Figure \ref{fig:tetra}.

If the 4 flags are in general position, then we can construct
a corresponding ``$(K-1)$-tetrahedron'' of lines in $\CE$:
\begin{equation}
\fL^{x,y,z,w} = A^x \cap B^y \cap C^z \cap D^w, \qquad x+y+z+w = K-1.
\end{equation}
Now consider the two regions $\CR_{ab}$ and $\CR_{cd}$.  These regions correspond to snakes $\CS\CN_{AB}$, $\CS\CN_{CD}$
drawn on edges of the tetrahedron.  Other regions correspond to other snakes, and in particular a
path from $\CR_{ab}$ to $\CR_{cd}$ corresponds to a particular path of snakes.
The path shown in Figure \ref{fig:tetra} corresponds before the flip to the composition of the two canonical paths
$\CS\CN_{AB} \to \CS \CN_{AC} \to \CS \CN_{CD}$, and  after the flip to the composition of the two canonical paths
$\CS\CN_{AB} \to \CS \CN_{BD} \to \CS \CN_{CD}$.  These two compositions can be homotoped into one another across
the $(K-1)$-tetrahedron.  Indeed, the variation of the spectral network during the flip
should give an explicit homotopy between them, with
the intermediate spectral networks corresponding to intermediate paths of snakes
from $\CS\CN_{AB}$ to $\CS\CN_{CD}$.

By studying the projective bases associated to these snakes, we would expect to find an equation of the form
\begin{equation}
T^{AB \to AC}(\CY_{\gamma}) T^{AC\to DC}(\CY_{\gamma}) = \left( \prod_{\alpha} \CK_{\gamma_\alpha} \right)\cdot
T^{AB \to BD}(\CY_{\gamma}) T^{BD\to DC}(\CY_{\gamma}),
\end{equation}
where the $\gamma_\alpha$ are the charges of the lifted dyons and the $T(\CY_\gamma)$ are the
Stokes matrices for crossing cables.
When we consider a spectral network with more than $2$ triangles, we will get various equations of this sort;
the mutual consistency of the resulting system should be guaranteed by identities following from lifts of pentagon identities,
involving homotopies of homotopies, related to 5-simplices, and so on.
We leave this to the future.

\section{Open problems and future directions}\label{sec:OpenProblems}

The present paper is (deliberately) left somewhat sketchy and leaves much work to be done.
Some of the more obvious directions in which this work could be continued
are the following:

\begin{enumerate}

\item In \S \ref{subsec:SpecialCurves} we describe how to perturb the
level $K$ lift of the $AD_1$ theory to produce spectral networks
which are ``essentially minimal'' in the sense of \S \ref{subsec:Def-Minimal-SWALL}.
We showed by example that this could be done for small $K$ but gave no systematic
procedure for doing it for all $K$.  We believe such a procedure
exists and it would be nice to fill in this gap.  We hasten to add that
finding such a procedure is not logically necessary for the validity of the
subsequent discussion of coordinates at arbitra.ry $K$.

\item In \S \ref{sec:Vary-Theta} we determined the spectrum generator
for the level $K$ lift of the $AD_1$ theory, in the region of the Coulomb
branch admitting minimal spectral networks.  The spectra of these theories
are essential building blocks in describing the BPS spectra of all level $K$
lifts, as follows from the amalgamation procedure
described in \S \ref{sec:Amalgamation} below.  We believe
there should be elegant formulae for the general transformation of coordinates
$r^\fv \to \hat r^\fv$ in terms of some simple algorithm involving
trees on the $(K-3)$-triangle of spaces. Moreover, to read off the BPS spectrum from the
spectrum generator requires one to factorize the spectrum generator
into elementary $\CK$-transformations
(given a point of the Coulomb branch and hence an
ordering of the phases of the periods).  One such factorization
has been found in \cite{le-kamnitzer}; it would be very interesting to know whether
this factorization corresponds to the BPS spectrum at
some point of the Coulomb branch.

\item Our discussion in the final three sections of the
paper --- the relation to cluster algebra amalgamation,
the BPS spectrum of level $K$
lifts of $A_1$ theories, and the spin lifts of wall-crossing
identities --- is especially sketchy. Much more
remains to be done here.  Among many open questions we mention
just one:  The physical interpretation of the amalgamation procedure
in terms of four-dimensional quantum field theory might be
very interesting. In the ultraviolet theory an analogous gluing
statement described in
\cite{Gaiotto:2009we,Moore:2011ee,Gaiotto:2011xs,Chacaltana:2010ks,Chacaltana:2011ze,Chacaltana:2012zy} is
extremely powerful and deep.  The amalgamation procedure should
be some kind of infrared version of that statement.
The inverse of amalgamation involves an RG flow, which decouples
some degrees of freedom. Thus a physical interpretation of amalgamation might involve
BPS domain walls between theories, such as the RG domain walls in \cite{Brunner2007,dgr}.

\item In \S \ref{sec:GeneralPicture} we described a general scenario
for constructing the BPS spectrum of any lifted $A_1$ theory of class S.
We did not show that there really do exist regions in the Coulomb branch
where the spectral network is an essentially minimal network
of, say, \Yin\ type in all the triangles of the $A_1$ theory.
We also did not show that the BPS states we exhibited exhaust all the
BPS states in such regions of the Coulomb branch. Of course, for any
given theory completeness can in principle be established by
varying $\wnet_{\vartheta}$ and applying the algorithm of
\cite{Gaiotto2012}, but it would be nice to have a general argument.

\item It would be interesting to use these methods
to compare in detail our results with the results of
\cite{Alim:2011ae,Alim2011a,Chen:2011gk,Fraser:1996pw}
on the BPS spectra of higher rank $\N=2$, $d=4$ theories.
We have not done so, but we conjecture that the BPS quiver
for the level $K$ lift of the $AD_1$ theory is
the following:  associate one node to each
vector $\fv\in T(K-3)$ with charge $\gamma^\fv$ and
then determine the edges
from equation \eqref{eq:Triangle-Intersections}. Similar remarks
apply to lifts of $AD_n$ theories.

\item While this paper focuses on some very special
spectral networks, we can apply some of the constructions
and lessons learned here to more general spectral networks. We think it could be
very fruitful to investigate these constructions at the
more general level. In particular, in this paper we show ---
in our special examples --- explicitly how to express the ``monodromy data''
(here, Stokes matrices) in terms of the ``Darboux coordinates''
$\CY_{\gamma}$ on the moduli space of flat connections.
But some of the key constructions of \S \ref{sec:FlagsFlatSections}
apply more generally. A general spectral
network provides a ``nonabelianization map'' from abelian flat connections on the Seiberg-Witten
curve $\Sigma$ (also known as the spectral curve, or the
infrared curve) to non-abelian flat connections on the
ultraviolet curve $C$.  Quite generally, the construction of
\cite{Gaiotto2012} produces three extra pieces of
data:

\begin{itemize}
\item A collection of $K$ lines $\fL_i^{\CR}$ of flat sections associated to each connected
component $\CR$ of the complement of the spectral network on $C$.

\item A collection of ``coplanarity relations'' associated to the $\CS$-walls
of the spectral network.
Indeed, the discussion of \S \ref{subsec:Trans-S-Wall} and \S \ref{subsec:Plane-And-BP} applies
quite generally to show that if an $\CS$-wall of type $ij$ separates two regions $\CR_1$ and $\CR_2$
then the $(K-1)$ lines $\fL_s^{\CR}$ with $s\not = i $ are the
same in the two regions, while $\fL_i^{\CR_1}$, $\fL_i^{\CR_2}$,
and $\fL_j^{\CR_1} = \fL_j^{\CR_2}$ lie in a common plane.

\item A collection of incidence relations determining how the lines are
related to each other in terms of flags attached to the punctures of $C$.

\end{itemize}

For the minimal spectral networks of this paper, the Darboux coordinates are
derived in two steps. First, one builds the collection of lines associated
to a non-abelian flat connection.  Next, the coplanarity relations on lines
associated with $\CS$-walls allow one to define canonical homs between
certain lines, discussed in \S\ref{subsec:Trans-S-Wall}
and \S\ref{subsec:CanHom}.  Third, composition of these canonical homs associated
with collections of lines forming a closed loop can be used to form coordinates,
as in \eqref{eq:MainResultOne}.

We conjecture that the three-step procedure described in the previous paragraph
can be generalized to arbitrary spectral networks.  Interestingly, the above
procedure naturally lends itself to a description in terms of a \emph{bipartite graph},
where lines are represented by white dots, coplanarity relations are represented by
black dots, and white dots are connected to black dots if they sit in the relevant
plane.  We believe this observation might provide a useful link to the recent
work of Goncharov \cite{goncharov} and Goncharov and Kontsevich \cite{goncharov-kontsevich}.
It might also be a link to upcoming work of Arkani-Hamed et al. on scattering
amplitudes in $\CN=4$ supersymmetric Yang-Mills theory \cite{grassmannian}.

\end{enumerate}

\section*{Acknowledgements}

We thank Sasha Goncharov for important discussions on his work with
Fock on defining cluster coordinates for moduli spaces of flat connections
with rank greater than $1$.  We also thank N. Arkani-Hamed for discussions
and P. Longhi for some suggestions on the draft.

GM and AN would like to thank the KITP, where some of this work was done, for hospitality.
The work of GM is supported by the DOE under grant
DE-FG02-96ER40959.  GM also gratefully acknowledges
partial support from the Institute for Advanced Study and
the Ambrose Monell Foundation.   The work of AN is supported by the NSF under grant numbers
DMS-1006046 and DMS-1151693.  The research of DG was supported in part by the NSF grant PHY-0503584.
The research of DG was supported in part by the Roger Dashen membership in the Institute for Advanced
Study. The research of DG was supported by the Perimeter Institute for Theoretical
Physics.  Research at Perimeter Institute is supported by the
Government of Canada through Industry Canada and by the Province of
Ontario through the Ministry of Economic Development and Innovation.

\appendix

\section{Linear algebra of three flags in general position}\label{app:LinearAlgebra}

Now we will review some constructions from Fock and Goncharov
\cite{MR2233852}, concerning three flags $A_{\bullet}, B_{\bullet}, C_{\bullet}$
in general position in a $K$-dimensional vector space.

In our main application, these three flags
will be determined by the $z \to \infty$ asymptotics in three regions
of the level $K$ lift of the $AD_1$ theory.

\subsection{\texorpdfstring{$m$-triangles}{m-triangles}}\label{app:m-triangles}

The \emph{$m$-triangle} is a triangulation of the triangle in $\IR^3$ with vertices $(m,0,0)$, $(0,m,0)$ and $(0,0,m)$.
The vertices of the $m$-triangle are all lattice points in the interior, i.e. all
triples $(x,y,z)$ of nonnegative integers satisfying
\begin{equation}
x+y+z = m.
\end{equation}
We sometimes denote the set of vertices in the $m$-triangle by $T(m)$.

\insfigscaled{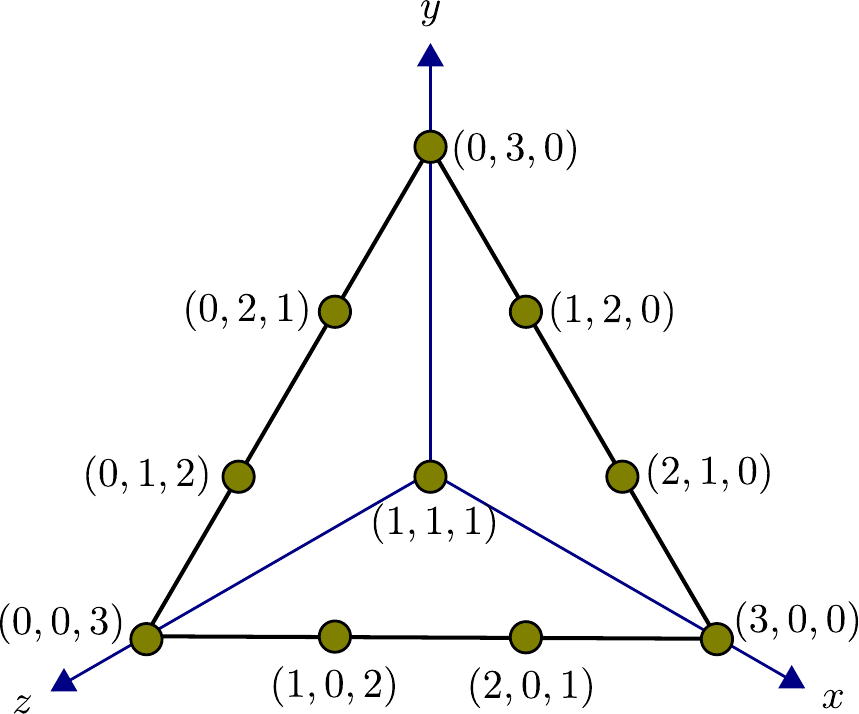}{0.80}{The lattice points of the $m$-triangle with $m=3$, embedded in $\IR^3$.}

\insfigscaled{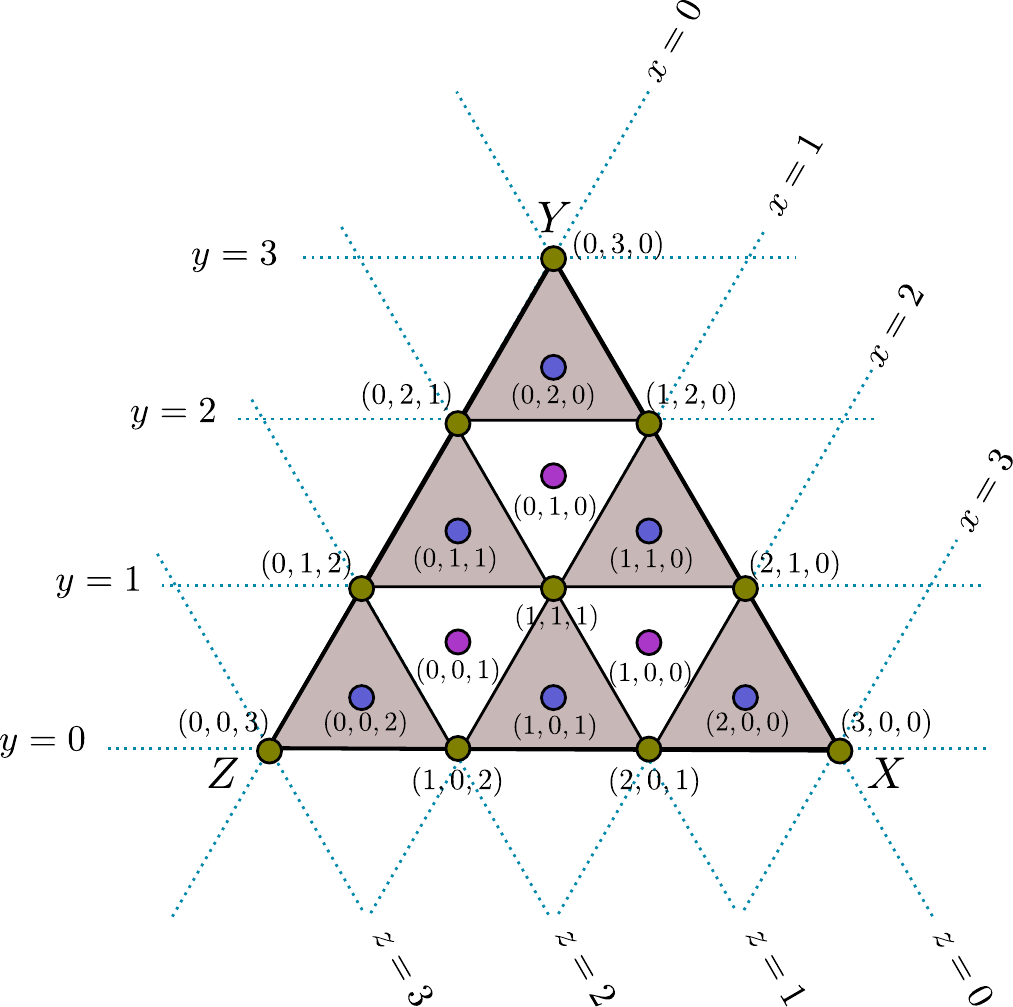}{1.07}{The $m$-triangle with $m=3$, projected
onto a plane along the $(1,1,1)$ axis.  The shaded upwards-pointing subtriangles are labeled by the lattice points of the
$2$-triangle (blue).  The unshaded downwards-pointing subtriangles are labeled by the lattice points of the
$1$-triangle (purple).}

The $m$-triangle lies on a plane orthogonal to the vector $(1,1,1)$.  The lines
with constant $x$, $y$ or $z$ are diagonals cutting through the triangle as indicated in
Figure \ref{fig:3-triangle-projected}.  Let us note a few properties:

\begin{enumerate}

\item Each edge of an $m$-triangle contains $m+1$ lattice points.  Altogether,
an $m$-triangle contains $\half (m+1)(m+2)$ lattice points.

\item The small upwards-pointing
subtriangles of the $m$-triangle, shown shaded in
Figure \ref{fig:3-triangle-projected} are in 1-1 correspondence with the
lattice points in the $(m-1)$-triangle.
The correspondence is induced by projecting both the $m$-triangle and the $(m-1)$-triangle
from $\IR^3$ to $\IR^2$ along the $(1,1,1)$-axis.  The lattice points of the
$(m-1)$-triangle then sit naturally at the center of the shaded up-triangles in the
$m$-triangle.

\item The unshaded downwards-pointing triangles in the $m$-triangle are in 1-1
correspondence with the vertices of the $(m-2)$-triangle. Again,
the correspondence is realized geometrically by projecting along the $(1,1,1)$
axis.

\item The lattice points of the $(m-3)$ triangle are in 1-1 correspondence with the
\emph{interior} lattice points of the $m$-triangle.  A simple way to see this
is that if $(x,y,z)$ are nonnegative and $x+y+z=m-3$ then $(x+1,y+1,z+1)$ is an
interior lattice point of an $m$-triangle. As before, this correspondence can
be realized geometrically by projecting along the $(1,1,1)$
axis.

\item Label the vertex $(m,0,0)$ in the $m$-triangle by $X$ and call the
opposite side the $X$-side.  Do the same for $y$ and $z$. Then the
$(x,y,z)$ coordinates of a point $P$ in the $m$-triangle can be easily computed by
the following observation:  consider any path from $P$ in the triangulation
to the $X$-side, always traveling toward the $X$-side (no backtracking.)
Then the $x$-coordinate of $P$ is the number of steps
taken by the path.  Analogous statements hold for the
$y,z$ coordinates.  We often refer to the $X,Y,Z$
vertices as the $A,B,C$ vertices, respectively.

\end{enumerate}

\subsection{\texorpdfstring{$m$-triangles}{m-triangles} associated with three flags}

Let us suppose we have a vector space $V$ of dimension $K$ and three
flags $A_\bullet, B_\bullet, C_\bullet$ in general position. In this
section we define a distinguished set of lines in $V$
labeled by lattice points of the $(K-1)$-triangle, planes in $V$ labeled by
lattice points of the $(K-2)$-triangle, and spaces in $V$ labeled by lattice points
of the $(K-3)$-triangle.\footnote{By \emph{space} in this subsection we mean a
three-dimensional vector space. Living as we do in spacetime with
only four macroscopic dimensions, our language has not evolved a
convenient term to distinguish ``space'' of three dimensions from ``space'' of
higher dimensions.}  The relations among the lattice points of the $(K-1)$, $(K-2)$, $(K-3)$-triangles
reflect the incidence relations among their attached lines, planes and spaces.

Our convention for a flag is that $A_n$ is an $n$-dimensional subspace of $V$, so
\begin{equation}
\{ 0 \} = A_0 \subset A_1 \subset A_2 \subset \cdots \subset A_{K-1} \subset A_K = V .
\end{equation}
It is very convenient to introduce a notation that emphasizes the
\emph{codimension}, so we define
\begin{equation}
A^{x} := A_{K-x}
\end{equation}
and so forth, so that
\begin{equation}
\{ 0 \} = A^K \subset A^{K-1} \subset A^{K-2} \subset \cdots \subset A^2 \subset A^1 \subset A^0 = V .
\end{equation}

\insfigscaled{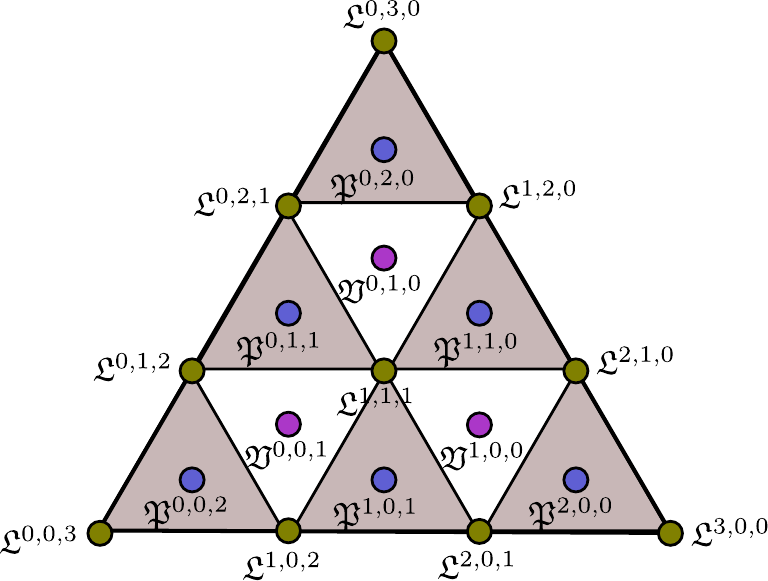}{1.07}{Incidence relations
among the lines $\fL^{x,y,z}$, planes $\fP^{x,y,z}$ and spaces $\fV^{x,y,z}$ in the case
$K=4$.  The vertices of the $3$-triangle
correspond to lines $\fL$. The shaded upwards pointing triangles correspond to
planes $\fP$, and the three corners of a shaded triangle are three lines
contained in the plane.  The unshaded downwards pointing triangles correspond
to spaces $\fV$, and the three shaded triangles abutting an unshaded triangle correspond
to three planes $\fP$ contained in the space $\fV$.}

The lines we consider are simply
\begin{equation}
\fL^{x,y,z} := A^x \cap B^y \cap C^z, \qquad  x+y+z = K-1 .
\end{equation}
Here $(x,y,z)$ is a point in the $(K-1)$-triangle. See Figure \ref{fig:incidence-triangle}
for the case $K=4$.
For flags in general position, $\fL^{x,y,z}$
is a one-dimensional subspace of $V$. In an analogous way we define planes
\begin{equation}
\fP^{x,y,z} := A^x \cap B^y \cap C^z, \qquad  x+y+z = K-2,
\end{equation}
labeled by points in the $(K-2)$-triangle, and spaces
\begin{equation}
\fV^{x,y,z} := A^x \cap B^y \cap C^z, \qquad  x+y+z = K-3,
\end{equation}
labeled by points of the $(K-3)$-triangle.  In particular, there are
$\half (K+1)K$ different lines, $\half K(K-1)$ different planes,
and $\half (K-1)(K-2)$ different spaces.

The relations between points of the $m$, $m-1$, $m-2$ triangles we observed
above have an elegant interpretation in terms of incidence relations among the
lines $\fL^{x,y,z}$, planes $\fP^{x,y,z}$ and spaces $\fV^{x,y,z}$.
We shade the small upwards-pointing subtriangles of the $(K-1)$-triangle.
Then the shaded subtriangles are in 1-1 correspondence with vertices of the
$(K-2)$-triangle.  The plane $\fP^{x,y,z}$ corresponding to each shaded subtriangle
contains the three lines $\fL^{x+1,y,z}$,
$\fL^{x,y+1,z}$, and $\fL^{x,y,z+1}$ at its three vertices.  Generically, these are the only such
incidence relations we expect.  That is, generically, we do not expect
other triples of lines $\fL^{x',y',z'}$ to sit in any plane, let alone one of the
special planes $\fP^{x,y,z}$.  Similarly, the unshaded
subtriangles are in 1-1 correspondence with the vertices of a $(K-3)$-triangle,
and with the spaces $\fV^{x,y,z}$.
Each unshaded triangle is abutted by 3
shaded triangles.  This corresponds to the fact that
$\fV^{x,y,z}$ contains the three planes
$\fP^{x+1,y,z}$, $\fP^{x,y+1,z}$, and $\fP^{x,y,z+1}$.
Generically, these are the only such
incidence relations we expect. That is, generically, we do not expect
any other planes of type $\fP^{x',y',z'}$ to sit in the space $\fV^{x,y,z}$.

\subsubsection{Canonical homs}\label{subsec:CanHom}

An edge $E$ oriented from a line $\fL_1$ to a line $\fL_2$ determines an element $x_E \in {\rm Hom}(\fL_1,\fL_2)$
as follows.  $E$ is a side of a unique shaded triangle,
and hence $E$ determines three lines $\fL_1, \fL_2, \fL_3$,
the vertices of that shaded triangle.
Given $v_1 \in \fL_1$ we define $x_E(v_1)$ to be the unique vector $v_2 \in \fL_2$ such that
\begin{equation}\label{eq:canhom3}
\begin{cases} v_1+v_2 \in \fL_3 & {\rm clockwise} \\
v_1 - v_2 \in \fL_3 & {\rm counterclockwise} \\
\end{cases}
\end{equation}
where the cases refer to whether $E$ is oriented clockwise or counterclockwise around
the shaded triangle.

Note that if $\bar E$
is the orientation reversal of $E$ then $x_E x_{\bar E} = -1$ and
$x_{\bar E} x_E = -1$. Similarly, if $E_1, E_2, E_3$ are three consecutive
edges going around a shaded up-triangle, then $x_{E_1} x_{E_2} x_{E_3} = -1$ if
they are oriented counterclockwise and $x_{E_1} x_{E_2} x_{E_3} =  1$
if oriented clockwise.

\subsection{Fock and Goncharov's triple ratio}

The space of flags in a $K$-dimensional complex vector space $V$ is $G/B$,
where $G = GL(K, \IC)$ and $B$ is the Borel subgroup of upper-triangular matrices.
It thus has dimension
$K^2 - \half K (K+1) = \half K (K-1)$.
We want to study the quotient of the space of triples of flags by the $G$-action:
\begin{equation}
\CC_3 = G \backslash (G/B)^3.
\end{equation}
Since only $SL(K,\IC)$ acts effectively, $\CC_3$ has dimension
\begin{equation}
\dim \CC_3 = \half (K-1)(K-2).
\end{equation}
In \cite{MR2233852} Fock and Goncharov introduced
a nice set of coordinates on this space. The coordinates are in 1-1
correspondence with points in the $(K-3)$-triangle of spaces in $V$.

Let us begin with the case $K=3$, in which case $\CC_3$ is 1-dimensional.
We fix the three flags:
\begin{equation}
\begin{split}
A_\bullet & = a_1 \IC \subset a_1 \IC \oplus a_2 \IC \subset V, \\
B_\bullet & = b_1 \IC \subset b_1 \IC \oplus b_2 \IC \subset V, \\
C_\bullet & = c_1 \IC \subset c_1 \IC \oplus c_2 \IC \subset V. \\
\end{split}
\end{equation}
where $a_i,b_i,c_i$ are vectors in $V$.
Then in  \cite{MR2233852}, p. 135, Section 4,
Fock and Goncharov introduce the triple ratio:
\begin{equation}\label{eq:tripleratio}
r(A_\bullet, B_{\bullet}, C_{\bullet}) : = \frac{ (a_1 \wedge a_2 \wedge b_1)}{(a_1\wedge a_2 \wedge c_1)}
\frac{(b_1 \wedge b_2 \wedge c_1)}{(b_1\wedge b_2 \wedge a_1)}\frac{ (c_1 \wedge c_2 \wedge a_1)}{(c_1\wedge c_2 \wedge b_1)}.
\end{equation}
The triple ratio $r$ only depends on the three flags, not on the choice of basis vectors.
To check this note first that the scales cancel out, and second that if we shift $x_2$ by a multiple of $x_1$,
where $x$ is one of $a,b,c$, then $r$ does not change.
Note that
\begin{equation}
r(A_\bullet, B_{\bullet}, C_{\bullet}) = \frac{1}{r(B_\bullet, A_{\bullet}, C_{\bullet})}, \qquad
r(A_\bullet, B_{\bullet}, C_{\bullet}) =  r( B_{\bullet}, C_{\bullet},A_{\bullet}).
\end{equation}

\subsubsection{Three alternative formulations of the triple ratio}

There are three useful ways to think about the triple ratio $r(A_\bullet, B_{\bullet}, C_{\bullet})$,
which we now describe.

\insfigscaled{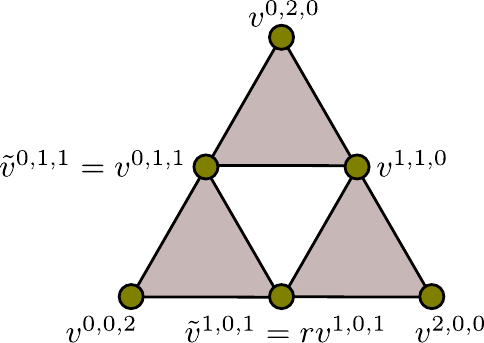}{1.05}{Basis vectors in the 2-triangle of lines in a 3-dimensional vector space,
making the canonical homs as simple as possible.}

First, consider
the $2$-triangle of lines defined by the three flags $A_\bullet, B_\bullet, C_\bullet$.
Let us attempt to choose basis vectors $v^{x,y,z} \in \fL^{x,y,z}$ so that the canonical homs
associated to edges are as simple as possible. See Figure \ref{fig:nice-basis-vectors}.
First we choose $v^{0,2,0}$, $v^{0,1,1}$ and $v^{1,1,0}$ such that
\begin{equation}\label{eq:2triangbas-a}
v^{0,2,0} + v^{0,1,1} + v^{1,1,0} = 0.
\end{equation}
Having chosen $v^{1,1,0}$, we are still at liberty to
choose $v^{1,0,1}$ and $v^{2,0,0}$ so that
\begin{equation}\label{eq:2triangbas-b}
v^{1,1,0} + v^{1,0,1} + v^{2,0,0} = 0.
\end{equation}
However, for the third triangle we are now stuck. We have already
chosen two of the relevant basis vectors $v^{0,1,1}$ and $v^{1,0,1}$.
There is of course a triplet of basis vectors $\tilde v^{0,1,1}$, $\tilde v^{1,0,1}$, $v^{0,0,2}$ such that
\begin{equation}\label{eq:2triangbas-c}
\tilde v^{0,1,1} + \tilde v^{1,0,1} + v^{0,0,2} = 0.
\end{equation}
Such a triplet is unique up to overall scale.  Now we fix this scale by choosing
$\tilde v^{0,1,1} = v^{0,1,1}$.  Then $\tilde v^{1,0,1} = r v^{1,0,1}$ for some $r \in \IC^\times$.
Now with
\begin{equation}
\begin{split}
A_\bullet & = \fL^{2,0,0} \subset \fP^{1,0,0} \subset V, \\
B_\bullet & = \fL^{0,2,0} \subset \fP^{0,1,0} \subset V, \\
C_\bullet & = \fL^{0,0,2} \subset \fP^{0,0,1} \subset V,
\end{split}
\end{equation}
a small computation shows that
\begin{equation}
r(A_\bullet, B_\bullet, C_\bullet) = r.
\end{equation}

A second way to understand $r$ is to consider the canonical homs associated with the
edges shown in Figure \ref{fig:move-type-2}  below.  An easy computation using
the above basis vectors reveals that
\begin{equation}\label{eq:TwoHomCom}
x_{\tilde E_1} x_{\tilde E_2} = r x_{E_1}x_{E_2}.
\end{equation}

A third, rather elegant, way to view the invariant $r$ associated with a
$2$-triangle of lines is to consider the composition of the canonical homs
associated to the three internal edges bounding the downward pointing unshaded triangle
in Figure \ref{fig:nice-basis-vectors}.
If we compose the three canonical homs going around the triangle counterclockwise,
\begin{equation}\label{eq:ThreeCanHom}
{\rm Hom}(\fL^{0,1,1}, \fL^{1,0,1}) \otimes {\rm Hom}(\fL^{1,0,1},\fL^{1,1,0})
\otimes  {\rm Hom}(\fL^{1,1,0},\fL^{0,1,1}) \to {\rm Hom}(\fL^{0,1,1}, \fL^{0,1,1}) \cong \IC,
\end{equation}
we obtain
\begin{equation}\label{eq:ThreeCanHom-b}
x_{E_1} x_{E_2} x_{E_3} = r.
\end{equation}
If we instead compose the three canonical homs going around the triangle clockwise,
we obtain
\begin{equation}
x_{\bar E_3} x_{\bar E_2} x_{\bar E_1} = -1/r.
\end{equation}
These statements are easily verified using the
basis vectors in \eqref{eq:2triangbas-a}-\eqref{eq:2triangbas-c}.

\subsubsection{Fock and Goncharov's triple ratios for \texorpdfstring{$K > 3$}{K>3}}

Now let us proceed to the case where $\dim V = K > 3$.
We want to assign a coordinate to each interior
point $(x,y,z)$ with $x+y+z = K-3$.
Around equation (9.10), p. 140, of  \cite{MR2233852},
Fock and Goncharov observe that the three flags $A_\bullet, B_{\bullet},C_{\bullet}$ in $V$ induce
three flags $\bar A_\bullet, \bar B_{\bullet},\bar C_{\bullet}$
in the three-dimensional space $\bar V := V/(A_x + B_y + C_z)$,
and they define their coordinate $r_{FG}^{x,y,z}$ to be the triple ratio of the induced flags in
this three-dimensional space.

More precisely, by the projected flag $\bar A_{\bullet}$ we mean the flag
\begin{equation}
\begin{split}
0 \cong (A_x + B_y + C_z)/W_{x,y,z} \subset (A_{x+1} + B_y + C_z)/W_{x,y,z}
\subset & (A_{x+2} + B_y + C_z)/W_{x,y,z} \subset \\
\subset  (A_{x+3} + B_y + C_z)/W_{x,y,z} \subset & \cdots
\end{split}
\end{equation}
where $W_{x,y,z} = A_x + B_y + C_z$.  For flags in generic position
$(A_{x+3} + B_y + C_z)/W_{x,y,z} = V/W_{x,y,z}$, and thus we get a flag
$\bar A_\bullet$ in three dimensions.  Defining similarly $\bar B_{\bullet}$
and $\bar C_{\bullet}$ and taking their triple ratio gives Fock and Goncharov's definition:
\begin{equation}
r_{FG}^{x,y,z} := r( \bar A_{\bullet}, \bar B_{\bullet},\bar C_{\bullet}).
\end{equation}

In preparation for the discussion in \S \ref{subsubsec:Coord-Relation}
we note that there is a dual construction which Fock and Goncharov
could equally well have employed.  In general if $W_1 \subset W_2 \subset V$
then there is a surjection $V/W_1 \rightarrow V/W_2 $ and hence a
dual injection $(V/W_2)^* \hookrightarrow (V/W_1)^*$. Therefore, we
can also consider the flag $A'_\bullet\subset V^*$ of length three:
\begin{equation}
\left( V/(A_{x+2} + B_y + C_z) \right)^*
\hookrightarrow
\left( V/(A_{x+1} + B_y + C_z) \right)^*
\hookrightarrow
\left( V/(A_{x} + B_y + C_z) \right)^*,
\end{equation}
and similarly $B'_{\bullet}$ and $C'_{\bullet}$.
A short computation shows that the triple ratio of these flags in the common
three-dimensional space $\left( V/(A_{x} \oplus B_y \oplus C_z) \right)^* $
is
\begin{equation} \label{eq:dual-FG-def}
r( A'_{\bullet}, B'_{\bullet}, C'_{\bullet}) =  1/r_{FG}^{x,y,z}.
\end{equation}

\subsubsection{An alternative coordinate system for \texorpdfstring{$K > 3$}{K>3}}

For $K=3$ we illustrated an alternative formulation of the triple ratio
using equations \eqref{eq:ThreeCanHom} and \eqref{eq:ThreeCanHom-b}.
For $K>3$ we can use the same idea to construct \emph{different}
coordinates $r^{x,y,z}$ on the moduli space $\CC_3$ of triples of flags, as follows.
We consider the down-triangle with vertices $\fL^{x+1,y+1,z}, \fL^{x,y+1,z+1}, \fL^{x+1,y,z+1}$, and take
the composition of canonical homs going around this triangle
counterclockwise:
\be
\fL^{x,y+1,z+1} \to \fL^{x+1,y,z+1} \to \fL^{x+1, y+1,z} \to \fL^{x,y+1,z+1}.
\ee
The composition of these homs is multiplication by
a scalar; we define $r^{x,y,z}$ to be this scalar.

Here is another construction of the same coordinate.  To the interior point $(x,y,z)$ with
$x+y+z = K-3$ we associate the three-dimensional space $\fV^{x,y,z}$. This space
contains the three planes $\fP^{x+1,y,z}, \fP^{x,y+1,z}, \fP^{x,y,z+1}$, the six lines
\begin{equation}\label{eq:six-lines-x,y,z}
\fL^{x+2,y,z},\ \fL^{x,y+2,z},\ \fL^{x,y,z+2},\ \fL^{x+1,y+1,z},\ \fL^{x,y+1,z+1},\ \fL^{x+1,y,z+1},
\end{equation}
and the three associated flags:
\begin{equation}
\begin{split}
\tilde A_{\bullet}: & \quad 0 \subset \fL^{x+2,y,z} \subset \fP^{x+1,y,z} \subset \fV^{x,y,z}, \\
\tilde B_{\bullet}: & \quad 0 \subset \fL^{x,y+2,z} \subset \fP^{x,y+1,z} \subset \fV^{x,y,z}, \\
\tilde C_{\bullet}: & \quad 0 \subset \fL^{x,y,z+2} \subset \fP^{x,y,z+1} \subset \fV^{x,y,z}. \\
\end{split}
\end{equation}
There is, accordingly, a triple ratio $r(\tilde A_{\bullet}, \tilde B_{\bullet}, \tilde C_{\bullet})$,
which coincides with our $r^{x,y,z}$.

\subsubsection{Relation between the coordinate systems}\label{subsubsec:Coord-Relation}

The coordinates $r^{x,y,z}$ and $r_{FG}^{x,y,z}$ are
\emph{not} the same when $K>3$. In this section we explain
the relation between them.

To begin, recall that if $W$ is any linear subspace of $V$
we can define
\begin{equation}
W^{\perp}:= \{ \ell\vert \ell(w) = 0 \quad \forall w\in W \} \subset V^*.
\end{equation}
Of course $(W^{\perp})^\perp = W$. We can use this to construct a
notion of a dual flag, as follows.  If
\be
W_\bullet\ : \quad 0 = W_0 \subset W_1 \subset \cdots \subset W_{K-1} \subset W_K = V
\ee
is a flag in $V$, with $\dim W_x = x$,  then we define
\be
\check W_x := (W_{K-x})^\perp = (W^x)^\perp.
\ee
Note that $\dim \check W_x = x$ and so
\be
\check W_{\bullet}\ : \quad 0=\check W_0 \subset \check W_{1} \subset \cdots \subset \check W_{K-1} \subset \check W_K = V^*
\end{equation}
is a flag, which we call the \emph{dual flag}.
There is an \emph{a priori} different notion of
dual flag:  we could have defined
\begin{equation}
0 \hookrightarrow (V/W_{K-1})^* \hookrightarrow (V/W_{K-2})^* \hookrightarrow \cdots
\hookrightarrow (V/W_2)^* \hookrightarrow (V/W_1)^* \hookrightarrow V^*
\end{equation}
to be the dual flag.  The two are in fact canonically isomorphic
since there is a canonical isomorphism $(V/W)^* \cong W^{\perp}$.\footnote{Indeed,
given $\ell \in W^{\perp}$, we may define $\bar\ell\in (V/W)^*$
by  $\bar\ell(\bar u):= \ell(u)$, where $u$ is any lift of $\bar u$.
The map $\ell \mapsto \bar \ell$ clearly has zero kernel, and hence,
by dimension counting, must be an isomorphism.}

So, given flags $A_{\bullet}$, $B_{\bullet}$, $C_{\bullet}$ in $V$, there are dual flags
$\check A_{\bullet}$, $\check B_{\bullet}$, $\check C_{\bullet}$ in $V^*$.
We are going to show that the Fock-Goncharov coordinates of these dual flags
are related to our coordinates by:
\begin{equation}\label{eq:Relation-FG-coord}
r_{FG}^{x,y,z}(\check A_{\bullet} ,\check B_{\bullet}, \check C_{\bullet}) = 1 / r^{x,y,z}(A_{\bullet}, B_{\bullet}, C_{\bullet}) .
\end{equation}
In order to prove \eqref{eq:Relation-FG-coord} we first note that
if $W_1$ and $W_2$ are any two subspaces of $V$
then $(W_1 + W_2)^\perp = W_1^\perp \cap W_2^\perp$. It follows that
\begin{equation}\label{eq:3-quot}
\left(V/(W_1 + W_2 + W_3 )\right)^* \cong W_1^\perp \cap W_2^\perp \cap W_3^{\perp} \subset V^*.
\end{equation}
We now apply \eqref{eq:3-quot} replacing $V$ by $V^*$ and taking $W_1, W_2,W_3$ to be
spaces in the dual flags, to obtain three canonical isomorphisms:
\be\label{eq:dual-abc}
\left( V^* /(\check A_x + \check B_y + \check C_z )\right)^*  \cong \check A_{x}^\perp \cap \check B_y^\perp
\cap \check C_z^\perp = A^{x}\cap B^y \cap C^z = \fV^{x,y,z},
\ee
\be\label{eq:dual-abc=ii}
\left(V^* /(\check A_{x+1} + \check B_y + \check C_z )\right)^*  \cong \check A_{x+1}^\perp \cap \check B_y^\perp
\cap \check C_z^\perp = A^{x+1}\cap B^y \cap C^z = \fP^{x+1,y,z},
\ee
\be\label{eq:dual-abc=iii}
\left(V^* /(\check A_{x+2} + \check B_y + \check C_z )\right)^*  \cong \check A_{x+2}^\perp \cap \check B_y^\perp
\cap \check C_z^\perp = A^{x+2}\cap B^y \cap C^z = \fL^{x+2,y,z}.
\ee
Now, the Fock-Goncharov coordinate $r_{FG}^{x,y,z}(\check A_{\bullet} ,\check B_{\bullet}, \check C_{\bullet})$
is, by \eqref{eq:dual-FG-def}, the reciprocal of the triple ratio associated to the flags
$\check A_{\bullet}', \check B_\bullet', \check C_\bullet'$ in three dimensions, defined by
\be
\check A_\bullet': \quad 0 \subset  \left(V^* /(\check A_{x+2} + \check B_y + \check C_z )\right)^* \hookrightarrow
 \left(V^* /(\check A_{x+1} + \check B_y + \check C_z )\right)^*
 \hookrightarrow  \left(V^* /(\check A_x + \check B_y + \check C_z )\right)^*
\ee
and similarly for $\check B_{\bullet}'$ and $\check C_{\bullet}'$.
Applying the isomorphisms \eqref{eq:dual-abc}, \eqref{eq:dual-abc=ii},
and \eqref{eq:dual-abc=iii} we identify $\check A'_\bullet$ with the flag
$\fL^{x+2,y,z} \subset \fP^{x+1,y,z}\subset \fV^{x,y,z}$, and similarly for $\check B'_\bullet$ and $\check C'_\bullet$.
Combining this with the observation \eqref{eq:dual-FG-def} we obtain \eqref{eq:Relation-FG-coord}.

We close with two remarks:

\begin{enumerate}

\item In \S \ref{subsec:rrcheck} below we give an algorithm for determining
explicit expressions for the $r_{FG}^{x,y,z}$ as functions of the $r^{x,y,z}$.
Explicit examples of such transformations are given in
\eqref{eq:explicit-rrcheck-K4} and \eqref{eq:explicit-rrcheck-K5} above
for $K=4,5$ respectively.

\item We have not demonstrated that the $r^{x,y,z}$ really do form an
independent set of coordinates on $\CC_3$.  One way to do this is to invoke
\eqref{eq:Relation-FG-coord} and the results of Fock and Goncharov.
Another way to proceed is to consider the perturbation of the
Stokes matrices defined in equations \eqref{eq:SN-BC-2} and
\eqref{eq:Snake-TMN} around the point $r^{x,y,z}=1$.  It can be shown
that the perturbations by the different $r^{x,y,z}$ are independent.

\end{enumerate}

\insfigscaled{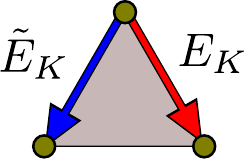}{0.78}{The type I elementary snake move of Fock-Goncharov.  The bottom
edge of the triangle shown is on the final destination edge of the snake. }

\insfigscaled{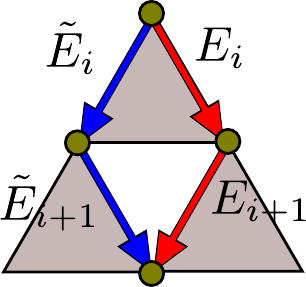}{0.78}{The type II elementary snake move of Fock-Goncharov. The triangle shown
is embedded into a larger $m$-triangle, and we only show the $i^{th}$ and $(i+1)^{th}$
edges of the snakes. The type II move replaces the chain $\cdots E_{i+1} E_i \cdots$ in the
first snake by $\cdots \tilde E_{i+1} \tilde E_i \cdots$ in the new snake.
The canonical homs are related by $x_{\tilde E_i} x_{\tilde E_{i+1}} = r x_{E_i}x_{E_{i+1}}$.}

\insfigscaled{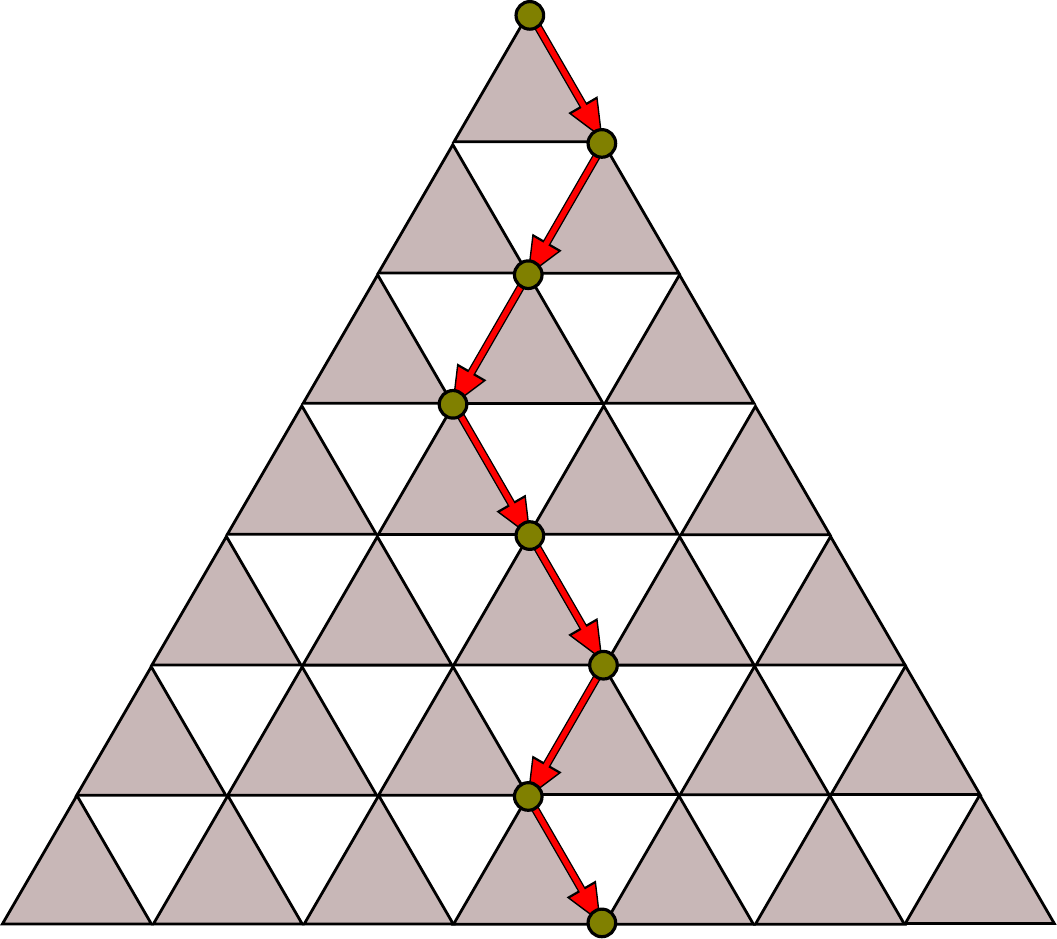}{0.55}{A typical snake, oriented from the $B$-vertex to the $B$-side.}

\subsection{Fock and Goncharov's snakes}\label{subsec:Snakes}

In Section 9, especially subsections 7, and 8, p. 138 et. seq.  of
\cite{MR2233852} Fock and Goncharov introduced the notion of
a snake associated to a vector space with three flags:

\textbf{Definition}: A \emph{snake} associated to a $K$-dimensional vector space $V$ with three flags
in general position is an oriented edge path of length $K-1$, which begins at a vertex of
the $(K-1)$-triangle of lines in $V$ and terminates at the opposite edge.
A typical example is shown in Figure \ref{fig:typical-snake}.

There are $2^{K-1}$ different snakes from a fixed vertex to the opposite side.
If we consider snakes from the $B$-vertex to the $B$-side,
we can label each snake by a word in two letters, $R$ and $L$, with the rightmost
letter indicating the direction of the first edge and the leftmost letter
indicating the direction of the last edge.
Any two snakes can be related
to each other by a series of elementary moves, of two basic types.
(See \cite{MR2233852}, p. 140, Figure 9.11.)
A move of type I is shown in Figure \ref{fig:move-type-1} and a move of type II is shown
in Figure \ref{fig:move-type-2}.  Note that type I moves can only be made
on the final edge of a snake.

To a snake we can associate a decomposition of $V$ into lines. For example
for a snake from the $B$-vertex we have:
\begin{equation}
V = \fL^{0,K-1,0} \oplus \bigoplus_{i=1}^{K-1} \fL^{t(E_i)},
\end{equation}
where $t(E_i)$ is the target vertex of the $i^{th}$ edge of the snake.
Further, as we have explained, each oriented edge $E$ on the $(K-1)$-triangle of
lines is associated with a canonical hom, $x_E$.  Therefore, to a snake
we can associate a projective basis of $V$, as follows.  We first choose a basis vector
$v_K$ in $\fL^{0,K-1,0}$ and then iterate:
\begin{equation}
v_{K-i} = v_{K+1-i}x_{E_i}, \qquad 1 \leq i \leq K-1.
\end{equation}
We write
\begin{equation}
\CB^{\CS\CN} = \begin{pmatrix} v_1 \\   v_2\\  \vdots \\ v_{K-1} \\  v_K \\ \end{pmatrix}
\end{equation}
for the basis determined by a snake $\CS\CN$.

As discussed in \cite{MR2233852}, given two snakes, it is of interest to
describe explicitly the linear transformation between the corresponding projective bases.
It suffices to describe the transformations
for the two types of moves. For type I moves it is easy to see that
only $v_{1}$ is changed:
\begin{equation}
v_1^{\CS\CN_2} = v_1^{\CS\CN_1} + v_{2}^{\CS\CN_1},
\end{equation}
and therefore\footnote{A word about conventions: Our linear transformations
act from the right. Therefore, the matrix of a linear transformation
$T$ in a basis $v_i$ is defined by $v_i T = \sum_j T_{ij} v_j$.
If we change from a basis $\{v_i\}$ to a basis $\{v_i T\}$,
the change of coordinates is
obtained by acting on a row vector of coordinates $x_i$ by right-multiplication
with the matrix $T_{ij}$.}
\begin{equation}
\CB^{\CS\CN_2} = (1 + e_{1,2})\CB^{\CS\CN_1}.
\end{equation}

\insfigscaled{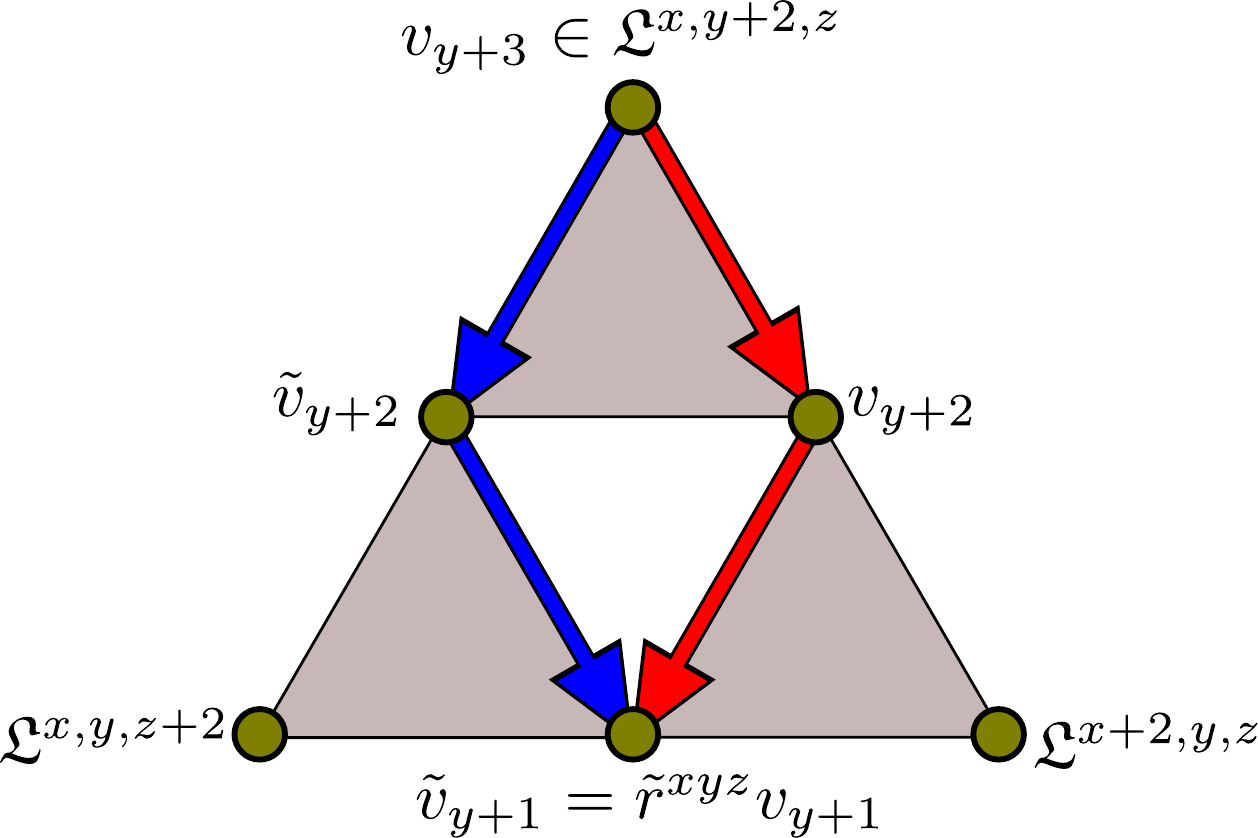}{0.55}{Computing the basis change from a type II move around the
point $(x,y,z)$ in the $(K-3)$-triangle of spaces. }

For a type II move centered on $\fV^{x,y,z}$, $x+y+z=K-3$, we first focus on the $2$-triangle of $6$ lines
\eqref{eq:six-lines-x,y,z}.  As indicated in Figure \ref{fig:move-type-2-sections}, we choose
basis vectors $v_{y+3}, v_{y+2}, v_{y+1}$ for the original snake $\CS\CN_1$.
Then we have
\begin{equation}
v_{y+2} + v_{y+3} - \tilde v_{y+2} = 0,
\end{equation}
and using the key formula \eqref{eq:TwoHomCom}, $\tilde v_{y+1} = r^{x,y,z} v_{y+1}$.
Thereafter the paths of the snakes $\CS\CN_1$ and $\CS\CN_2$ are the same, so the
full change of basis is
\begin{equation}\label{eq:SN-BC-1}
\CB^{\CS\CN_2} = M(x,y+1,z) \CB^{\CS\CN_1},
\end{equation}
where
\begin{equation}\label{eq:SN-BC-2}
M(x,y+1,z) := \left(r^{x,y,z} \sum_{i=1}^{y+1} e_{ii} + \sum_{i=y+2}^{K} e_{ii} \right) \left(1 + e_{y+2,y+3}\right).
\end{equation}
We have labeled $M$ by a point $(x,y+1,z)$ on the $(K-2)$-triangle of planes.
This is convenient, since if we define $r=0$ for $y=-1$ then the formulae
\eqref{eq:SN-BC-1} and \eqref{eq:SN-BC-2} also hold for $y=-1$, in which case
we are considering a move of type I.
In the application to spectral networks, the transformations \eqref{eq:SN-BC-1}, \eqref{eq:SN-BC-2}
correspond to $\CS$-wall factors of type $S_{y+2,y+3}$, for $-1 \leq y \leq K-3$.

There are two canonical snakes beginning at the $B$-vertex, $\CS\CN_{BA}$ and $\CS\CN_{BC}$,
given respectively by the right and the left sides of the triangle of lines.
The change of projective basis relating these canonical snakes is of particular
importance for the spectral networks of the level $K$ lift of the $AD_1$ theory.
One can choose a path of $\half K (K-1) + 1$ snakes with
each successive snake related to the previous one by a move of type I or type II.
The total transformation between the two bases can therefore be written as
\begin{equation}\label{eq:Snake-TMN}
\CB^{\CS\CN_{BC}} = \prod_{(x,y,z)\in T(K-2)} M(x,y,z) \CB^{\CS\CN_{BA}},
\end{equation}
where the product runs over the $\half K (K-1)$ points $(x,y,z)$ on the triangle $T(K-2)$ of planes.
It is ordered so that later moves are further to the left.
The formula \eqref{eq:Snake-TMN} is equivalent to Proposition 9.2, p. 142, of \cite{MR2233852}.

Of course, there are many ways of using type I and type II moves to transform the
snake $\CS\CN_{BA}$ into the snake $\CS\CN_{BC}$, and any two such paths of snakes
must lead to the same overall transformation \eqref{eq:Snake-TMN}.
Nevertheless, it is useful to define a \emph{canonical path of snakes}
\begin{equation} \label{eq:CanSN-Path}
\CS\CN_{BA} \to \CS\CN_{BC}
\end{equation}
as follows.  We begin with the type I move
with $y=-1$ on $\CS\CN_{BA}$, and then proceed with type II moves up
the right edge of the $(K-3)$-triangle of spaces to $y=K-3$.
The result is a snake that goes one step to the left to $(0,K-2,1)$,
and then proceeds parallel to the right hand side down to $(K-2,0,1)$.
Next we begin again with a type I move at $y=-1$ and then proceed again up the edge
to the left with type II moves to $y=K-4$, and so on. See Figure 9.14, p.142, of \cite{MR2233852}
for an illustration.

In terms of $\CS$-transformations the canonical path of snakes is described by first implementing
\begin{equation}
S_{K-1,K} S_{K-2,K-1} \cdots S_{1,2}.
\end{equation}
(Remember that, since we are making a passive change of basis, this product is to be
read from right to left; thus the first move is $S_{1,2}$, a type I move at the end of the snake, and the remaining factors are type II moves going successively up to the beginning of the snake.)
Then the next set of moves is
\begin{equation}
S_{K-2,K-1} S_{K-3,K-2} \cdots S_{1,2},
\end{equation}
and so on with the length of factors getting shorter
by removing one $\CS$-factor successively from the left until we have just $S_{1,2}$.
The resulting product of $\CS$-factors corresponds
precisely to those in the product of $\CS$-wall factors for the $b$-cable in
\eqref{eq:b-cable-factors-1} and \eqref{eq:b-cable-factors-2}.

The change of
orientation from $\CS\CN_{BC} \to \CS \CN_{CB}$ multiplies the canonical homs
along the snake by $(-1)$ (because of the change of sign in the two lines
in \eqref{eq:canhom3}).
If we fix the overall scale by $v_K^{\CS\CN_{CB}} = v_1^{\CS\CN_{BC}}$, the
effect of this orientation reversal is the transformation $W_0^{(K)}$
of \eqref{eq:W0-DEF}:
\begin{equation}
v_i^{\CS\CN_{CB}}  = (-1)^{K+1-i} v_{K+1-i}^{\CS\CN_{BC}}.
\end{equation}
Let us denote the total transformation in \eqref{eq:Snake-TMN} by $M^{BA\to BC}$. Then
it is clear that we have
\begin{equation}\label{eq:GoingRoundA}
M^{AC \to AB} W_0 M^{CB\to CA} W_0 M^{BA\to BC} = \kappa W_0,
\end{equation}
where $\kappa$ is a constant. We must allow for such a constant because
a snake only determines a \emph{projective} basis.
We claim that the constant is
\begin{equation}\label{eq:GoingRoundB}
\kappa = (-1)^{K-1} \prod_{(x,y,z)\in T(K-3)} r^{x,y,z},
\end{equation}
where on the right hand side we take the product over the points in the $(K-3)$-triangle of
spaces.  To prove \eqref{eq:GoingRoundB} we take the determinant
of \eqref{eq:GoingRoundA}.  That determines $\kappa$ up to a $K^{th}$ root of
unity. It is clear from the definition of the matrices that $\kappa$ must also be a polynomial in the
$r^{x,y,z}$.  Therefore the root of unity is independent of the $r^{x,y,z}$, and
can be determined by considering the special case
$r^{x,y,z}=1$.  We write out the matrices explicitly for $r^{x,y,z}=1$ in
\S \ref{subsec:r-is-one}, and then \eqref{eq:explct-r-1} shows that the
root of unity is $(-1)^{K-1}$.

The formula \eqref{eq:SN-BC-2} provides an explicit solution of the monodromy
constraint \eqref{eq:N-Odd-Monodromy}, when the mass parameters are equal to $1$.

\subsubsection{Examples of snakes for \texorpdfstring{$K=3$}{K=3} and \texorpdfstring{$K=4$}{K=4}} \label{app:snakes-ex}

As examples we consider the first two cases $K=3$ and $K=4$.
Referring to the
basis vectors defined in \eqref{eq:2triangbas-a}, \eqref{eq:2triangbas-b}, \eqref{eq:2triangbas-c}
(with $\tilde v^{011} = v^{011}$), we have the bases defined by the four snakes:
\begin{equation}
\begin{split}
\CB^{RR}: & \quad \{ v^{020}, v^{110}, v^{200} \} \\
\CB^{LR}: & \quad \{ v^{020}, v^{110}, - v^{101} \} \\
\CB^{RL}: & \quad \{ v^{020}, - v^{011}, -\tilde v^{101} \} \\
\CB^{LL}: & \quad \{ v^{020}, - v^{011},  v^{002} \} \\
\end{split}
\end{equation}
Then we have the basis changes:
\begin{equation}
\CB^{LR} = \CB^{RR} \begin{pmatrix} 1 & & \\ 1 & 1 &  \\  & &  1\\ \end{pmatrix},
\end{equation}
\begin{equation}
\CB^{RL} = \CB^{LR} \begin{pmatrix} r &  & \\  & 1 &   \\  & 1 & 1 \\ \end{pmatrix},
\end{equation}
\begin{equation}
\CB^{LL} = \CB^{RL} \begin{pmatrix} 1 &  & \\ 1 & 1 &  \\  & & 1\\ \end{pmatrix}.
\end{equation}

Next let us consider the case $K=4$. There are three sub-2-triangles we should
focus on. In the top 2-triangle we can choose basis vectors
\begin{equation}
\begin{split}
v^{030} + v^{021} + v^{120} & = 0 \\
v^{120} + v^{111} + v^{210} & = 0 \\
v^{021} + \tilde v^{111} + v^{012} & = 0 \\
\end{split}
\end{equation}
with $\tilde v^{111} = r^{010} v^{111}$. On the
bottom right 2-triangle we choose
\begin{equation}
\begin{split}
v^{120} + v^{111} + v^{210}  & = 0 \\
 v^{210} + v^{201} + v^{300}  & = 0 \\
v^{111} +   v^{102} + \tilde v^{201}  & = 0 \\
\end{split}
\end{equation}
with $\tilde v^{201} = r^{100} v^{201}$. Finally, on the bottom
left 2-triangle we choose:
\begin{equation}
\begin{split}
v^{021} + v^{012} + \tilde v^{111}  & = 0 \\
 \tilde v^{111} + \hat v^{201} +\hat  v^{102}  & = 0 \\
v^{012} +   v^{003} + \tilde {\hat v}^{102}  & = 0 \\
\end{split}
\end{equation}
with $\tilde {\hat v}^{102} = r^{001} \hat v^{102}$ and also
$\hat v^{201} = r^{010} \tilde v^{201}$ and $\hat v^{102}=r^{010} v^{102}$.
In terms of these vectors we can write the $7$ snakes in the canonical
path of snakes from $RRR$ to $LLL$:\footnote{Of course there are $8$
snakes in total, but one of them does not occur in this canonical
path. In general there are $2^{K-1}$ snakes but only $1+ \half K(K-1)$
snakes in the canonical path. The overall change of basis is of course independent
of which path of snakes from $RRR$ to $LLL$ we take.  This fact
is reflected in the spectral network, since one can freely
pass an $\CS$-wall of type $ij$ through $\CS$-walls and branch points of
type $kl$ as long as the indices do not overlap.}
\begin{equation}
\begin{split}
\CB^{RRR}: & \quad \{ v^{030}, v^{120}, v^{210}, v^{300} \} \\
\CB^{LRR}: & \quad \{ v^{030}, v^{120}, v^{210}, -v^{201} \} \\
\CB^{RLR}: & \quad \{ v^{030}, v^{120}, - v^{111}, - \tilde v^{201} \} \\
\CB^{RRL}: & \quad \{ v^{030}, - v^{021},-\tilde  v^{111}, - \hat v^{201} \} \\
\CB^{LRL}: & \quad \{ v^{030}, - v^{021},-\tilde  v^{111},  \hat v^{102} \} \\
\CB^{RLL}: & \quad \{ v^{030}, - v^{021},  v^{012}, \tilde{\hat v}^{102} \} \\
\CB^{LLL}: & \quad \{ v^{030}, - v^{021},  v^{012}, - v^{003} \}
\end{split}
\end{equation}
From the linear relations among these bases, one can check that the changes of basis are as expected:
\begin{equation}
\CB^{LRR} = \CB^{RRR} \begin{pmatrix} 1 &  & & \\ 1 & 1 &  &   \\  & & 1 &  \\   & & & 1 \end{pmatrix},
\end{equation}
\begin{equation}
\CB^{RLR} = \CB^{LRR} \begin{pmatrix} r^{100} &  & & \\  & 1 &  &   \\  & 1 & 1 &  \\   & & & 1 \end{pmatrix},
\end{equation}
\begin{equation}
\CB^{RRL} = \CB^{RLR} \begin{pmatrix} r^{010} & & & \\  & r^{010} &    &   \\  & & 1 &  \\   & & 1 & 1 \end{pmatrix},
\end{equation}
\begin{equation}
\CB^{LRL} = \CB^{RRL} \begin{pmatrix} 1 & & & \\ 1 & 1 &  &   \\  & & 1 &  \\   & & & 1  \end{pmatrix},
\end{equation}
\begin{equation}
\CB^{RLL} = \CB^{LRL} \begin{pmatrix} r^{001} &  & & \\  & 1 & &  \\  & 1 & 1 &  \\ & & & 1 \end{pmatrix},
\end{equation}
\begin{equation}
\CB^{LLL} = \CB^{RLL} \begin{pmatrix} 1 &  & & \\ 1 & 1 & & \\  & & 1 & \\ & & & 1 \end{pmatrix}.
\end{equation}

\subsection{Duality}\label{subsec:rrcheck}

Finally we consider a basic and important question:  how are the coordinates $r^{x,y,z}$ of a triple
$(A_\bullet, B_\bullet, C_\bullet)$ related to the coordinates $\check{r}^{x,y,z}$ of the dual triple
$(\check A_\bullet, \check B_\bullet, \check C_\bullet)$?

To answer this question, first note that the coordinates $r^{xyz}$ can be read off\footnote{This is clear from the examples below. It seems likely one can give
a rigorous proof by induction, but we did not do that.} from the
cable transformation $M = M^{BA \to BC}$.
For the dual flags we also have a cable transformation $\check{M} = M^{\check B \check A \to \check B \check C}$, from which we could similarly read off the dual coordinates $\check{r}^{x,y,z}$.

Let us define matrices $\alpha$, $\beta$ by
\begin{equation}
 v_i^{\CS\CN^{\check B \check A}} \cdot v_j^{\CS\CN^{BA}} = \alpha_{ij}, \qquad v_i^{\CS\CN^{\check B \check C}} \cdot v_j^{\CS\CN^{BC}} = \beta_{ij}.
\end{equation}
Then substituting $ v_i^{\CS\CN^{BC}} = M_{ij} v_j^{\CS\CN^{BA}}$ and the dual relation
into the formula for $\beta$ reveals
\begin{equation} \label{eq:duality-relation}
 \beta = M \alpha \check M^{t}.
\end{equation}

On the other hand, it follows directly from the definition of the snakes that
$\alpha$, $\beta$ are antidiagonal. To see this note that
$v_j^{\CS\CN^{BA}}$ belongs to $A^{K-j} \cap B^{j-1}$,
while $v_i^{\CS\CN^{\check B \check A}}$ belongs to $\check{A}^{K-i} \cap \check{B}^{i-1}$;
$\check{B}^{i-1}$ annihilates $B^{j-1}$ unless $i+j-2 < K$, i.e. $i+j < K+2$,
while $\check{A}^{K-i}$ annihilates $A^{K-j}$ unless $2K - i - j < K$,
i.e. $i+j > K$.  Thus
$v_i^{\CS\CN^{\check B \check A}} \cdot v_j^{\CS\CN^{BA}} = 0$ unless $i+j=K+1$.

Equation \eqref{eq:duality-relation} can be rewritten as
\be\label{eq:cMtoM}
\check M = \tilde \beta M^{-1,t} \tilde \alpha
\ee
where $\tilde\beta$ and $\tilde \alpha$ are antidiagonal.
This equation turns out to be strong enough to determine $\check r^{x,y,z}$ as a
function of $r^{x,y,z}$, as well as determining $\tilde \alpha$ and $\tilde \beta$
up to an overall scale.  To check that this is reasonable, let us count the parameters.
$M$ is an upper-triangular $K \times K$ matrix, with entries depending only on the
$\half (K-1)(K-2)$ independent variables $r^{x,y,z}$ --- in other words its entries obey
$2K-1$ nontrivial relations.  The matrix $\check M$ given by \eqref{eq:cMtoM} is
upper-triangular for any $\alpha$, $\beta$.  For general $\tilde\alpha$, $\tilde\beta$ though,
$\check M$ would not obey the same $2K-1$ relations; requiring that it does is just enough to
fix $\tilde\alpha$, $\tilde\beta$ up to scale.  The remaining
$\half (K-1)(K-2)$ equations in \eqref{eq:cMtoM} are then enough
to fix $\check r^{x,y,z}$.

In practice it is useful to rewrite \eqref{eq:cMtoM} slightly
as follows.  We define a normalized matrix $T(r^\fv)$ by left-multiplying
$M$ by a diagonal matrix so that the diagonal elements of $T(r^\fv)$
are all $1$, and similarly construct $T(\check r^\fv)$ from $\check M$. Then we obtain
\begin{equation}
T(\check r^\fv) = \tilde{\tilde\beta} \left( T(r^\fv)\right)^{-1,t} \tilde{\tilde\alpha}
\end{equation}
This easily implies that $\tilde{\tilde\beta} = \tilde{\tilde\alpha}^{-1}$.  We can thus
rewrite it as
\begin{equation} \label{eq:explicit-sg}
T(\check r^\fv) = D W_0 \left( T(r^\fv)\right)^{-1,t} W_0^{-1} D^{-1}
\end{equation}
where $W_0=W_0^{(K)}$ and
$D$ is some diagonal matrix (uniquely determined up to scale as a function
of $r^\fv$ by \eqref{eq:explicit-sg}).

\section{Explicit form of cable transformations}\label{app:ExplicitCable}

In this section we give explicit cable transformations
in some tractable special cases.

\subsection{\texorpdfstring{$K=3$}{K=3}} \label{app:cable3}

Writing out \eqref{eq:SN-BC-2} explicitly we find that $M^{BA\to BC}$ is a product of three factors:
\be\label{eq:Explcit-b-cableK4}
\begin{split}
M^{BA\to BC}(1,0,0) & = (1 + e_{12}), \\
M^{BA\to BC}(0,1,0) & =(r^{000} e_{11} + e_{22} + e_{33} )(1 + e_{23}), \\
M^{BA\to BC}(0,0,1) & = (1 + e_{12}),
\end{split}
\ee
where we have ordered the list so that the product from left to right is the product of factors
from bottom to top. The explicit matrix is easily multiplied out. We find that $M^{BA \to BC}$ is
\be \label{eq:M-K3}
\begin{pmatrix} r^{000} & 1 + r^{000} & 1 \\ 0 & 1 & 1 \\ 0 & 0 & 1 \end{pmatrix}.
\ee
The other cable transformations are identical, and we can multiply them out to find
\be
M^{AC\to AB}W_0^{(3)} M^{CB\to CA} W_0^{(3)} M^{BA\to BC} = \kappa W_0^{(3)}
\ee
with $\kappa = r^{000}$.

Now let us work out the relation between $r$ and $\hat{r}$ in this example.
Writing out \eqref{eq:explicit-sg} longhand gives
\begin{equation}
\begin{pmatrix} \check r^{000} & 1 + \check r^{000} & 1 \\ 0 & 1 & 1 \\ 0 & 0 & 1 \end{pmatrix} =
D W_0 \begin{pmatrix} r^{000} & 1 + r^{000} & 1 \\ 0 & 1 & 1 \\ 0 & 0 & 1 \end{pmatrix}^{-1,t} W_0^{-1} D^{-1}
\end{equation}
from which we easily read off
\begin{equation}
 \hat r^{000} = 1 / r^{000}.
\end{equation}

\subsection{\texorpdfstring{$K=4$}{K=4}} \label{app:cable4}

Writing out \eqref{eq:SN-BC-2} explicitly we find that $M^{BA\to BC}$ is a product of
the following six factors:
\begin{equation}\label{eq:K4BABC}
\begin{split}
M^{BA\to BC}(2,0,0) & = (1 + e_{12}), \\
M^{BA\to BC}(1,1,0) & =(r^{100} e_{11} + \sum_{i=2}^4 e_{ii})(1 + e_{23}), \\
M^{BA\to BC}(0,2,0) & =\left(r^{010} (e_{11} + e_{22}) + (e_{33}+e_{44}) \right)(1 + e_{34}), \\
M^{BA\to BC}(1,0,1) & = (1 + e_{12}), \\
M^{BA\to BC}(0,1,1) & =(r^{001} e_{11} + \sum_{i=2}^4 e_{ii})(1 + e_{23}), \\
M^{BA\to BC}(0,0,2) & = (1 + e_{12}).
\end{split}
\end{equation}
We have ordered the list so that the product from left to right is the product of factors
from bottom to top.

The explicit matrix is easily multiplied out.
For example, $M^{BA \to BC}$ is
\be\label{eq:explicitk4bcable}
\begin{pmatrix}
 r^{001} r^{010} r^{100} & (r^{001}+1) r^{010}+r^{001}
   r^{100} r^{010} & (r^{001}+1) r^{010}+1 & 1 \\
 0 & r^{010} & r^{010}+1 & 1 \\
 0 & 0 & 1 & 1 \\
 0 & 0 & 0 & 1
\end{pmatrix}.
\ee

Similarly, $M^{CB\to CA}$ is
\be\label{eq:Explcit-c-cableK4}
\begin{split}
M^{CB\to CA}(0,2,0) & = (1 + e_{12}), \\
M^{CB\to CA}(0,1,1) & = (r^{010} e_{11} + \sum_{i=2}^4 e_{ii})(1 + e_{23}), \\
M^{CB\to CA}(0,0,2) & = \left(r^{001} (e_{11} + e_{22}) + (e_{33}+e_{44}) \right)(1 + e_{34}), \\
M^{CB\to CA}(1,1,0) & = (1 + e_{12}), \\
M^{CB\to CA}(1,0,1) & = (r^{100} e_{11} + \sum_{i=2}^4 e_{ii})(1 + e_{23}), \\
M^{CB\to CA}(2,0,0) & = (1 + e_{12}).
\end{split}
\ee
Note particularly that we are using the same coordinates for the triangle $T(K-2)$,
rather than a rotated set of coordinates which would put the $C$ vertex on top.
Also, note that the matrices \eqref{eq:K4BABC}
and \eqref{eq:Explcit-c-cableK4} are the same except for a permutation of the coordinates $r^{\fv}$
induced by a rotation of the triangle.

Similarly, we find the $a$-cable is a product of
\be\label{eq:Explcit-a-cableK4}
\begin{split}
M^{AC\to AB}(0,0,2) & = (1 + e_{12}), \\
M^{AC\to AB}(1,0,1) & = (r^{001} e_{11} + \sum_{i=2}^4 e_{ii})(1 + e_{23}), \\
M^{AC\to AB}(2,0,0) & = \left(r^{100} (e_{11} + e_{22}) + (e_{33}+e_{44}) \right)(1 + e_{34}), \\
M^{AC\to AB}(0,1,1) & = (1 + e_{12}), \\
M^{AC\to AB}(1,1,0) & = (r^{010} e_{11} + \sum_{i=2}^4 e_{ii})(1 + e_{23}), \\
M^{AC\to AB}(0,2,0) & = (1 + e_{12}).
\end{split}
\ee

The explicit matrices are easily multiplied out; we find that
\be
M^{AC\to AB}W_0^{(4)} M^{CB\to CA} W_0^{(4)}  M^{BA\to BC} = \kappa W_0^{(4)}
\ee
with $\kappa = - r^{100} r^{010} r^{001}$.

The formula \eqref{eq:explicit-sg} yields in this case
\be
\begin{split}
\hat r^{100} & = \frac{1 + r^{100} + r^{100} r^{010}}{r^{010} \left(1 + r^{001} + r^{001}r^{100}\right)},\\
\hat r^{010} & = \frac{1 + r^{010} + r^{010} r^{001}}{r^{001} \left(1 + r^{100} + r^{100}r^{010}\right)},\\
\hat r^{001} & = \frac{1 + r^{001} + r^{001} r^{100}}{r^{100} \left(1 + r^{010} + r^{010}r^{001}\right)}.
\end{split}
\ee

\subsection{The point \texorpdfstring{$r^{x,y,z}=1$}{r(xyz)=1}}\label{subsec:r-is-one}

At the point of $\CC_3$ where all $r^{x,y,z}=1$, it is possible to
write explicit formulae for the cable factors $T_a, T_b, T_c$, valid for
all $K$.

We find that $M^{BA \to BC}$, which can be written as
\be
M^{BA \to BC} = S_{1,2} (S_{2,3} S_{1,2}) (S_{3,4} S_{2,3} S_{1,2}) \cdots
\ee
with $S_{i,j} = 1 + e_{i,j}$,
has Pascal's triangle embedded within it and is explicitly:
\begin{equation}\label{eq:a-cable-me}
M^{BA \to BC} =\tau(K) =  \sum_{i,j=1}^K \binom{K-i}{K-j} e_{i,j}.
\end{equation}
The matrices $M^{CB \to CA}$ and $M^{AC \to AB}$ are also identically given by
$\tau(K)$.
The proof of \eqref{eq:a-cable-me} is easily given by induction after noting that
\be
S_{K,K+1} S_{K-1,K} \cdots S_{1,2} = 1 + \sum_{j=1}^K e_{j,j+1}.
\ee

We can also compute
\be
W_0^{(K)} \tau(K) W_0^{(K)} = \sum_{i,j=1}^K \binom{i-1}{j-1} (-1)^{K+1-i-j} e_{i,j},
\ee
and now the monodromy identity
\be\label{eq:explct-r-1}
(W_0^{(K)} \tau(K) W_0^{(K)} ) \tau(K) (W_0^{(K)} \tau(K) W_0^{(K)} )  = (-1)^{K-1} W_0^{(K)}
\ee
boils down to an identity on sums of binomial coefficients:
\begin{equation}\label{eq:binid}
\sum_{s,t=0}^n \binom{n-s}{a} \binom{t}{s} \binom{b}{n-t} (-1)^{s+t} = \delta_{a+b,n} (-1)^a,
\end{equation}
for $a,b,n$ nonnegative integers.  Equation \eqref{eq:binid} is proved by defining
the left-hand side of \eqref{eq:binid} as $M(a,b)$ and then summing
$\sum_{a=0}^n x^a M(a,b)$ successively using the binomial theorem to
produce $(-1)^{n+b}x^{n-b}$.

\bibliographystyle{utphys}

\bibliography{snakes-paper}

\end{document}